\documentclass[fleqn,usenatbib]{mnras}

\usepackage{newtxtext,newtxmath}

\usepackage[T1]{fontenc}

\DeclareRobustCommand{\VAN}[3]{#2}
\let\VANthebibliography\thebibliography
\def\thebibliography{\DeclareRobustCommand{\VAN}[3]{##3}\VANthebibliography}

\usepackage{graphicx}	
\usepackage{amsmath}	
\usepackage{subcaption}

\newcommand{\vect}[1]{\boldsymbol{#1}}
\newcommand{\diff}{\text{d}}

\title[Non-linear tides in BNS]{
Beyond the linear tide: impact of the non-linear tidal response of neutron stars on gravitational waveforms from binary inspirals
}

\author[H. Yu et al.]{
Hang Yu$^{1}$\thanks{E-mail: hang.yu2@montana.edu}, 
Nevin N. Weinberg$^{2}$, 
Phil Arras$^{3}$,  
James Kwon$^{4}$ and
Tejaswi Venumadhav$^{4, 5}$ 
\\
$^{1}$Kavli Institute for Theoretical Physics, University of California at Santa Barbara, Santa Barbara, CA 93106, USA, \\
$^{2}$Department of Physics, University of Texas at Arlington, Arlington, TX 76019, USA, \\
$^{3}$Department of Astronomy, University of Virginia, P.O. Box 400325, Charlottesville, VA 22904, USA, \\
$^{4}$Department of Physics, University of California at Santa Barbara, Santa Barbara, CA 93106, USA, \\
$^{5}$International Centre for Theoretical Sciences, Tata Institute of Fundamental Research, Bangalore 560089, India.
}

\date{Accepted XXX. Received YYY; in original form ZZZ}

\pubyear{2022}

\begin{document}
\label{firstpage}
\pagerange{\pageref{firstpage}--\pageref{lastpage}}
\maketitle

\begin{abstract}
Tidal interactions in coalescing binary neutron stars modify the dynamics of the inspiral and hence imprint a signature on their gravitational wave (GW) signals in the form of an extra phase shift. 
We need accurate models for the tidal phase shift in order to constrain the supranuclear equation of state from observations.
In previous studies, GW waveform models were typically constructed by treating the tide as a linear response to a perturbing tidal field. 
In this work, we incorporate non-linear corrections due to hydrodynamic three- and four-mode interactions and show how they can improve the accuracy and explanatory power of waveform models.
We set up and  numerically solve the coupled differential equations for the orbit and the modes and analytically derive solutions of the system's equilibrium configuration. 
Our analytical solutions agree well with the numerical ones up to the merger and involve only algebraic relations, allowing for fast phase shift and waveform evaluations for different equations of state over a large parameter space. 
We find that, at Newtonian order, non-linear fluid effects can enhance the tidal phase shift by $\gtrsim 1\,{\rm radian}$ at a GW frequency of 1000 Hz, corresponding to a $10-20\%$ correction to the linear theory. 
The scale of the additional phase shift near the merger is consistent with the difference between numerical relativity and theoretical predictions that account only for the linear tide. 
Non-linear fluid effects are thus important when interpreting the results of numerical relativity and in the construction of waveform models for current and future GW detectors. 
\end{abstract}




\section{Introduction}

Neutron stars (NSs) are astrophysical laboratories for physics at extreme conditions. A NS in a coalescing binary driven by gravitational-wave (GW) radiation can be tidally deformed, and the deformation and the associated change in the binary dynamics leave imprints in the associated GW waveform. These effects have been incorporated into the analysis of GW170817, the first binary NS (BNS) event detected by GW observation~\citep{GW170817, GW170817prop}, and enabled valuable constraints on the supranuclear equation of state (EoS)~\citep{GW170817eos}. With more BNSs detected~\citep{GW190425} and even more to be expected, especially when future GW detectors like the Einstein Telescope~\citep{Hild:10, Sathyaprakash:12} and the Cosmic Explorer~\citep{Evans:17, Evans:21} become operational, it is imperative to develop more sophisticated theoretical waveform models to maximize the information we can extract. 

Tidal effects within BNS systems can, in principle, lead to rich phenomenology.
The dominant effect is the interaction between the tidal field and the fundamental mode (f-mode) of the NS, which characterizes the star's large-scale deformation of the NS. This was first studied in the adiabatic limit, i.e., assuming the tidal driving frequency is much smaller than the eigenfrequency of the f-mode~\citep{Lai:93, Lai:94a, Lai:94b, Flanagan:08, Bini:14, Bernuzzi:15}. In this limit, the tidal response can be well characterized by a single coefficient known as the Love number $k_2$~(or equivalently, the NS deformability $\Lambda =2 k_2 R^5/3$ with $R$ the radius of the NS; \citealt{Damour:09, Binnington:09, Hinderer:10}). However, the driving frequency can become comparable to the eigenfrequency of the f-mode near the merger, leading to important corrections to the tidal response due to finite-frequency effects~\citep{Hinderer:16, Steinhoff:16, Andersson:21}. If the NS spins significantly in a retrograde manner with respect to the orbit, the f-mode can even be resonantly excited~\citep{Ho:99, Ma:20, Steinhoff:21}. 

At a more detailed level, as the orbit decays due to GW radiation, the tide can resonantly excite gravity modes (\citealt{Reisenegger:94, Lai:94c, Yu:17a, Yu:17b, Kuan:21, Kuan:21b}) and interface modes~\citep{Tsang:12, Pan:20, Passamonti:21} within the neutron stars; if the stars are spinning about their individual axes, inertial modes can also be resonantly excited~\citep{Ho:99,Flanagan:07, Xu:17, Poisson:20, Ma:21, Gupta:21}.

An accurate waveform model incorporating these dynamics is essential to properly use the tidal signature in the data to constrain the EoS~\citep{Read:09, Damour:12, DelPozzo:13, Lackey:15, Andersson:18, Landry:19, Matas:20, Pratten:22}. If the NS EoS is known, BNS events can be further used to test the theory of general relativity (GR; \citealt{Saffer:21}) and probe the cosmological expansion history~\citep{Messenger:12}. 

Previous theoretical tidal models show good agreement with numerical relativity~\citep{Hotokezaka:15, Foucart:19} for most of the binary's inspiral; however, there is a discrepancy of about 1 radian between the phases of analytical and numerical waveforms near the final merger~\citep{Hinderer:16, Nagar:18, Steinhoff:21}
whose origin is yet unclear. Understanding the discrepancy would be of great theoretical interest for correctly interpreting the results of numerical relativity, and constraining the binary dynamics in the highly relativistic regime. Moreover, it would enable us to construct accurate waveform templates for efficient parameter space exploration and data analysis. 

\begin{figure}
   \centering
   \includegraphics[width=0.45\textwidth]{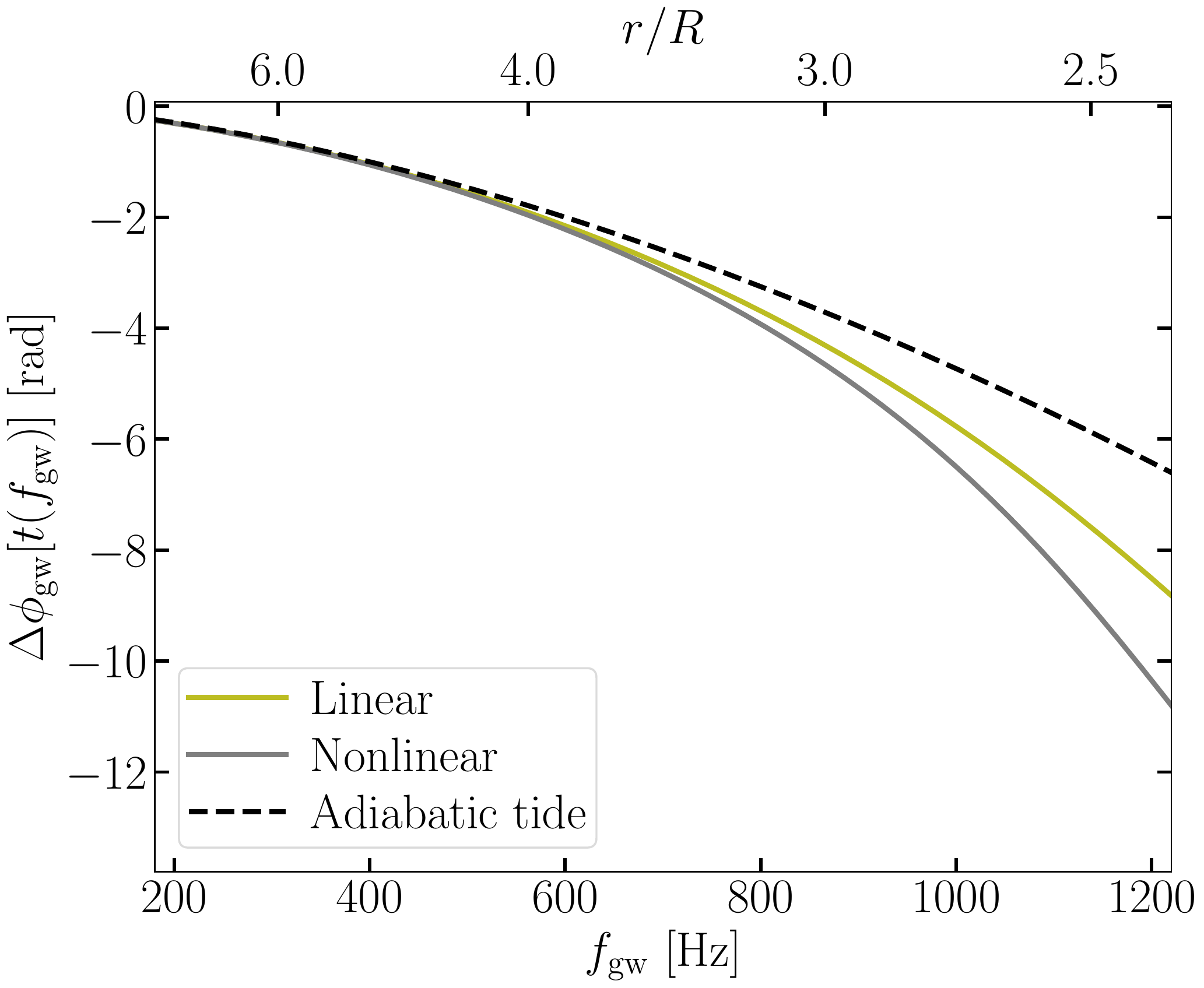} 
   \caption{Tidal phase shift due to a NS with $M=1.3\,M_\odot$, $R=12\,{\rm km}$, and $k_2=0.26$. The companion is assumed to be a point particle with $M'=M$. We show the result including just the linear tide (olive curve) and when including non-linear fluid effects (gray curve). As a reference, the phase shift in the adiabatic limit is shown in the black-dashed line using expressions in \citet{Hinderer:10}. 
   The non-linear tide corrects the phase by $\gtrsim 1\,{\rm rad}$ at $f_{\rm gw}>1000\,{\rm Hz}$ (at Newtonian order). This is consistent with the discrepancy between analytical results with linear tides and numerical relativity~\citep{Hinderer:16}. 
   }
   \label{fig:dPhi_evol}
\end{figure}

In this work, we investigate the effects of the non-linear tide including interactions among NS modes as well as between modes and the non-linear tidal driving. Our main result is illustrated in Fig.~\ref{fig:dPhi_evol}. Here we present the tidal phase shift of the GW waveform. The olive curve shows the result obtained using the linear dynamical tide theory and the gray curve includes the effects of the non-linear tide which we will examine in detail in this work. In particular, the non-linear tide creates an additional phase shift of $\mathcal{O}(1)\, {\rm rad}$ near the merger compared to the linear result, which is consistent with the discrepancy between previous theoretical and numerical works. This suggests that non-linear effects could be (at least in part) the cause of the discrepancy, and that they should be a key component of future waveform modeling.

We note that our result is greater than the prior estimate of  non-linear hydrodynamic corrections in \citet{Hinderer:10} for a few reasons. First, we perform a first-pinciple calculation of the non-linear coupling following \citet{Weinberg:12} and find that the coupling strength is greater than the estimation of \citet{Hinderer:10} (see later in Table~\ref{tab:coup_coeff}). Also, \citet{Hinderer:10} ignored the non-linear part of the tidally-induced NS mass quadrupole $Q_{\rm ns}^{ij}$ [see Eq.~(\ref{eq:Qlm})] and therefore did not account for the non-linear tidal driving [i.e., the term $\propto U_{ablm}$ in Eq.~(\ref{eq:ode_mode_amp_general})]. Lastly, the finite-frequency correction to the f-mode was underestimated in \citet{Hinderer:10} (though its significance was later realized in \citealt{Hinderer:16, Steinhoff:16}). Our study finds that the major non-linear hydrodynamic correction is in fact a shift to the f-mode's frequency, which strengthens the mode's finite-frequency response [see Eq.~(\ref{eq:C_ab_leading}) and Fig.~\ref{fig:domega_vs_dVa}]. These reasons explain the difference between our result and the estimate of \citet{Hinderer:10}. 

Before we proceed, we also note that the non-linear tide we investigate here is different from the non-linear pg-instability~\citep{Weinberg:13, Venumadhav:14, Weinberg:16, Essick:16, GW170817pg}. The pg-instability describes the coupling between the tide and high-order pressure and gravity modes and it modifies the orbital evolution by fluid dissipation. In contrast, our focus in this work will be on the interactions among low-order modes as well as their non-linear couplings with the tidal potential. The interaction is conservative when the GW radiation is ignored. 

In the rest of the paper, we will explain the details leading to Fig.~\ref{fig:dPhi_evol}. In particular, we will first introduce the equations governing the evolution of NS modes in Section~\ref{sec:dyn_modes}. Approximate solutions of the modes will be presented in Sections~\ref{sec:lin_sol} and \ref{sec:mode_nl} at the linear and non-linear orders, respectively. We will then describe the evolution of the orbit in Section~\ref{sec:dyn_orb} including both tidal back reaction and radiation reaction. This is followed by analytical solutions to the system's equilibrium configuration in Section~\ref{sec:eq_config}. Lastly, we conclude and discuss our results in Section~\ref{sec:discussion}. The appendices contain a section with a simple intuitive explanation of non-linear corrections to the tide (Appendix~\ref{appx:toy}), as well as important technical details relevant to the calculations (Appendices \ref{appx:anharm}-\ref{appx:kap4_del_Phi}). Throughout the paper, we will use geometrical units with $G=c=1$.

\section{Dynamics of the modes}
\label{sec:dyn_modes}

In this Section, we study the motion of perturbed fluid in a NS in terms of the NS's eigenmodes. We derive a set of differential equations governing the amplitude of each mode, that include leading-order non-linear interactions corresponding to three-mode and four-mode couplings. This set of equations can be  integrated numerically or solved analytically with approximations (Sections~\ref{sec:lin_sol} and \ref{sec:mode_nl}). When combined with the equations governing the orbit (Section~\ref{sec:dyn_orb}), we can obtain a complete description of the system.

Suppose $\vect{\xi}$ is the Lagrangian displacement of a tidally perturbed NS. We perform a phase space decomposition following~\citet{Schenk:02} as
\begin{equation}
   \begin{bmatrix} 
      \vect{\xi} (\vect{x}, t) \\
      \dot{\vect{\xi}}(\vect{x}, t)  \\
   \end{bmatrix}
=\sum_a c_a(t) 
  \begin{bmatrix} 
      \vect{\xi}_a (\vect{x}) \\
      -i \omega_a \vect{\xi}_a(\vect{x})  \\
   \end{bmatrix},
   \label{eq:expansion}
\end{equation}
where a mode $a$ is labeled by two angular quantum numbers $(l_a, m_a)$ (for its angular pattern governed by the spherical harmonic $Y_{l_a m_a}$), one radial order $n_a$ (with $n_a=0$ for the f-mode and $n_a>0$ for p-modes), and a sign of its eigenfrequencies (either positive or negative). Following~\citet{Schenk:02}, one can show that a mode with $(-\omega_a, -m_a)$ is the complex conjugate of the mode with $(\omega_a, m_a)$. When we discuss a mode in the subsequent text, we will restrict to the positive-frequency one if we do not explicitly mention its sign. In our equations, on the other hand, the summations run over all the modes including both signs of eigenfrequencies. We will add $\omega_a>0$ above the summation symbol if we explicitly pair a mode and its complex conjugate first and then restrict the summation over only half of the modes.

Consider the leading order non-linear effect, the conservative part of the amplitude equation of a mode $a$ is (\citealt{Weinberg:12, Venumadhav:14, Weinberg:16})
\begin{align}
&\dot{c}_a + i \omega_a  c_a 
	= i \omega_a \left[U_a \nonumber \right. \\
	&\left. + \sum_{b,lm} U_{ablm}^\ast c_{b}^\ast + \sum_{bc}\kappa_{a b c} c_b^\ast c_c^\ast
	+\sum_{bcd}\eta_{abcd}c_b^\ast c_c^\ast c_d^\ast 
	\right], \label{eq:ode_mode_amp_general}
\end{align}
where the left hand side describes a harmonic oscillator and the right hand side various driving terms which will describe in detail shortly. 
In principle, there will also be a four-mode counterpart to $U_{ablm}$, yet we argue in Appendix~\ref{appx:anharm} that it should be subdominant and would not significantly modify the results obtained in this study. 
To obtain the above equation, we normalize each mode so that
\begin{equation}
    2\omega_a^2 \int \diff^3x \rho \vect{\xi}_a^\ast\cdot \vect{\xi}_b=\delta_{ab}E_0,
\end{equation}
where $E_0=M^2/R$. 
Dissipation due to, e.g., Urca reactions is estimated to be small~\citep{Arras:19, Alford:21} and hence ignored in our analysis (but see Section~\ref{sec:dyn_orb} and Appendix~\ref{appx:quadrupole} for the damping due to GW radiation). 
In particular, the term $U_a$ describes the linear tidal driving and  is given by
\begin{equation}
    U_a=\frac{M'}{M}W_{l_a m_a}I_{alm}\left(\frac{R}{r}\right)^{l_a+1}e^{-im_a\phi}
    =V_a e^{-im_a\phi},
\end{equation}
where $r$ and $\phi$ are respectively the orbital separation and phase, and $W_{lm}=4\pi(2l+1)^{-1}Y_{lm}(\pi/2, 0)$. For the $l=2$ tide, the non-zero values of $W_{lm}$ are $W_{22}=W_{2-2}=\sqrt{3\pi/10}$ and $W_{20}=-\sqrt{\pi/5}$. We used for the spatial coupling $I_{alm}=I_a \delta_{l l_a}\delta_{m m_a}$, where $I_{a}$ is the linear tidal coupling coefficient (also known as the tidal overlap), 
\begin{equation}
    I_a = \frac{1}{MR^{l_a}}\int \diff^3 x \rho \vect{\xi}_a^\ast \cdot \nabla (r^{l_a} Y_{l_a m_a}).
\end{equation}
It is evaluated using eq. (A15) in \citet{Weinberg:12}. 
Under the adiabatic limit $\Omega\equiv\dot{\phi}\ll \omega_a$, $I_a$ of the $l_a=2$ f-mode is related to the Love number $k_2$ by (Appendix~\ref{appx:quadrupole}; see also, e.g., \citealt{Andersson:20, Passamonti:22})
\begin{equation}
    k_2 \simeq \frac{4\pi}{5} I_a^2.
    \label{eq:love_vs_Ia_ad}
\end{equation}
where we ignore other modes' contributions to the Love number since they are negligible for a NS. 
The $U_{ablm}$ term is due to non-linear tidal driving. We write
\begin{equation}
    U_{ablm} = \frac{M'}{M}W_{lm}J_{ablm}\left(\frac{R}{r}\right)^{l+1}e^{-im\phi}
    =V_{ablm} e^{-im\phi}
\end{equation}
where the coefficient $J_{ablm}$ is defined as 
\begin{equation}
    J_{ablm} = \frac{1}{MR^l}\int \diff^3 x \rho \vect{\xi}_a \cdot \left(\vect{\xi}_b\cdot \nabla\right)\nabla (r^l Y_{lm}), 
\end{equation}
and we compute it numerically according to eq. (A23) of \citet{Weinberg:12}.
The $\kappa_{abc}$ term describes the coupling between three eigenmodes of the star and it is computed according to eqs. (A55)-(A62) of \citet{Weinberg:12}. 
Lastly, the $\eta_{abcd}$ term describes the four-mode coupling and its computation is described by appendix C in \citet{Weinberg:16}. For f-modes, the perturbed gravity is significant and the Cowling approximation~\citep{Cowling:41} should not be adopted. We derive in Appendix~\ref{appx:kap4_del_Phi} the additional contributions to $\eta_{abcd}$ due to terms involving perturbed gravity. 
We find they can modify the results of \citet{Weinberg:16} made under the Cowling approximation by 70\% for f-modes.

We present numerical values for key coupling coefficients in Table~\ref{tab:coup_coeff}.\footnote{Note that when normalizing the eigenfunction of each mode by a constant, we have a freedom in choosing the sign. Changing the sign convention will change the signs of $I_a$ and $\kappa_{abc}$, yet the physical results (see, e.g., Section~\ref{sec:mode_nl} below) will not be affected because they depends on the product $\kappa_{abc}I_a$. When $J_{ablm}$ appears alone, it will be due to a mode coupling with its complex conjugate, so it is not affected by the choice of the normalization sign, either.} In our study, we assume the background NS is described by a $P\propto \rho^\Gamma$ polytrope with $\Gamma=2$. We set its mass to $M=1.3\,M_\odot$ and radius to $R=12\,{\rm km}$, corresponding to a compactness $M/R=0.16$. Other natural units of this model are $E_0=3.7\times10^{53}\,{\rm erg}$ and $\omega_0/2\pi = 1.6\times10^{3}\,{\rm Hz}$ where $\omega_0^2\equiv M/R^3$. 
The modes are computed using the stellar oscillation code \texttt{GYRE}~\citep{Townsend:13, Townsend:18}. 
We further assume the background NS is non-spinning, neutrally stratified (with the Brunt-V{\"a}is{\"a}l{\"a} frequency $\mathcal{N}=0$, i.e.,  no g-modes in our model), and under \emph{Newtonian} hydrostatic equilibrium. While we include the quadrupole GW radiation, other GR effects will be ignored in the current study for simplicity. 

For future convenience, we introduce $C_a=c_a\exp[i m_a \phi]$. Prior to resonance, $c_a$ oscillates at the same rate as the driving potential at $m_a\Omega$, where $\Omega = \pi f_{\rm gw}$ with $f_{\rm gw}$ the GW frequency. Thus by using $C_a$, we factor out the fast-oscillating part of $c_a$ and the remaining temporal changes are due to the GW-driven orbital decay only. 

In our numerical calculations, we include the $l=2$, $m=0, \pm 2$ f-modes which dominate the linear tidal responses. The p-modes have little contribution to the result because of their small overlap with the tidal potential (Table~\ref{tab:coup_coeff} compares the f-mode and the $n_a=1$ p-mode). When computing the non-linear tidal driving $U_{ablm}$, we focus on the contributions from $l=2$.
We further include the first $l=0$ (radial) mode and the $l=4$, $m=0,\pm2,\pm4$ f-modes which can couple with a pair of $l=2$ f-modes via the three-mode coupling channel. They are critical in determining the anharmonic frequency shift of an $(l, m)=(2, 2)$ \emph{free} oscillator~\citep{Yu:21, Yu:22a} together with the four-mode couplings among the $l=2$ modes. Yet as we will see in the later discussions, because the $l=2$ modes are \emph{continuously forced} by the tidal potential (thus not freely oscillating) in our case, the leading-order non-linear correction comes from their mutual couplings and the anharmonic effect of a free oscillator is small.  An $l=3$, $|m|=1 \text{ or } 3$ mode with $W_{3m}\neq0$ cannot couple with a pair of $l=2$ modes as it violates the angular selection rule~\citep{Weinberg:12} and therefore does not contribute to the non-linear tide at the leading order. Since the linear $l=3$ tide has been well studied~\citep{Hinderer:16}, we ignore it here for simplicity. 

\begin{table}
 \caption{Coupling coefficients of the $\Gamma=2$ polytrope model assumed in our study. The $l=2$ f-modes in our model have $\omega_a=1.2\omega_0\simeq 1.95\times10^3 \,{\rm Hz}$. }
 \label{tab:coup_coeff}
 \begin{tabular}{cccc}
  \hline
  Quantity & $(l_a, m_a)$ & $n_a$ & Value \\
  \hline
  $I_a$ & (2, $\pm2$ or 0) & 0 & 0.32 \\
        & (2, $\pm2$ or 0) & 1 & $-5.5\times 10^{-3}$ \\
  \hline
  $J_{ablm}$ & (2,+2), (2,-2), (2,0) & 0, 0, tide & -0.21(=$J_2$) \\
             & (2,0),  (2,0),  (2,0) & 0, 0, tide & 0.21(=$J_0$) \\
             & (2,+2), (2,-2), (2,0) & 0, 1, tide & $1.2\times 10^{-3}$ \\
  \hline
  $\kappa_{abc}$ & (2,+2), (2,-2), (2, 0) & 0, 0, 0 & -0.45(=$\kappa_2$) \\
                 & (2, 0), (2, 0), (2, 0) & 0, 0, 0 & 0.45(=$\kappa_0$) \\
                 & (2,+2), (2,-2), (2, 0) & 0, 0, 1 & -0.04 \\
                 & (2,+2), (2,+2), (4,-4) & 0, 0, 0 & 0.19 \\
                 & (2,+2), (2,-2), (4, 0) & 0, 0, 0 & 0.02 \\
                 & (2,+2), (2,-2), (0, 0) & 0, 0, 1 & 0.88 \\
                 & (2, 0), (2, 0), (4, 0) & 0, 0, 0 & 0.13 \\                 
                 & (2, 0), (2, 0), (0, 0) & 0, 0, 1 & 0.88 \\
  \hline
  $\eta_{abcd}$ & (2,+2), (2,+2), (2,-2), (2,-2) & 0,0,0,0 & -1.75(=$\eta_{22}$) \\
                  & (2,+2), (2,-2), (2, 0), (2, 0) & 0,0,0,0 & -0.89(=$\eta_{20}$) \\
                  & (2, 0), (2, 0), (2, 0), (2, 0) & 0,0,0,0 & -0.89 \\
  \hline
 \end{tabular}
\end{table}

\subsection{Linear solution}
\label{sec:lin_sol}

Let $b_a$ be the solution of the linear problem and define $B_a {=} b_a \exp[i m_a \phi]$. By Eq.~(\ref{eq:ode_mode_amp_general}), the equation for $B_a$ is given by 
\begin{equation}
    \dot{B}_a + i(\omega_a - m_a \Omega)B_a = i\omega_a V_a.
    \label{eq:ode_Ba}
\end{equation}
Since $B_a$ varies slowly in time prior to resonance (i.e., when $m_a \Omega < \omega_a$), we can obtain a zeroth order solution by ignoring the $\dot{B}_a$ term, which leads to 
\begin{equation}
    B_a^{(0)}=\frac{\omega_a}{\omega_a-m_a\Omega}V_a.
    \label{eq:Ba_0}
\end{equation}
The zeroth order solution is then plugged back into Eq.~(\ref{eq:ode_Ba}) to obtain the first order correction $B_a^{(1)}$. 
Again dropping the $\dot{B}_a^{(1)}$ term, we have
\begin{align}
    B_a^{(1)} &= \frac{i}{(\omega_a-m_a \Omega)}\dot{B}_a^{(0)} \nonumber \\
    &= \frac{i\omega_a}{(\omega_a-m_a \Omega)^2} \left[\frac{2}{3}(l+1)+\frac{m_a\Omega}{\omega_a-m_a\Omega}\right]\frac{\dot{\Omega}}{\Omega} V_a. 
    \label{eq:Ba_1}
\end{align}
The solution of $B_a$ is thus obtained as $B_a=B_a^{(0)} + B_a^{(1)}+...$. 



\subsection{Including non-linear effects}
\label{sec:mode_nl}

We now consider a system including the leading-order non-linear corrections. In this subsection, we will let mode $(a, b, c)$ respectively have $(l_a, m_a)=(2, 2)$, $(l_b, m_b)=(2, -2)$, and $(l_c, m_c)=(2, 0)$. In other words, we let mode $a$ ($b$) be the prograde (retrograde) mode specifically in this section. We consider their mutual couplings as well as the coupling with the $l=2$ tidal potential (via the $U_{ablm}$ term). This set of interactions covers all the non-linear corrections to the linear mode amplitude formally at the $(R/r)^3$ order. 
To make the problem explicit, we  write out all the allowed couplings 
\begin{align}
    \dot{C}_a + i(\omega_a - m_a\Omega) &C_a = i\omega_a [V_a + V_{aa^\ast20} C_a \nonumber \\
    +& V_{ab20} C_b^\ast  
    + V_{ac2{-}2} C_c^\ast + V_{ac^\ast 2{-}2} C_c\nonumber \\
    +& 2\kappa_{aa^\ast c}C_a C_c^\ast + 2\kappa_{aa^\ast c^\ast}C_a C_c \nonumber \\
    +& 2\kappa_{abc}C_b^\ast C_c^\ast + 2\kappa_{abc^\ast}C_b^\ast C_c],
    \label{eq:ode_q_a_l2_only_a}
    \\
    \dot{C}_b + i(\omega_b - m_b \Omega) &C_b = i\omega_b [V_b + V_{ab20} C_a^\ast \nonumber \\
    +& V_{bb^\ast20} C_b
    + V_{bc22} C_c^\ast + V_{bc^\ast 22} C_c\nonumber \\
    +& 2\kappa_{abc}C_a^\ast C_c^\ast + 2\kappa_{ab c^\ast}C_a^\ast C_c \nonumber \\
    +& 2\kappa_{bb^\ast c}C_b C_c^\ast + 2\kappa_{bb^\ast c^\ast}C_b C_c],
    \label{eq:ode_q_a_l2_only_b}
    \\
    \dot{C}_c + i\omega_c  C_c = i\omega_c &[V_c + V_{ac2{-}2} C_a^\ast + V_{a^\ast c22} C_a   \nonumber \\
    +& V_{bc22} C_b^\ast + V_{b^\ast c2{-}2} C_b
    + V_{cc^\ast20}C_c + V_{cc20}C_c^\ast\nonumber \\
    +& 2\kappa_{aa^\ast c}C_a C_a^\ast + 2\kappa_{bb^\ast c}C_b C_b^\ast  \nonumber \\
    +& 2\kappa_{abc}C_a^\ast C_b^\ast + 2\kappa_{a^\ast b^\ast c} C_a C_b \nonumber \\
    +& \kappa_{ccc}C_c^\ast C_c^\ast + \kappa_{cc^\ast c^\ast} C_c C_c + 2\kappa_{ccc^\ast}C_c C_c^\ast
    \label{eq:ode_q_a_l2_only_c}
    ].
\end{align}
Furthermore, we note that both $\kappa_{abc}$ and $J_{ablm}$ are symmetric with respect to permutations of the mode indices. Moreover, the $l=2$ modes have the same eigenfrequency ($\omega_a=\omega_b=\omega_c>0$) and the same reduced eigenfunction (after separating out the angular part described by each mode's specific spherical harmonic). We thus have
\begin{align*}
   & V_a=V_b = V_2, \quad V_c = V_0; \\
   & V_{aa^\ast 20}=\text{similar terms} = V_{20}; \\
   & V_{ac2{-}2} = \text{similar terms} =V_{22}; \\
   & V_{cc20} = \text{similar terms} = V_{00}; \\
   & J_{ab20} = J_{ac2-2} = \text{similar terms} = J_2;\\
   & J_{cc20} = \text{similar terms} = J_0;\\
   & \kappa_{abc}=\text{similar terms}=\kappa_2; \\
   & \kappa_{ccc}=\text{similar terms}=\kappa_0.
\end{align*}
Numerically, we see from Table~\ref{tab:coup_coeff} that
\begin{equation}
    J_2=-J_0 = -0.21 \text{ and } \kappa_2=-\kappa_0=-0.45.
    \label{eq:coup_coeff_values}
\end{equation}
The difference between $J_2$ and $J_0$ (and similarly between $\kappa_2$ and $\kappa_0$) is purely due to the angular overlap. Therefore, $J_2=-J_0$ and $\kappa_2=-\kappa_0$ hold independent of the choice of EoS. 

The non-linear terms in, e.g.,  the right hand side of Eq.~(\ref{eq:ode_q_a_l2_only_a}) have two effects. The terms containing $C_a$ correspond to an effective shift of the mode's eigenfrequency $\omega_a$ while terms that are independent of $C_a$ modify the driving potential. Using the linear solutions obtained in Section~\ref{sec:lin_sol}, we can define the leading-order frequency shift\footnote{Note the partial cancellation between $J_{ablm}$ and $2\kappa_{abc}I_c$ described in section 5.2 of \citet{Weinberg:12} is not significant in our case. The partial cancellation arises when using method 2 in section 2.1.2 of \citet{Weinberg:12}, or considering the linear tide coupling with two eigenmodes. In this description, the inhomogeneous piece of the linear tide will lead to an extra piece in the three-mode coupling that cancels the contribution to $J_{ablm}$ from the horizontal mode displacements [the term containing $a_h b_h$ in eq. (A23) of \citet{Weinberg:12}]. This cancellation is significant when considering the coupling with high-order g-modes whose displacements are predominantly horizontal but less significant for the coupling with f-modes whose motions are mainly radial. 
Moreover, we adopt method 1 in section 2.1.1 of \citet{Weinberg:12} and describe the tide in terms of eigenmodes. This avoids the inhomogeneous piece in the coupling coefficient yet we will be subject to truncation errors due to ignoring high-order ($|n_a|\geq 1$) modes. Nevertheless, from Table~\ref{tab:coup_coeff} we see that $I_a$, $J_{ablm}$, and $\kappa_{abc}$ are all strongly dominated by the f-modes (one or two orders of magnitudes above the values involving p-modes). The g-modes, when present, also have small contributions to the tidal response (see, e.g., \citealt{Lai:94c}). Consequently, the truncation error due to ignoring high-order modes is expected to be small.  } 
\begin{align}
    \frac{\Delta \omega_a}{\omega_a} &= -V_{20} - 4\kappa_2 {\rm Re}\left[B_c\right], \nonumber \\
    &=\sqrt{\frac{\pi}{5}}\left(J_2 + 4\kappa_2 I_a \right) R^3 \frac{M'}{M} \frac{\Omega^2}{M_{\rm t}},  
    \label{eq:domega_a_leading}
\end{align}
where $M_{\rm t}=M'+M$. Note $\Delta \omega_a/\omega_a <0$ (Table~\ref{tab:coup_coeff}). The origin of this frequency shift and its sign can be understood with a toy model described in Appendix~\ref{appx:toy}. Meanwhile, the modifications of the driving forces are 
\begin{align}
    &\Delta V_a = V_{20} B_b^\ast + 2V_{22}{\rm Re}[B_c] + 4\kappa_2 {\rm Re}[B_c] B_b^\ast, \nonumber \\
               &\quad \simeq -\frac{\pi}{5} \sqrt{\frac{3}{2}}\left[2 J_2 I_a + \frac{\omega_a}{\omega_a + 2\Omega}\left(J_2 + 4 \kappa_2 I_a\right)I_a\right]
                  R^6 \frac{M'^2}{M^2}\frac{\Omega^4}{M_{\rm t}^2}, \\
    &\Delta V_b = V_{20} B_a^\ast + 2V_{22}{\rm Re}[B_c] + 4\kappa_2 {\rm Re}[B_c] B_a^\ast, \nonumber \\
               &\quad \simeq -\frac{\pi}{5} \sqrt{\frac{3}{2}}\left[2 J_2 I_a + \frac{\omega_a}{\omega_a - 2\Omega}\left(J_2 + 4 \kappa_2 I_a\right)I_a\right]
                  R^6 \frac{M'^2}{M^2}\frac{\Omega^4}{M_{\rm t}^2}, \\
    &\Delta V_c = 2 {\rm Re} \left[V_{22} B_a + V_{22}B_b + V_{00}B_c + \right. \nonumber \\
                &\quad \quad +\left. \kappa_2 |B_a|^2 + \kappa_2 |B_b|^2 + 2 \kappa_2 B_a B_b + 2 \kappa_0 |B_c|^2\right] \nonumber \\
                &\quad \simeq 2\left[
                  \frac{3\pi}{10} \frac{2\omega_a^2}{\omega_a^2 - 4\Omega^2} I_a 
                    \left(J_2 + \frac{2\omega_a^2}{\omega_a^2 - 4\Omega^2} \kappa_2 I_a\right)
                    \right.
                    \nonumber \\
                &   \hspace{0.8cm}\left. 
                + \frac{\pi}{5} I_a \left(J_0 + 2\kappa_0 I_a\right)
                \right]
                R^6\frac{M'^2}{M^2}\frac{\Omega^4}{M_{\rm t}^2}.
                \label{eq:dV_c_leading}
\end{align}
We can thus obtain the amplitude of each mode as
\begin{align}
    C_{a, b} &= \frac{\omega_a}{\omega_a + \Delta \omega_a - m_{a, b} \Omega} \left(V_2 + \Delta V_{a, b}\right),
    \label{eq:C_ab_leading}
    \\
    C_c &= V_0 + \Delta V_c, \label{eq:C_c_leading}
\end{align}
with the linear potentials given by
\begin{equation}
    V_{2(\text{or } 0)}  = W_{22(\text{or } 20)} \frac{M'}{M} I_a R^3 \frac{\Omega^2}{M_t}.
\end{equation}
In the equations above, we have used the point-particle (PP) Keplerian orbit to replace $r^3$ by $M_{\rm t}/\Omega^2$. As we will see in Section~\ref{sec:eq_config}, the errors introduced by using the Keplerian orbit are of higher order than the leading-order non-linear tide we consider here and can thus be dropped. 

The above equations show that to get the leading-order non-linear corrections due to hydrodynamics, we only need to compute two additional EoS-dependent coupling coefficients, $\kappa_2$ and $J_2$ [see the discussion below Eq.~(\ref{eq:coup_coeff_values})], which describe, respectively, the coupling between three f-modes and the coupling between two f-modes and the tidal driving potential. They can be determined from an isolated NS, similar to the determination of $I_a$ (or effectively, the Love number $k_2$). Once the coupling coefficients are known, we can then express the mode amplitude in terms of the orbital frequency $\Omega$, allowing them to be easily evaluated with algebraic relations only. 

\begin{figure}
   \centering
   \includegraphics[width=0.45\textwidth]{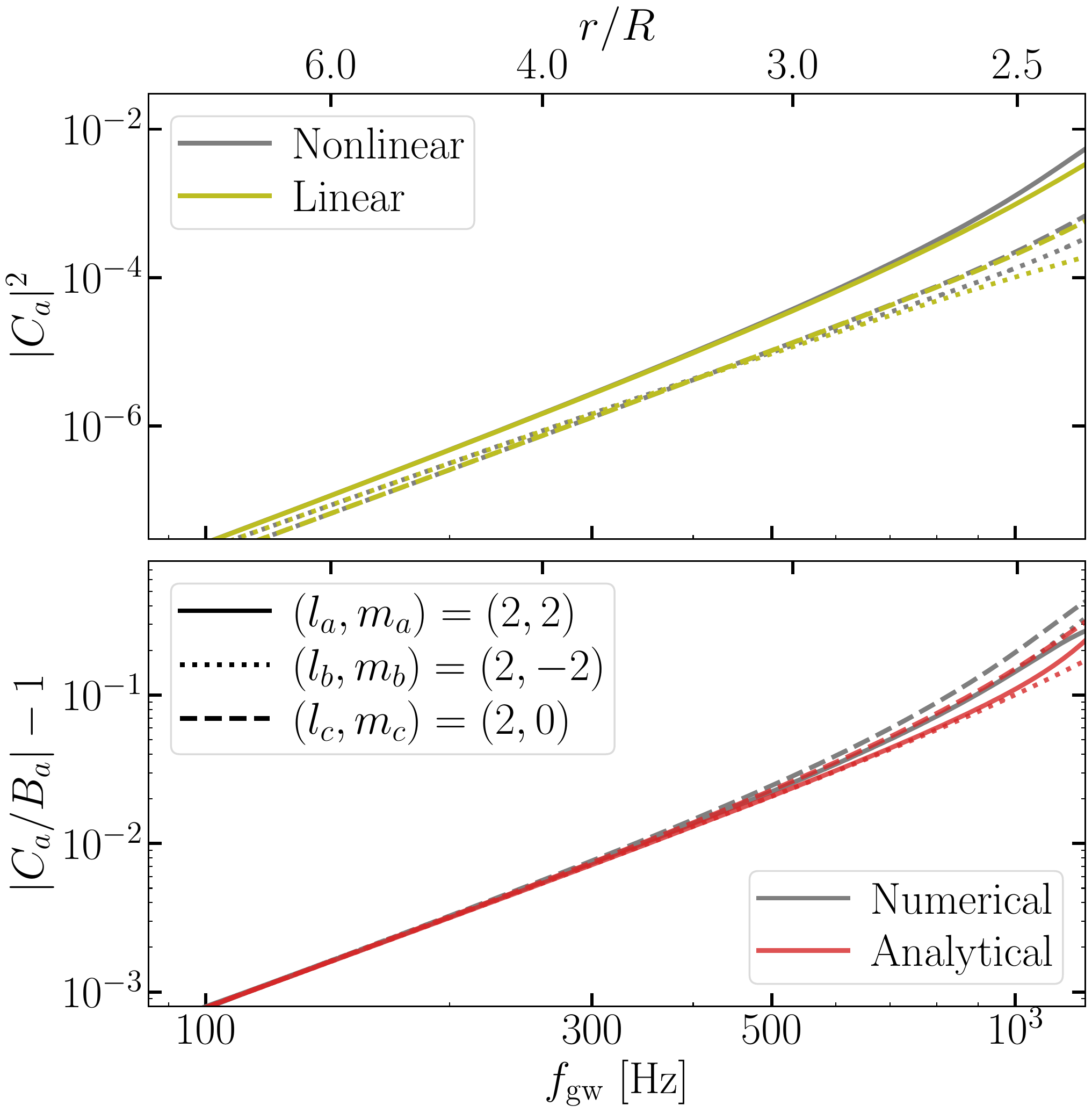} 
   \caption{Top: energy of each mode in units of $E_0$ as a function of the GW frequency.
   Bottom: fractional corrections to the mode amplitude due to non-linear effects. 
   The analytical approximation (red curves) obtained from Eqs.~(\ref{eq:domega_a_leading})-(\ref{eq:C_c_leading}) shows good agreement with the numerical solution (gray curves). 
   }
   \label{fig:mode_vs_freq}
\end{figure}

Fig.~\ref{fig:mode_vs_freq} shows that the analytical approximation computed using Eqs.~(\ref{eq:domega_a_leading}) to (\ref{eq:C_c_leading}) is in good agreement with the full numerical solution to the differential equations [Eq.~(\ref{eq:ode_mode_amp_general}) coupled to Eqs. (\ref{eq:ddr}) and (\ref{eq:ddphi})] up to the merger defined as where the NS's perturbed surface would contact the companion ($r\simeq 2.3R$ or $f_{\rm gw}\simeq 1250\,{\rm Hz}$). 
The simple expressions we obtained under only one iteration of perturbation holding well is a consequence of the fact that the next-order perturbation from the three-mode terms cancels partially with the four-mode couplings included in Eq.~(\ref{eq:ode_mode_amp_general}). 
Without this cancellation, the system would evolve into an unphysical amplitude instability as described in appendix D of \citet{Wu:98}. 
We will illustrate this point further in~Appendix~\ref{appx:anharm}.

\begin{figure}
   \centering
   \includegraphics[width=0.45\textwidth]{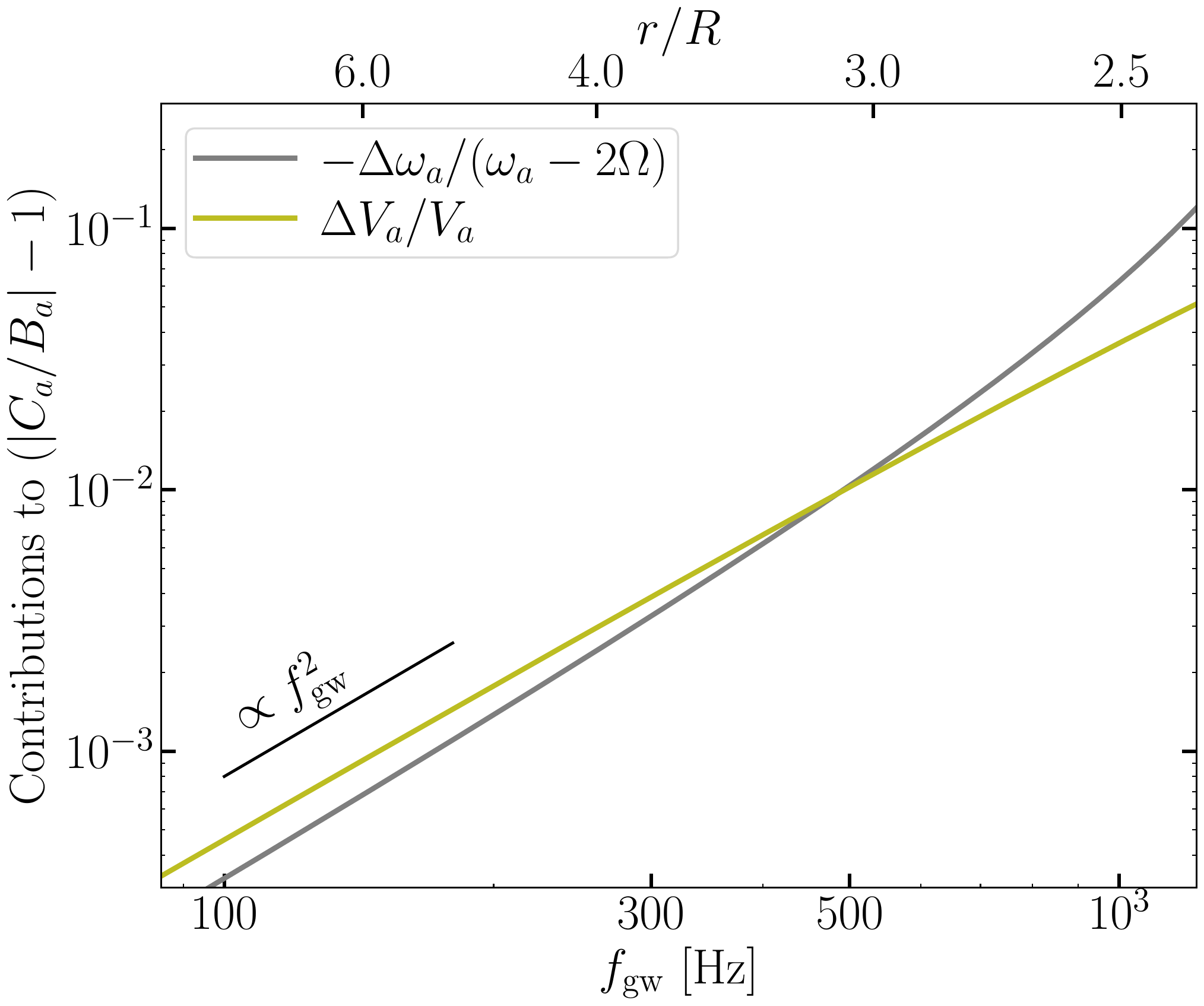} 
   \caption{Comparison of different non-linear corrections to the mode amplitude [Eq.~(\ref{eq:frac_nl_corr_mode_amp})]. For the $l_a=m_a=2$, the dominant correction near the merger comes from the non-linear frequency shift $\Delta \omega_a$. As the frequency is lowered by non-linear interactions, the finite-frequency response of the mode is amplified, allowing the non-linear correction to be greater than $(R/r)^3\propto f_{\rm gw}^2$. 
   }
   \label{fig:domega_vs_dVa}
\end{figure}

It is interesting to note that the fractional correction to the mode amplitude, 
\begin{equation}
    \frac{C_a}{B_a}-1 \simeq -\frac{\Delta \omega_a}{\omega_a - m_a \Omega} + \frac{\Delta V_a}{V_a},
    \label{eq:frac_nl_corr_mode_amp}
\end{equation}
is greater than $(R/r)^3\propto f_{\rm gw}^2$ because of the finite frequency response of the modes. This is illustrated in Fig.~\ref{fig:domega_vs_dVa}. For the $|m|=2$ modes (i.e. $a$ and $b$), the main non-linear correction is the shift of the mode frequency towards lower values [gray curve in Fig.~\ref{fig:domega_vs_dVa}; see also Eq.~(\ref{eq:domega_a_leading}) and Table~(\ref{tab:coup_coeff}), as well as the toy model in Appendix~\ref{appx:toy}]. 
The finite-frequency effect makes it greater than the correction to the driving potential $\propto \Delta V_a/V_a$ (olive curve) near the merger. Since we will need to sum over modes to get the physical tidal correction [see, e.g., Eq.~(\ref{eq:r_vs_omega})], it is also convenient to write the correction to the sum of the $|m|=2$ modes as (focusing on the $\Delta \omega_a$ term)
\begin{equation}
    \frac{C_a + C_b}{B_a+B_b} - 1 \simeq \frac{-2\omega_a^2}{\omega_a^2 - 4 \Omega^2} \frac{\Delta \omega_a}{\omega_a}. 
\end{equation}
For mode $c$, the correction is also amplified by the finite frequency response as $\Delta V_c$ contains terms like $B_a + B_b$ and $|B_a|^2 + |B_b|^2$. See Eq.~(\ref{eq:dV_c_leading}). 

We note that the frequency shift, Eq.~(\ref{eq:domega_a_leading}) is different from the standard anharmonic behavior of a \emph{free} oscillator where the frequency shift is proportional to the energy  the mode (or $\propto f_{\rm gw}^4$ in the adiabatic limit; \citealt{Landau:82, Kumar:94, Yu:21}). The anharmonicity originates from the oscillating mode deforming the background star~\citep{Lai:96}, which then creates a frequency shift of the mode. 
In contrast,  in Eq.~(\ref{eq:domega_a_leading}) the frequency shift goes as the amplitude (of a different mode) instead of the energy because of the continuous tidal forcing, and its origin can be understood from an intuitive toy model we present in Appendix~\ref{appx:toy}. 
While in our case we find the anharmonic effect to be subdominant, it could nonetheless be significant if the $l=m=2$ f-mode has a significantly greater amplitude than the other $l=2$ f-modes due to, e.g., a strong resonance with the orbit. The resonant excitation of the $m=2$ mode could be possible if the NS is rapidly spinning~\citep{Ma:20, Steinhoff:21} or if the orbit is eccentric~\citep{Chirenti:17, Parisi:18, Yang:18, Yang:19, Vick:19, Wang:20}. Thus we also demonstrate the appearance of the standard anharmonic frequency shift in the modal picture we adopt in this study in Appendix~\ref{appx:anharm}.

\section{Dynamics of the orbit}
\label{sec:dyn_orb}

Having described the evolution of the eigenmodes in the previous section, we now turn to the dynamics of the orbit including the effects due to tidal back-reactions and GW radiation.

The orbital evolution can be computed by (see, e.g., \citealt{Flanagan:07})
\begin{align}
    &\ddot{r} - r\dot{\phi}^2 + \frac{(M+M')}{r^2} = g_{r}, \label{eq:ddr}\\
    &r\ddot{\phi} + 2\dot{r}\dot{\phi} = g_{\phi}, \label{eq:ddphi}
\end{align}
where $g_{r}=g_{r}^{\rm (tide)} + g_{r}^{\rm (gw)}$ describes the radial acceleration acting on the orbit. It contains a conservative part due to the tidal back-reaction, $g_r^{\rm (tide)}$ and a dissipative part due to GW radiation, $g_r^{\rm (gw)}$. The tangential part, $g_\phi$, can be decomposed in a similar way. 

To derive the tidal back-reactions, we start from the interaction Hamiltonian given by~\citep{Weinberg:12, Yu:20a}
\begin{align}
    H_{\rm int} = -E_0 &\sum_{lm}\left[\sum_{a}^{\omega_a>0}(U_a c_a^\ast + U_a^\ast c_a) \right. \nonumber \\
   &\left. +\frac{1}{2}\sum_{ab}^{\omega_{a}>0}(U_{ab}c_ac_b + U_{ab}^\ast c_a^\ast c_b^\ast )\right].
   \label{eq:H_int}
\end{align}
Note that we explicitly write out mode $a$ and its complex conjugate $a^\ast$, so the summation runs over only modes with positive frequencies. In the non-linear tide term, the summation of mode $b$ still runs over both signs of frequency. 

From the Hamiltonian, we can derive the radial and tangential acceleration exerted by the mode on the orbit, 
\begin{align}
    &g_r^{\rm (tide)} = - \frac{1}{\mu} \frac{\partial H_{\rm int}}{\partial r}=-\frac{E_0}{\mu r} \nonumber \\
    &\times \sum_{lm}(l+1)
    \left[ 2\sum_{a,\omega_a>0}^{m_a=m}{\rm Re}\left(U_a c_a^\ast \right)
    +\sum_{ab, \omega_a>0}^{m_a+m_b=-m} {\rm Re}\left(U_{ab}c_ac_b\right)\right],
    \label{eq:ar_tide} \\
    &g_\phi^{\rm (tide)} = - \frac{1}{\mu r} \frac{\partial H_{\rm int}}{\partial \phi}=\frac{E_0}{\mu r}\nonumber \\
    &\times \sum_{lm} m 
    \left[2\sum_{a, \omega_a>0}^{m_a=m}{\rm Im}\left(U_a c_a^\ast \right)
    +\sum_{ab,\omega_{a>0}}^{m_a+m_b=-m} {\rm Im}\left(U_{ab}c_ac_b\right)
    \right].
    \label{eq:aphi_tide}
\end{align}


The $g_{r(, \phi)}^{\rm (gw)}$ terms describes the Burke-Throne dissipation and they can be computed by (using tensor notations in a Cartesian coordinate with Einstein summation; \citealt{Poisson:14})
\begin{equation}
    g_{\rm (gw)}^i = -\frac{2}{5} r_j \frac{\diff^5}{\diff t^5} Q_{\rm tot}^{\langle ij \rangle},
\end{equation} 
where $r_j$ is the displacement vector of the orbit and $Q_{\rm tot}^{\langle ij \rangle}$ is the total mass quadrupole of the system, $Q_{\rm tot}^{\langle ij \rangle}=Q_{\rm orb}^{\langle ij \rangle} + Q_{\rm ns}^{\langle ij \rangle}$. The angular bracket denotes a symmetric, trace-free (STF) tensor. In other words, $Q_{\rm tot}^{\langle ij \rangle}$ is the linear sum of the orbital quadrupole, $Q_{\rm orb}^{\langle ij \rangle}=\mu r^i r^j - \mu r^2 \delta^{ij}/3$, and the NS quadrupole $Q_{\rm ns}^{\langle ij \rangle}$ (see Appendix \ref{appx:quadrupole}). 
In the point-particle (PP) limit, the Burke-Throne terms are given by~\citep{Flanagan:07} 
\begin{align}
    &g_r^{\rm (gw, pp)} = \frac{16MM'}{5r^3}\dot{r}\left[\dot{r}^2 + 6 r^2\dot{\phi}^2+\frac{4M_{\rm t}}{3r}\right], 
    \label{eq:a_BT_r_pp}\\
    &g_\phi^{\rm (gw, pp)} = \frac{8MM'}{5r^2}\dot{\phi}\left[9\dot{r}^2-6r^2\dot{\phi}^2+\frac{2M_{\rm t}}{r}\right].
    \label{eq:a_BT_phi_pp}
\end{align}
Note that $\dot{r}\ll r\dot{\phi}$, and consequently, $g_r^{\rm (gw, pp)} \ll g_\phi^{\rm (gw, pp)}$. 

Prior to resonance, the tidally induced quadrupole $Q_{\rm ns}^{ij}$ oscillates in phase with the orbit and accelerates the GW radiation~(\citealt{Lai:94a, Flanagan:08}; see also Appendix \ref{appx:quadrupole} for detailed derivations). In particular, two additional terms need to be included in $g_\phi^{\rm (gw)}$. The first is  due to $Q_{\rm ns}^{\langle ij \rangle}$, leading to 
\begin{align}
    &g_{\phi}^{\rm (gw, ns)}\simeq 
    -\frac{128}{5}\sqrt{\frac{2\pi}{15}}M R^2 r \Omega^5  \nonumber \\
    &\times \sum_{m=\pm 2}
    \left(
        \sum_{a, \omega_a>0}^{m_a=m}I_a{\rm Re}\left[C_a\right] 
        +\frac{1}{2}\sum_{ab, \omega_a>0}^{m_a+m_b=-m} J_{ab2m}{\rm Re}\left[C_a C_b\right]
    \right). 
\end{align}
Meanwhile, the tidal back reaction modifies the relation between $r$ and $\dot{\phi}\equiv \Omega$ [see later in Eq.~(\ref{eq:r_vs_omega})], causing a correction
\begin{align}
    &g_{\phi}^{\rm (gw, br)}\simeq -\frac{96}{5} M M'\left(\frac{R}{r}\right)^2 \Omega^3 \nonumber \\
    &\times    \sum_a^{\omega_a>0}\left[W_{l_a m_a}I_a {\rm Re}[C_a] 
        + \left(\frac{1}{2}\sum_{b, lm}^{m_a+m_b=-m} W_{lm}J_{ablm}{\rm Re}[C_a C_b]\right)
        \right].
        \label{eq:g_br}
\end{align}
Note that its sum with the PP part leads to the intuitive result
\begin{equation}
    g_{\phi}^{\rm (gw, pp)} + g_{\phi}^{\rm (gw, br)} = -\frac{32}{5}\mu r^3 \Omega^5. 
\end{equation}

Similarly, the Burke-Throne dissipation would also act on the mode. This modifies the dynamics of the mode as [cf. Eq.~(\ref{eq:ode_mode_amp_general})]
\begin{equation}
    \dot{c}_a + i\omega_a c_a = i \omega_a \left[Z_a + (\text{conservative terms})\right],
    \label{eq:BT_in_mode}
\end{equation}
where 
\begin{equation}
    Z_a\simeq i \frac{2}{15}W_{22}\frac{M'}{M_{\rm t}} \left(\frac{R}{r}\right)^3 (m_a r \Omega)^5 \left( I_a + \sum_b^{m_b=0} J_{ab2-m_a} c_b^\ast \right) e^{-im_a\phi},
    \label{eq:Z_a_Qorb}
\end{equation}
for the $l_a=|m_a|=2$ modes. The effect of $Z_a$ can be ignored for other modes at the order we are interested in. 
Note $Z_a$ leads to an imaginary part to $C_a$, which then becomes a torque on the orbit [Eq.~(\ref{eq:aphi_tide})] and contributes to the orbital decay. See Appendix \ref{appx:quadrupole} for more discussions. 

We have now outlined all the components in the differential equations we solve numerically. As a brief summary, the quantities we integrate are $(r, \dot{r}, \phi, \dot{\phi}{=}\Omega, C_a)$ and they are governed by Eqs.~(\ref{eq:ddr}) and (\ref{eq:ddphi}) for the orbital part and Eqs.~(\ref{eq:ode_mode_amp_general}) and (\ref{eq:Z_a_Qorb}) for the eigenmodes. We start the numerical integration at $f_{\rm gw}=\Omega/\pi=50\,{\rm Hz}$ with $\phi=0$. We set the initial values of $r$ using the PP Keplerian orbit and $\dot{r}$ by the PP GW decay. The modes are initialized with their linear solution $B_a^{(0)}$ (Section~\ref{sec:lin_sol}). Note that this choice of initial condition does not affect the results at $f_{\rm gw}\gtrsim 500\,{\rm Hz}$ where the tidal effects are significant. This is because all the tidal effects have a sharp power-law dependence on $f_{\rm gw}$ which we will see explicitly in Section~\ref{sec:eq_config}. We terminate our integration at $r/R\simeq 2.3$, corresponding to $r/2\simeq R+\xi^r(R)$ with $\xi^r(R)$ the radial component of the perturbed fluid evaluated at the surface of the NS and on the equator. In other words, our integration terminates approximately when the two NSs come into contact. In comparison, the innermost stable circular orbit is located at a smaller separation of $r=6(M+M')\simeq 1.9 R$. 

Our main numerical result is shown in Fig.~\ref{fig:dPhi_evol} where we compare the tidal phase shift of the time-domain GW waveform with (gray curve) and without (olive curve) the non-linear tide. We derive in the next section the equilibrium configuration of the system which will allow us analytically understand Fig.~\ref{fig:dPhi_evol}. 

\section{Equilibrium configuration}
\label{sec:eq_config}

The equilibrium configuration of the system can be obtained by assuming the GW decay is a slow process. Consequently, terms caused by the GW decay can be dropped from the equation of motion. Our goal is to derive $r$, $E$, and $\dot{E}$ in terms of mode amplitude $C_a$ and GW frequency (or equivalently, $\Omega$). Since we derived the analytical solutions to the mode amplitudes in Section~\ref{sec:mode_nl}, we have a complete description of the orbit with algebraic expressions only. 

\begin{figure}
   \centering
   \includegraphics[width=0.45\textwidth]{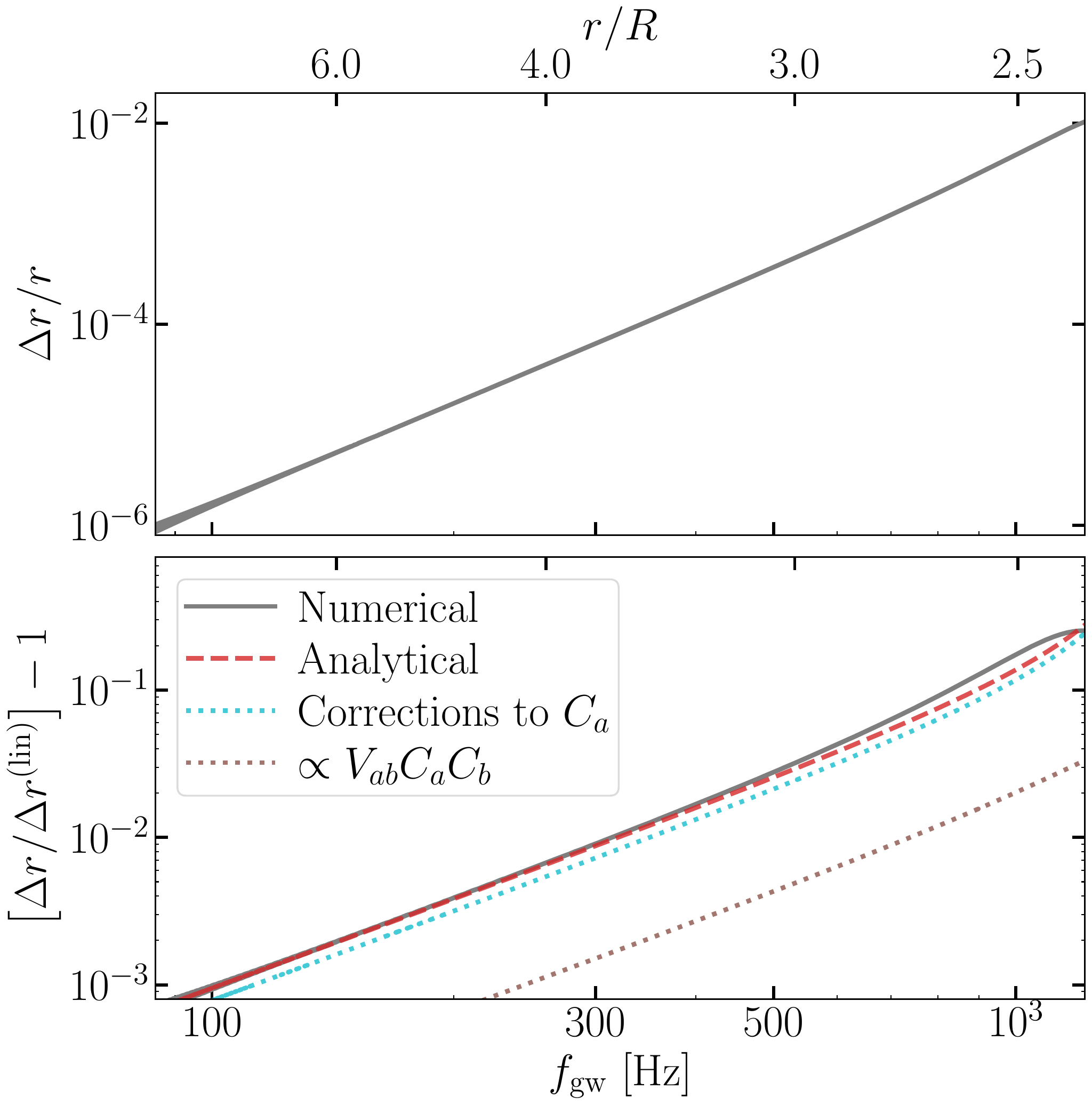} 
   \caption{Top: tidal modification of the PP Keplerian orbit, $\Delta r/r$, as a function of GW frequency.
   Bottom: fractional corrections to the linear theory's prediction on $\Delta r$ due to non-linear tide. The cyan-dotted curve shows the contribution due to non-linear corrections to the mode amplitude $C_a$ and the brown-dotted curve shows the corrections due to the $V_{ab}C_aC_b$ term. 
   }
   \label{fig:Delta_r}
\end{figure}

First, the tidal interaction modifies the Keplerian orbit. We can derive the relation between $r$ and $\Omega$ from Eq.~(\ref{eq:ddr}). Ignoring the GW decay and thus the $\ddot{r}$ term, we arrive at
\begin{equation}
    r^3 = \frac{M_{\rm t}}{\Omega^2} - \frac{r^2 g_r^{\rm (tide)}}{\Omega^2},
\end{equation}
as the modified Kepler's law. 
We can write $r=r_0 + \Delta r$ with $r_0^3 = M_{\rm t}/{\Omega^2}$,
leading to
\begin{align}
    &\frac{\Delta r}{r} \simeq -\frac{1}{3} \frac{r^2 g_r^{\rm (tide)}}{M_{\rm t}}
    = \frac{2}{3}\sum_{lm} (l+1)W_{lm}\left(\frac{R}{r}\right)^l \nonumber \\
    & \times \left(\sum_{a,\omega_a>0}^{m_a=m}I_a {\rm Re} [C_a] + \frac{1}{2}\sum_{ab, \omega_a>0}^{m_a+m_b=-m} J_{ablm} {\rm Re} [C_a C_b]\right).
    \label{eq:r_vs_omega}
\end{align}
The linear correction is obtained by evaluating the right hand side of Eq.~(\ref{eq:r_vs_omega}) in terms of $r_0$ and including the linear tide only in $C_a$. 
\begin{align}
    \frac{\Delta r^{\rm (lin)}}{r} 
    &= \frac{2}{3}\frac{M'}{M}\sum_a^{\omega_a > 0} (l+1) W_{lm}^2 I_a^2 R^{2l+1} M_{\rm t}^{-(2l+1)/3} \nonumber \\
    &\times  \frac{\omega_a \Omega^{(4l+2)/3}}{\omega_a - m_a\Omega}.
\end{align}
For $l=l_a=2$ and in the adiabatic limit ($\Omega \ll \omega_a$), this further simplifies to~\citep{Lai:93, Lai:94a, Lai:94b, Flanagan:08, Hinderer:10}
\begin{equation}
    \frac{\Delta r^{\rm (lin)}}{r} \simeq 2 k_2 \frac{M'}{M} M_{\rm t}^{-5/3}R^5 \Omega^{10/3},
    \label{eq:r_vs_omega_ad}
\end{equation}
where we have used Eq.~(\ref{eq:love_vs_Ia_ad}) for for love number $k_2$. 

We can also compute the epicyclic frequency $K$ of the system~\citep{Choudhuri:10} under the linear, adiabatic limit, which reads, 
\begin{align}
    K^2 &= 4 \Omega^2 + r \frac{d \Omega^2}{dr}=\Omega^2\left[1-54 k_2 \frac{M'}{M}\left(\frac{R}{r}\right)^6\right]. 
\end{align}
For $M'\simeq M$ and $r\simeq 2R$, we have $K^2\simeq 0.8 \Omega^2>0$. Including the dynamical response and the non-linear tide will modify $K^2$ by order unity and we still have $K^2>0$. In other words, we do not expect to see a tidally-induced plunge throughout the inspiral and the quasi-circular approximation holds. 

To compute the next-order corrections, we note that the first non-linear correction to $C_a$ and the non-linear tide piece  in $g_r^{\rm (tide)}$ ($\propto {\rm Re}[V_{ab} V_{a} V_{b}]$) will both give corrections to $\Delta r/r$ at the order $(R/r)^3\left[\Delta r^{\rm (lin)}/r\right]\propto (R/r)^8$. In comparison, errors due to linearization of Eq.~(\ref{eq:r_vs_omega}) and evaluating its right hand side at $r_0$ instead of $r_0+\Delta r^{(\rm lin)}$ are both on the order $\mathcal{O}\left[\Delta r^{\rm (lin)}/r\right]^2\propto (R/r)^{10}$. Therefore, we can obtain the leading-order non-linear correction by including the non-linear tide pieces in $C_a$ and $g^{\rm (tide)}_r$ while dropping terms $\propto \left[\Delta r^{\rm (lin)}/r\right]^2$. This provides us with an analytical approximation accurate to $(R/r)^8$. 

The result of the modification to the Keplerian $r-\Omega$ relation is shown in Fig.~\ref{fig:Delta_r}. In the top panel, we show the total correction to $\Delta r/r$ including both linear and non-linear tides. The fractional correction to the linear theory's prediction due to the non-linear tide  is presented in the bottom panel. The gray curve is extracted from the numerical solution to the differential equations described in Sections~\ref{sec:dyn_modes} and \ref{sec:dyn_orb}. The analytical approximation (the red-dashed curve), obtained by substituting the non-linear solution of $C_a$ given by Eq.~(\ref{eq:domega_a_leading})-(\ref{eq:C_c_leading}) to Eq.~(\ref{eq:r_vs_omega}) and keeping the $V_{ab}$ terms,  agrees well with the numerical result. At $f_{\rm gw}\gtrsim 1000\,{\rm Hz}$ or $(r/R)\lesssim 2.5$, the non-linear tide modifies the linear result by more than $10\%$. The main effect is due to the non-linear corrections to the mode amplitudes $C_a$,\footnote{Please note that in this section, the subscript $a$ stands for a mode in general and does not correspond to the specific $l=m=2$ mode we discussed in Section~\ref{sec:mode_nl}. } which is greater than the contribution from the $V_{ab}$ term by about a factor of 5.



\begin{figure}
   \centering
   \includegraphics[width=0.45\textwidth]{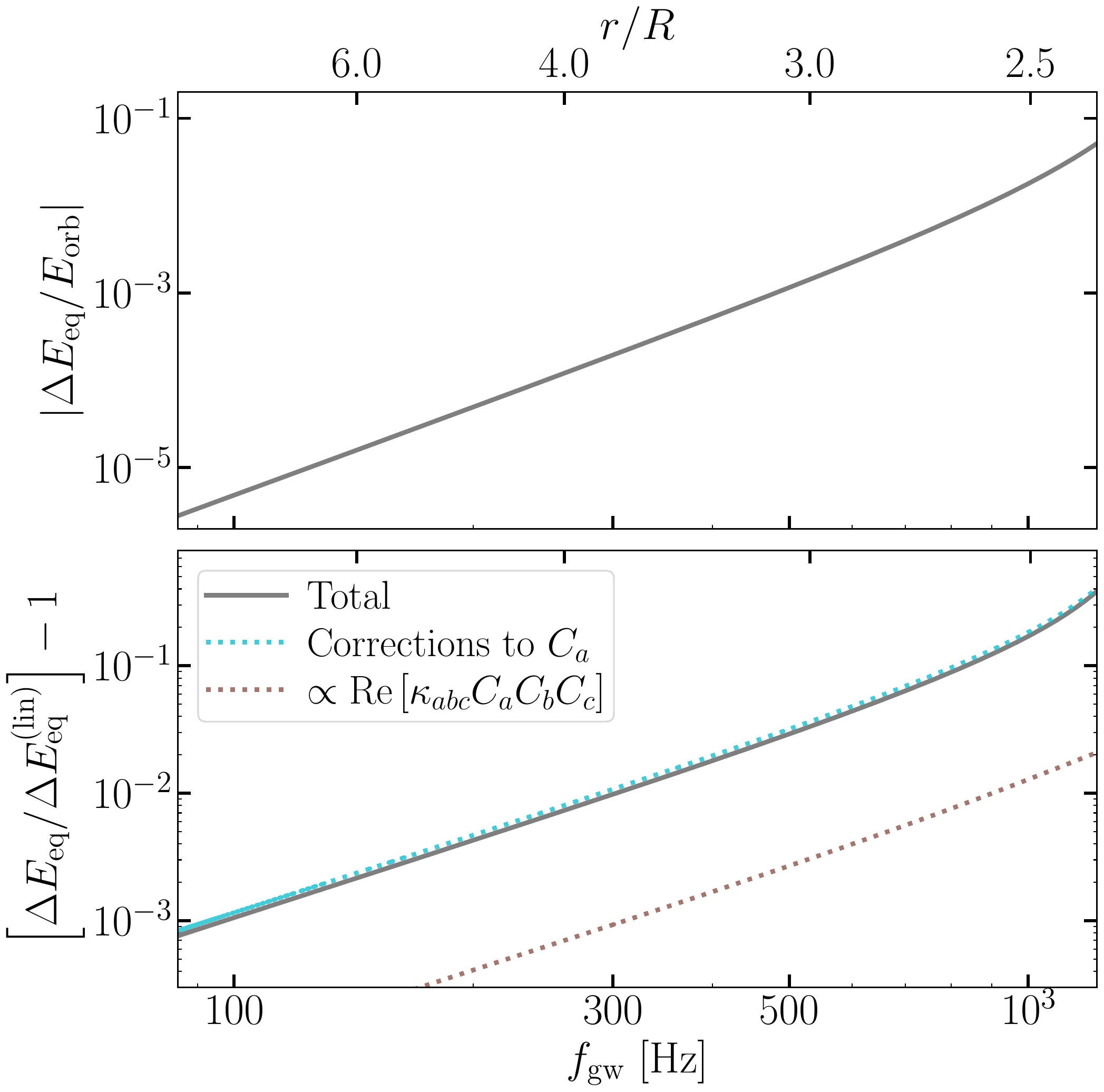} 
   \caption{Similar to Fig.~\ref{fig:Delta_r} but for tidally induced changes in the equilibrium energy, $\Delta E_{\rm eq}$. In the bottom panel, we see the dominant non-linear effect comes from the corrections to the mode amplitude $C_a$ while the energy in the non-linear interaction ($\propto {\rm Re}\left[\kappa_{abc} C_a C_b C_c\right]$) is subdominant. 
   }
   \label{fig:Delta_E}
\end{figure}

To compute the total energy of the system $\Delta E_{\rm eq}= \Delta E_{\rm orb} + E_{\rm int} + E_{\rm mode}$, we also need to include the energy due to mode-orbit interaction [$E_{\rm int}=H_{\rm int}$; see Eq.~(\ref{eq:H_int})] and the energy of modes 
\begin{equation}
    \frac{E_{\rm mode}}{E_0} = \sum_a^{\omega_a>0} |c_a|^2 -\frac{1}{3}\sum_{abc}^{\omega_{a}>0} \kappa_{abc}\left( c_a c_b c_c + c_a^\ast c_b^\ast c_c^\ast \right). 
    \label{eq:E_mode}
\end{equation}

The modified $r-\Omega$ relation [Eq.~(\ref{eq:r_vs_omega})] modifies the energy of the orbit. The kinetic energy of a quasi-circular orbit is given by 
\begin{equation}
    E_{\rm orb, k}=\frac{1}{2}\mu r^2 \dot{\phi}^2,
\end{equation}
and the potential energy
\begin{equation}
    E_{\rm orb, p}=-\frac{MM'}{r}.
\end{equation}
It is a well-known result that $E_{\rm orb,k}=-E_{\rm orb, p}/2=-E_{\rm orb}$. When tide is present, however, we have
\begin{align}
    &\frac{\Delta E_{\rm orb, k}}{E_{\rm orb, k}} = -\frac{\Delta E_{\rm orb, k}}{E_{\rm orb}}=2\frac{\Delta r}{r}, \\
    &\frac{\Delta E_{\rm orb, p}}{E_{\rm orb, p}} = 2\frac{\Delta E_{\rm orb, p}}{E_{\rm orb}}=-\frac{\Delta r}{r}. 
\end{align}
Consequently, 
\begin{equation}
    \frac{\Delta E_{\rm orb}}{E_{\rm orb}} = \frac{\Delta E_{\rm orb, k}+\Delta E_{\rm orb, p}}{E_{\rm orb}}= -4 \frac{\Delta r}{r}. 
\end{equation}

For $l=2$, we also have
\begin{equation}
    \frac{E_{\rm int}}{E_{\rm orb}} = 2\frac{\Delta r}{r} = -\frac{1}{2} \frac{\Delta E_{\rm orb}}{E_{\rm orb}}, 
\end{equation}
allowing us to easily obtain the interaction energy. 

At the linear order, we get
\begin{align}
    \frac{\Delta E_{\rm eq}}{E_{\rm orb}} &= -2\frac{r}{R}\frac{M}{M'}\sum_a^{\omega_a>0} \left[ 2{\rm Re}\left(V_a C_a^\ast \right) + |C_a|^2 
    \right],\nonumber \\
    &= -2 \frac{M'}{M} M_{\rm t}^{-5/3}R^5 \Omega^{10/3}  \nonumber \\
    &\times \sum_a^{\omega_a>0} W_{lm}^2 I_a^2 \left[2\left(\frac{\omega_a}{\omega_a - m_a \Omega}\right) + \left(\frac{\omega_a}{\omega_a - m_a \Omega}\right)^2\right]. 
\end{align}
In the adiabatic limit, this further reduces to 
\begin{equation}
    \frac{\Delta E_{\rm eq}}{E_{\rm orb}} =-6 k_2 \frac{M'}{M} M_{\rm t}^{-5/3}R^5 \Omega^{10/3}. 
\end{equation}

As explained before, to obtain the next order corrections, we include the non-linear tide but drop $({\Delta E_{\rm eq}}/{E_{\rm orb}})^2 \sim (\Delta r/r)^2$ terms. Since we have shown the accuracy of our analytical results in Figs.~\ref{fig:mode_vs_freq} and \ref{fig:Delta_r}, we show in Fig.~\ref{fig:Delta_E} the change of the equilibrium energy using the analytical results only. Similar to Fig.~\ref{fig:Delta_r}, we show in the top panel of Fig.~\ref{fig:Delta_E} the total correction to the equilibrium energy of the system including both linear and non-linear tides, and in the bottom panel the fractional correction to the linear theory's prediction.  We again note a $>10\%$ modification to the linear tide at $f_{\rm gw}\gtrsim 1000\,{\rm Hz}$ and it is mostly due to the non-linear correction to mode amplitudes. 

The last piece we need is the energy flux $\dot{E}=- \langle \dddot{Q}_{\rm tot}^{\langle ij \rangle} \dddot{Q}_{\rm tot}^{\langle ij \rangle} \rangle/5$ (see, e.g., \citealt{Poisson:14}).
It is enhanced by the tidal interaction due to two main effects. 
The first one comes from the coupling between the tidal and orbital quadrupoles which can be written as 
\begin{equation}
    \Delta \dot{E}_{\rm ns-orb} = -\frac{2}{5}\left\langle \dddot{Q}_{\rm ns}^{\langle ij \rangle} \dddot{Q}_{\rm orb}^{\langle ij \rangle} \right\rangle.
\end{equation}
This piece has the same origin as the dissipative accelerations due to $g_\phi^{\rm (gw, ns)}$ and $Z_a$ in the differential equations (Section~\ref{sec:dyn_orb}).
Using techniques described in Appendix~\ref{appx:quadrupole}, 
we find
\begin{align}
    \Delta \dot{E}_{\rm ns-orb} 
    &=-\frac{4}{15}\sum_{m}W_{2m}
    \left\langle
    \dddot{Q}_{2m}^{\rm ns} \dddot{Q}{^{\rm orb}_{2m}}^\ast
    \right\rangle,
\end{align}
where $Q_{2m}^{\rm ns}$ and $Q_{2m}^{\rm orb}$ are respectively the mass quadrupole of the NS and the orbit with spherical degree $(2, m)$ (see Appendix~\ref{appx:quadrupole}). 

When the GW decay is slow, we have 
\begin{align}
    &\dddot{Q}_{2m}^{\rm ns} \simeq 
    (- i m \Omega)^3 Q_{2m}^{\rm ns} 
      \nonumber \\
    =& (-i m \Omega)^3 MR^2 \left[\sum_a^{m_a=m} I_a C_a  
    + \frac{1}{2}\sum_{ab}^{m_a+m_b=-m} J_{ab2m}C_a^\ast C_b^\ast \right] e^{-im\phi},
\end{align}
and 
\begin{equation}
    \dddot{Q}{^{\rm orb}_{2m}}^\ast \simeq (i m \Omega)^3 {Q_{2m}^{\rm orb}}^\ast =  (i m \Omega)^3 \mu r^2 e^{im\phi}.
\end{equation}
Thus 
\begin{align}
    &\Delta \dot{E}_{\rm ns{-}orb}=-\frac{8}{15} \mu MR^2 M_{\rm t}^{2/3} \Omega^{14/3}  \sum_{m=\pm 2} m^6 W_{2m} \nonumber \\
    & \times 
    \left(
    \sum_{a, \omega_a>0}^{m_a=m} I_a {\rm Re}\left[C_a\right] 
    + \frac{1}{2} \sum_{ab, \omega_a>0}^{m_a+m_b=-m} J_{ab2m} {\rm Re}\left[C_a C_b\right]
    \right).
    \label{eq:del_dE_ns_orb}
\end{align}
Note the summation runs over positive-frequency modes for $a$ while it runs over both signs of frequency for mode $b$ in the non-linear term.


In the linear, adiabatic limit, the result reduces to \citep{Lai:94a}
\begin{equation}
    \frac{\Delta \dot{E}_{\rm ns-orb}}{\dot{E}_{\rm pp}} 
    =4 k_2\frac{M_{\rm t}}{M}R^5 M_{\rm t}^{-5/3}\Omega^{10/3}, 
\end{equation}
where we have used 
\begin{equation}
    \dot{E}_{\rm pp} = -\frac{32}{5} \mathcal{M}_c^{10/3}\Omega^{10/3},
\end{equation}
for the PP GW radiation with $\mathcal{M}_c= \mu^{3/5}M_{\rm t}^{2/5}$ the chirp mass. 

The second effect that enhances $\dot{E}$ is the tide-modified $r-\Omega$ relation, which enhances the  $\left\langle \dddot{Q}_{\rm orb}^{\langle ij \rangle } \dddot{Q}_{\rm orb}^{\langle ij \rangle} \right\rangle$ term compared to the PP case [corresponding to the $g_\phi^{\rm (gw, br)}$ term; Eq.~(\ref{eq:g_br})]. 
Since $Q_{\rm orb}\sim r^2$, the correction to the energy flux due to modified $r-\Omega$ relation given by
\begin{equation}
    \frac{\Delta \dot{E}_{r-\Omega}}{\dot{E}_{\rm pp}} = 4 \frac{\Delta r}{r}. 
    \label{eq:del_dE_r_O}
\end{equation}

The total enhancement of the energy loss is thus given by 
$\Delta \dot{E} = \Delta \dot{E}_{\rm ns{-}orb} + \Delta \dot{E}_{r{-}\Omega}$. This ignores the contribution from $\langle \dddot{Q}_{\rm ns}^{\langle ij \rangle} \dddot{Q}_{\rm ns}^{\langle ij \rangle} \rangle$, which is a higher-order correction than the non-linear tide we consider in this study [see the discussion around Eq.~(\ref{eq:Z_a_Qmode})]. 

\begin{figure}
   \centering
   \includegraphics[width=0.45\textwidth]{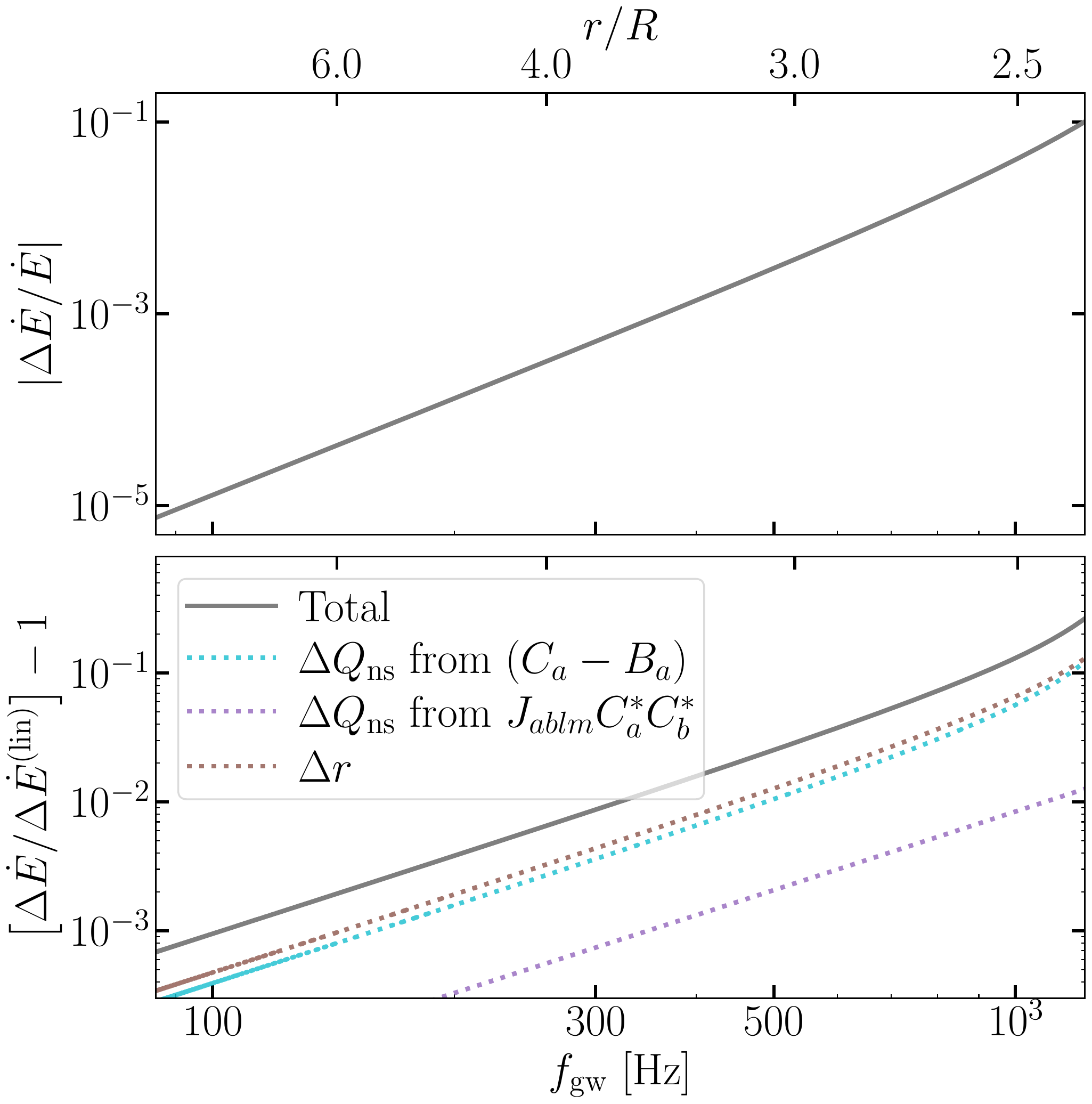} 
   \caption{Similar to Fig.~\ref{fig:Delta_r} but for the energy flux $\dot{E}$. In the bottom panel, we show non-linear corrections to the energy flux from both the modifications of the NS quadrapole (cyan and purple curves) and the non-linear $r-\Omega$ relation (fig.~\ref{fig:Delta_r}). 
   }
   \label{fig:Delta_dEdt}
\end{figure}

We present in Fig.~\ref{fig:Delta_dEdt} the results of $\Delta \dot{E}$. We note similar non-linear corrections to $\Delta \dot{E}$ (lower panel) from the interaction between NS and orbital quadurpoles [Eq.~(\ref{eq:del_dE_ns_orb})] and from the modification to the $r-\Omega$ relation [Eq.~(\ref{eq:del_dE_r_O})]. In both effects, the non-linear correction comes mainly from its correction to the mode amplitudes $(C_a-B_a)$ whereas the $\propto J_{ablm}$ term is subdominant. 

    
\begin{figure}
   \centering
   \includegraphics[width=0.45\textwidth]{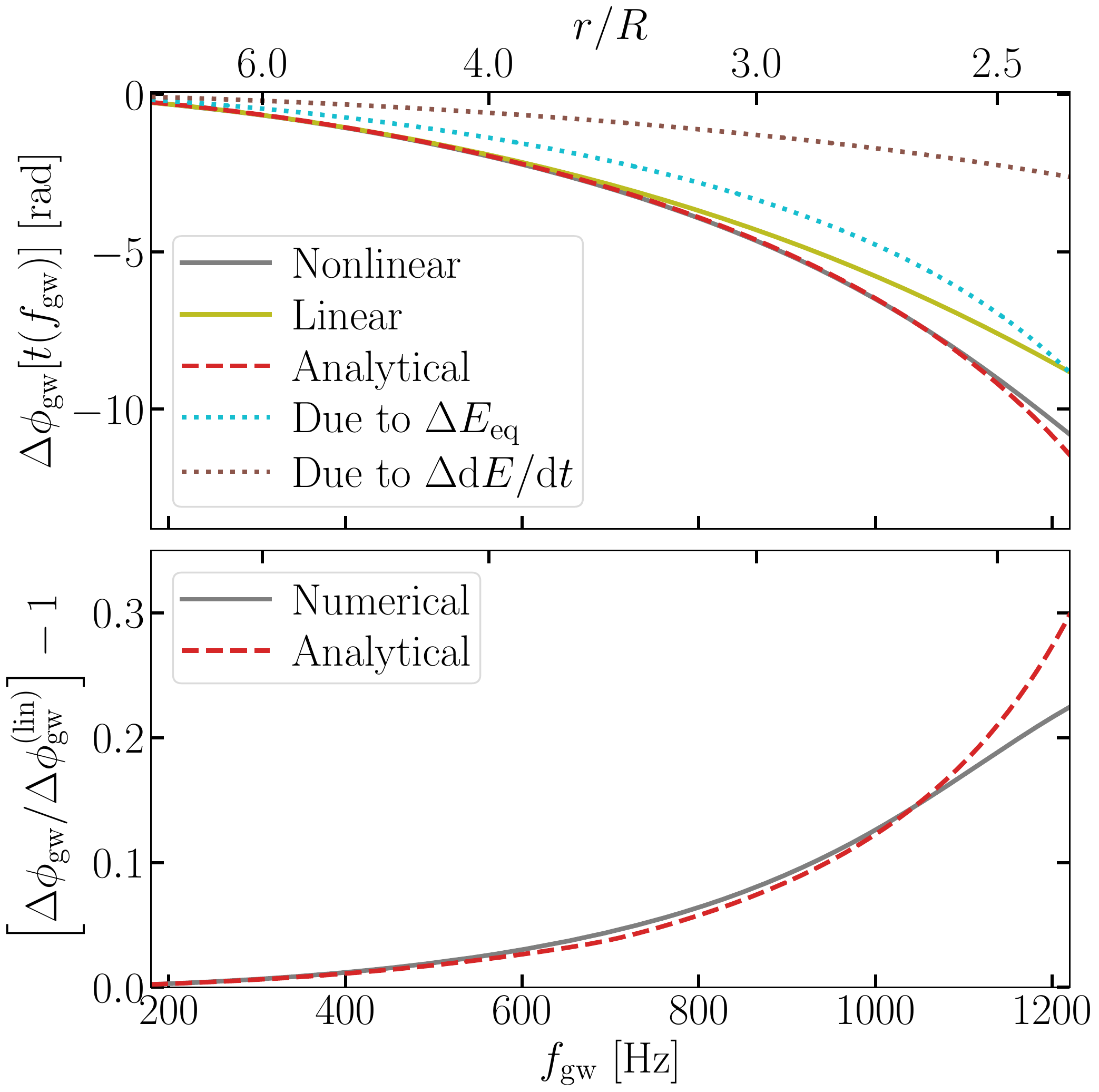} 
   \caption{GW phase shift as a function of GW frequency (bottom axis) and orbital separation (top axis). The top panel is similar to Fig.~(\ref{fig:dPhi_evol}). We further include the analytical approximation from Eq.~(\ref{eq:dphidf_ana}) in the red-dashed curve and it agrees well with the numerical result in the gray curve. The contribution to the phase shift from the modified equilibrium energy and the enhanced GW radiaiton are shown in the cyan-dotted  and brown-dotted lines. 
   In the bottom panel, we show the fraction correction to the phase shift predicted by linear tide. The correction is greater than $10\%$ at $f_{\rm gw}\gtrsim 1000\,{\rm Hz}$. 
   }
   \label{fig:dPhi_num_ana}
\end{figure}

From $E_{\rm eq}$ and $\dot{E}$, we can get the phase of the GW signal as~\citep{Lai:94b, Hinderer:10}
\begin{equation}
    \frac{\diff \phi_{\rm gw}}{\diff f_{\rm gw}} = 2\pi f_{\rm gw}\frac{\diff E/\diff f_{\rm gw}}{\dot{E}}.
\end{equation}
Note that $\phi_{\rm gw}=2\phi$ is the phase of the time-domain waveform expressed as a function of $f_{\rm gw}$. The phase shift due to the tidal effect is 
\begin{equation}
    \frac{\diff \Delta \phi_{\rm gw}}{\diff f_{\rm gw}} = \frac{2\pi f_{\rm gw}}{\dot{E}} \left( \frac{\diff \Delta E_{\rm eq}}{\diff f_{\rm gw}} - \frac{\diff E_{\rm eq}}{\diff f_{\rm gw}} \frac{\Delta \dot{E}}{\dot{E}}\right).
    \label{eq:dphidf_ana}
\end{equation}
As noted in the discussion below Eq.~(\ref{eq:r_vs_omega_ad}), the corrections due to the second-order expansion [i.e., terms like $\left(\Delta E_{\rm eq}/E_{\rm eq}\right)^2$] are smaller than the non-linear tide we consider by a factor of $(R/r)^2$ and are thus ignored in the expression. 

Our final result is presented in Fig.~\ref{fig:dPhi_num_ana} (also in Fig.~\ref{fig:dPhi_evol}). In the top panel, we show the total phase shift due to tidal effects in the NS $M$ (while treating $M'$ as a point-particle). The olive curve is the prediction assuming just the linear tide while the gray curve also includes the non-linear tide. Both curves are obtained by numerically solving the differential equations described in Sections~\ref{sec:dyn_modes} and \ref{sec:dyn_orb}. For comparison, the red-dashed curve shows the analytical phase shift, Eq.~(\ref{eq:dphidf_ana}). We note a good agreement between the analytical and numerical results. The effects due to the modified equilibrium energy and the modified GW radiation are respectively shown in the cyan-dotted and brown-dotted curves. The fractional correction to the linear tidal phase shift is shown in the bottom panel. 

For a Newtonian NS, non-linear tide could introduce an additional $2\,{\rm rad}$ of phase shift, corresponding to about 20\% enhancement of the tidal effect near the final merger. GR is likely to reduce the result because the NS will be ``softer'' (harder to perturb). As we argue in Section~\ref{sec:discussion}, after the GR correction, the excess phase shift due to non-linear tide should be around 1 rad. 
This is consistent with the discrepancy between the state-of-art theoretical models assuming linear tides in GR and numerical relativity~\citep{Hinderer:16, Steinhoff:21}. Therefore, our result suggests that including non-linear tidal interactions could explain the discrepancy and allow the theoretical models to be accurate all the way to the final merger.

\section{Conclusion and Discussions}
\label{sec:discussion}

In this work, we investigated tidal interactions in coalescing BNS including leading-order non-linear corrections. We discussed the dynamics of the NS eigenmodes and the orbit in Sections~\ref{sec:dyn_modes}
and \ref{sec:dyn_orb}, respectively. Utilizing analytical approximations to the mode amplitudes in Section~\ref{sec:mode_nl}, we derived algebraic solutions governing the binary's evolution track and the corresponding GW phase in Section~\ref{sec:eq_config}. For a Newtonian NS approximated by a $\Gamma=2$ polytrope, we found the non-linear tide could lead to an additional $\sim 2\, {\rm rad}$ phase shift in the GW waveform near the binary's final merger. While this is likely an overestimation of the non-linear effect because a Newtonian NS is stiffer than its GR counterpart, our result suggests that the non-linear tide is a critical component to be included in the waveform modeling and it could improve the agreement between theoretical models and numerical relativity, especially near the final merger.

Incorporating GR corrections is thus one of the major future steps to obtain a robust theoretical estimate of the non-linear tide. In particular, there are two major GR effects to be considered and they act in opposite directions. First, we note that the relativistic value of the love number, $k_2\propto I_a^2$, is smaller than its Newtonian counterpart by a factor of 2-3~\citep{Damour:09, Binnington:09, Hinderer:10}, reducing the linear tidal phase shift accordingly. The phase shift induced by non-linear tidal corrections may scale as $k_2^2$ because the non-linear correction to the mode amplitude is sourced by the square of the linear amplitude (assuming GR has similar effects on the linear tidal overlaps and the non-linear coupling coefficients). This could reduce the non-linear correction to the phase by a factor of $\sim 5-10$. 

While GR reduces the spatial coupling, it nonetheless enhances the finite-frequency effect by lowering the mode frequency $\omega_a$. When perturbing the \emph{same} background model, the GR oscillation equations typically result in smaller eigenfrequencies than the Newtonian result~\citep{Yu:17b} due to the redshift of the NS itself $\propto M/R$. Moreover, the orbit will further redshift $\omega_a$ to a lower value by a factor $\propto M_{\rm t}/r$ ($\sim 20\%$ near the merger; see, e.g., \citealt{Steinhoff:16, Steinhoff:21}). For $|m|=2$ modes, lowering $\omega_a$ enhances the finite-frequency response via of $\omega_a/(\omega_a+\Delta \omega_a - m_a \Omega)$ [Eq.~(\ref{eq:C_ab_leading})], while for $m=0$, $\Delta V_c$ will be greater [Eq.~(\ref{eq:dV_c_leading})]. The non-linear corrections to $\Delta r/r$ (and similarly to $\Delta E_{\rm eq}/E_{\rm eq}$ and $\Delta \dot{E}/\dot{E}$) goes approximately as $\omega_a^2/(\omega_a^2-4\Omega^2)$. Reducing $\omega_a$ by $35\%$ will amplify $\omega_a^2/(\omega_a^2-4\Omega^2)$ by about a factor of 2 when $f_{\rm gw}=\Omega/\pi=1000\,{\rm Hz}$. This partially compensates for the reduction of the phase shift due to the reduced coupling strength. After considering both GR effects, our estimation of the non-linear tide's contribution to the phase shift  becomes $|\Delta \phi_{\rm gw}|\simeq 1\,{\rm rad}$ (including contributions from both NSs) near the merger.  

Exploring the non-linear contribution to a wide range of EoSs will be another important extension. The non-linear tide is likely to exhibit a stronger dependence on the EoS than the linear tide because it is sourced by the square of the linear tidal amplitude. It could therefore strengthen the constraints on the NS EoS potentially, though this remain to be shown by future studies. Along the same line, it would be interesting to examine if a universal relation exists between the non-linear coupling strength and other properties of the NS (in analogy to the universal relation between the love number and NS mass quadrapole;~\citealt{Yagi:13}).

We assumed a non-spinning NS in our analysis. If the NS has a retrograde spin relative to the orbit, the f-mode could be shifted to a lower frequency due to both the Doppler effect and modifications to the NS structure. If the spin rate is sufficiently high, the $l_a=m_a=2$ f-mode could even be resonantly excited~\citep{Ho:99, Ma:20, Steinhoff:21}. Alternatively, the f-mode could be resonantly excited if the orbit has some residual eccentricity when the binary enters the sensitivity band of a ground-based GW detector~\citep{Chirenti:17, Parisi:18, Yang:18, Yang:19, Vick:19, Wang:20}. Since the non-linear correction is amplified by the finite-frequency response of the NS, we may thus expect the non-linear tide to play an even more significant role in those systems. The standard anharmonic frequency shift could also be significant since $m=2$ f-mode can have much greater amplitude than other $l=2$ modes (see Appendix~\ref{appx:anharm}).  In turn, as the non-linear frequency shift lowers the mode frequency, it makes the resonance more likely, which enhances the overall tidal signatures. Thus incorporating NS spin and orbital eccentricities are also potentially interesting extensions to the current study.

Another simplification assumed in this work is that we modeled the fluid inside the NS as a normal fluid. In reality, we would expect the NS to be cold and its core is likely in the superfluid state~\citep{Yakovlev:99}. \citet{Passamonti:22} showed that the correction due to superfluidity is small for f-modes, and we thus expect our main results to hold in realistic NSs. Nonetheless, a careful calculation incorporating superfluidity would be worthwhile. 

\section*{Acknowledgements}
HY's work at KITP is supported by the National Science Foundation (NSF PHY-1748958) and by the Simons Foundation (216179, LB).  NNW and PA acknowledge support from  NSF AST-2054353. JK and TV acknowledge support from NSF PHY-2012086.

\section*{Data Availability}
The main data underlying this article are available in the article.
Additional information may be requested from the authors.



\bibliographystyle{mnras}
\bibliography{ref} 

\begin{thebibliography}{}
\makeatletter
\relax
\def\mn@urlcharsother{\let\do\@makeother \do\$\do\&\do\#\do\^\do\_\do\%\do\~}
\def\mn@doi{\begingroup\mn@urlcharsother \@ifnextchar [ {\mn@doi@}
  {\mn@doi@[]}}
\def\mn@doi@[#1]#2{\def\@tempa{#1}\ifx\@tempa\@empty \href
  {http://dx.doi.org/#2} {doi:#2}\else \href {http://dx.doi.org/#2} {#1}\fi
  \endgroup}
\def\mn@eprint#1#2{\mn@eprint@#1:#2::\@nil}
\def\mn@eprint@arXiv#1{\href {http://arxiv.org/abs/#1} {{\tt arXiv:#1}}}
\def\mn@eprint@dblp#1{\href {http://dblp.uni-trier.de/rec/bibtex/#1.xml}
  {dblp:#1}}
\def\mn@eprint@#1:#2:#3:#4\@nil{\def\@tempa {#1}\def\@tempb {#2}\def\@tempc
  {#3}\ifx \@tempc \@empty \let \@tempc \@tempb \let \@tempb \@tempa \fi \ifx
  \@tempb \@empty \def\@tempb {arXiv}\fi \@ifundefined
  {mn@eprint@\@tempb}{\@tempb:\@tempc}{\expandafter \expandafter \csname
  mn@eprint@\@tempb\endcsname \expandafter{\@tempc}}}

\bibitem[\protect\citeauthoryear{{Abbott} et~al.,}{{Abbott}
  et~al.}{2017}]{Evans:17}
{Abbott} B.~P.,  et~al., 2017, \mn@doi [Classical and Quantum Gravity]
  {10.1088/1361-6382/aa51f4}, \href
  {http://adsabs.harvard.edu/abs/2017CQGra..34d4001A} {34, 044001}

\bibitem[\protect\citeauthoryear{{Alford}, {Haber}, {Harris}  \&
  {Zhang}}{{Alford} et~al.}{2021}]{Alford:21}
{Alford} M.~G.,  {Haber} A.,  {Harris} S.~P.,   {Zhang} Z.,  2021, \mn@doi
  [Universe] {10.3390/universe7110399}, \href
  {https://ui.adsabs.harvard.edu/abs/2021Univ....7..399A} {7, 399}

\bibitem[\protect\citeauthoryear{{Andersson} \& {Ho}}{{Andersson} \&
  {Ho}}{2018}]{Andersson:18}
{Andersson} N.,  {Ho} W. C.~G.,  2018, \mn@doi [\prd]
  {10.1103/PhysRevD.97.023016}, \href
  {https://ui.adsabs.harvard.edu/abs/2018PhRvD..97b3016A} {97, 023016}

\bibitem[\protect\citeauthoryear{{Andersson} \& {Pnigouras}}{{Andersson} \&
  {Pnigouras}}{2020}]{Andersson:20}
{Andersson} N.,  {Pnigouras} P.,  2020, \mn@doi [\prd]
  {10.1103/PhysRevD.101.083001}, \href
  {https://ui.adsabs.harvard.edu/abs/2020PhRvD.101h3001A} {101, 083001}

\bibitem[\protect\citeauthoryear{{Andersson} \& {Pnigouras}}{{Andersson} \&
  {Pnigouras}}{2021}]{Andersson:21}
{Andersson} N.,  {Pnigouras} P.,  2021, \mn@doi [\mnras]
  {10.1093/mnras/stab371}, \href
  {https://ui.adsabs.harvard.edu/abs/2021MNRAS.503..533A} {503, 533}

\bibitem[\protect\citeauthoryear{{Arras} \& {Weinberg}}{{Arras} \&
  {Weinberg}}{2019}]{Arras:19}
{Arras} P.,  {Weinberg} N.~N.,  2019, \mn@doi [\mnras] {10.1093/mnras/stz880},
  \href {https://ui.adsabs.harvard.edu/abs/2019MNRAS.486.1424A} {486, 1424}

\bibitem[\protect\citeauthoryear{{Bernuzzi}, {Nagar}, {Dietrich}  \&
  {Damour}}{{Bernuzzi} et~al.}{2015}]{Bernuzzi:15}
{Bernuzzi} S.,  {Nagar} A.,  {Dietrich} T.,   {Damour} T.,  2015, \mn@doi
  [\prl] {10.1103/PhysRevLett.114.161103}, \href
  {https://ui.adsabs.harvard.edu/abs/2015PhRvL.114p1103B} {114, 161103}

\bibitem[\protect\citeauthoryear{{Bini} \& {Damour}}{{Bini} \&
  {Damour}}{2014}]{Bini:14}
{Bini} D.,  {Damour} T.,  2014, \mn@doi [\prd] {10.1103/PhysRevD.90.124037},
  \href {https://ui.adsabs.harvard.edu/abs/2014PhRvD..90l4037B} {90, 124037}

\bibitem[\protect\citeauthoryear{{Binnington} \& {Poisson}}{{Binnington} \&
  {Poisson}}{2009}]{Binnington:09}
{Binnington} T.,  {Poisson} E.,  2009, \mn@doi [\prd]
  {10.1103/PhysRevD.80.084018}, \href
  {https://ui.adsabs.harvard.edu/abs/2009PhRvD..80h4018B} {80, 084018}

\bibitem[\protect\citeauthoryear{{Chirenti}, {Gold}  \& {Miller}}{{Chirenti}
  et~al.}{2017}]{Chirenti:17}
{Chirenti} C.,  {Gold} R.,   {Miller} M.~C.,  2017, \mn@doi [\apj]
  {10.3847/1538-4357/aa5ebb}, \href
  {https://ui.adsabs.harvard.edu/abs/2017ApJ...837...67C} {837, 67}

\bibitem[\protect\citeauthoryear{Choudhuri}{Choudhuri}{2010}]{Choudhuri:10}
Choudhuri A.~R.,  2010, Astrophysics for Physicists.
Cambridge University Press, \mn@doi{10.1017/CBO9780511802218}

\bibitem[\protect\citeauthoryear{{Cowling}}{{Cowling}}{1941}]{Cowling:41}
{Cowling} T.~G.,  1941, \mn@doi [\mnras] {10.1093/mnras/101.8.367}, \href
  {https://ui.adsabs.harvard.edu/abs/1941MNRAS.101..367C} {101, 367}

\bibitem[\protect\citeauthoryear{{Damour} \& {Nagar}}{{Damour} \&
  {Nagar}}{2009}]{Damour:09}
{Damour} T.,  {Nagar} A.,  2009, \mn@doi [\prd] {10.1103/PhysRevD.80.084035},
  \href {https://ui.adsabs.harvard.edu/abs/2009PhRvD..80h4035D} {80, 084035}

\bibitem[\protect\citeauthoryear{{Damour}, {Nagar}  \& {Villain}}{{Damour}
  et~al.}{2012}]{Damour:12}
{Damour} T.,  {Nagar} A.,   {Villain} L.,  2012, \mn@doi [\prd]
  {10.1103/PhysRevD.85.123007}, \href
  {https://ui.adsabs.harvard.edu/abs/2012PhRvD..85l3007D} {85, 123007}

\bibitem[\protect\citeauthoryear{{Del Pozzo}, {Li}, {Agathos}, {Van Den Broeck}
   \& {Vitale}}{{Del Pozzo} et~al.}{2013}]{DelPozzo:13}
{Del Pozzo} W.,  {Li} T. G.~F.,  {Agathos} M.,  {Van Den Broeck} C.,   {Vitale}
  S.,  2013, \mn@doi [\prl] {10.1103/PhysRevLett.111.071101}, \href
  {https://ui.adsabs.harvard.edu/abs/2013PhRvL.111g1101D} {111, 071101}

\bibitem[\protect\citeauthoryear{{Essick}, {Vitale}  \& {Weinberg}}{{Essick}
  et~al.}{2016}]{Essick:16}
{Essick} R.,  {Vitale} S.,   {Weinberg} N.~N.,  2016, \mn@doi [\prd]
  {10.1103/PhysRevD.94.103012}, \href
  {https://ui.adsabs.harvard.edu/abs/2016PhRvD..94j3012E} {94, 103012}

\bibitem[\protect\citeauthoryear{{Evans} et~al.,}{{Evans}
  et~al.}{2021}]{Evans:21}
{Evans} M.,  et~al., 2021, arXiv e-prints, \href
  {https://ui.adsabs.harvard.edu/abs/2021arXiv210909882E} {p. arXiv:2109.09882}

\bibitem[\protect\citeauthoryear{{Flanagan} \& {Hinderer}}{{Flanagan} \&
  {Hinderer}}{2008}]{Flanagan:08}
{Flanagan} {\'E}.~{\'E}.,  {Hinderer} T.,  2008, \mn@doi [\prd]
  {10.1103/PhysRevD.77.021502}, \href
  {https://ui.adsabs.harvard.edu/abs/2008PhRvD..77b1502F} {77, 021502}

\bibitem[\protect\citeauthoryear{{Flanagan} \& {Racine}}{{Flanagan} \&
  {Racine}}{2007}]{Flanagan:07}
{Flanagan} {\'E}.~{\'E}.,  {Racine} {\'E}.,  2007, \mn@doi [\prd]
  {10.1103/PhysRevD.75.044001}, \href
  {https://ui.adsabs.harvard.edu/abs/2007PhRvD..75d4001F} {75, 044001}

\bibitem[\protect\citeauthoryear{{Foucart} et~al.,}{{Foucart}
  et~al.}{2019}]{Foucart:19}
{Foucart} F.,  et~al., 2019, \mn@doi [\prd] {10.1103/PhysRevD.99.044008}, \href
  {https://ui.adsabs.harvard.edu/abs/2019PhRvD..99d4008F} {99, 044008}

\bibitem[\protect\citeauthoryear{{Gupta}, {Steinhoff}  \& {Hinderer}}{{Gupta}
  et~al.}{2021}]{Gupta:21}
{Gupta} P.~K.,  {Steinhoff} J.,   {Hinderer} T.,  2021, \mn@doi [Physical
  Review Research] {10.1103/PhysRevResearch.3.013147}, \href
  {https://ui.adsabs.harvard.edu/abs/2021PhRvR...3a3147G} {3, 013147}

\bibitem[\protect\citeauthoryear{{Hild}, {Chelkowski}, {Freise}, {Franc},
  {Morgado}, {Flaminio}  \& {DeSalvo}}{{Hild} et~al.}{2010}]{Hild:10}
{Hild} S.,  {Chelkowski} S.,  {Freise} A.,  {Franc} J.,  {Morgado} N.,
  {Flaminio} R.,   {DeSalvo} R.,  2010, \mn@doi [Classical and Quantum Gravity]
  {10.1088/0264-9381/27/1/015003}, \href
  {http://adsabs.harvard.edu/abs/2010CQGra..27a5003H} {27, 015003}

\bibitem[\protect\citeauthoryear{{Hinderer}, {Lackey}, {Lang}  \&
  {Read}}{{Hinderer} et~al.}{2010}]{Hinderer:10}
{Hinderer} T.,  {Lackey} B.~D.,  {Lang} R.~N.,   {Read} J.~S.,  2010, \mn@doi
  [\prd] {10.1103/PhysRevD.81.123016}, \href
  {https://ui.adsabs.harvard.edu/abs/2010PhRvD..81l3016H} {81, 123016}

\bibitem[\protect\citeauthoryear{{Hinderer} et~al.,}{{Hinderer}
  et~al.}{2016}]{Hinderer:16}
{Hinderer} T.,  et~al., 2016, \mn@doi [\prl] {10.1103/PhysRevLett.116.181101},
  \href {https://ui.adsabs.harvard.edu/abs/2016PhRvL.116r1101H} {116, 181101}

\bibitem[\protect\citeauthoryear{{Ho} \& {Lai}}{{Ho} \& {Lai}}{1999}]{Ho:99}
{Ho} W. C.~G.,  {Lai} D.,  1999, \mn@doi [\mnras]
  {10.1046/j.1365-8711.1999.02703.x}, \href
  {https://ui.adsabs.harvard.edu/abs/1999MNRAS.308..153H} {308, 153}

\bibitem[\protect\citeauthoryear{{Hotokezaka}, {Kyutoku}, {Okawa}  \&
  {Shibata}}{{Hotokezaka} et~al.}{2015}]{Hotokezaka:15}
{Hotokezaka} K.,  {Kyutoku} K.,  {Okawa} H.,   {Shibata} M.,  2015, \mn@doi
  [\prd] {10.1103/PhysRevD.91.064060}, \href
  {https://ui.adsabs.harvard.edu/abs/2015PhRvD..91f4060H} {91, 064060}

\bibitem[\protect\citeauthoryear{{Kuan}, {Suvorov}  \& {Kokkotas}}{{Kuan}
  et~al.}{2021a}]{Kuan:21}
{Kuan} H.-J.,  {Suvorov} A.~G.,   {Kokkotas} K.~D.,  2021a, \mn@doi [\mnras]
  {10.1093/mnras/stab1898}, \href
  {https://ui.adsabs.harvard.edu/abs/2021MNRAS.506.2985K} {506, 2985}

\bibitem[\protect\citeauthoryear{{Kuan}, {Suvorov}  \& {Kokkotas}}{{Kuan}
  et~al.}{2021b}]{Kuan:21b}
{Kuan} H.-J.,  {Suvorov} A.~G.,   {Kokkotas} K.~D.,  2021b, \mn@doi [\mnras]
  {10.1093/mnras/stab2658}, \href
  {https://ui.adsabs.harvard.edu/abs/2021MNRAS.508.1732K} {508, 1732}

\bibitem[\protect\citeauthoryear{{Kumar}, {Goldreich}  \& {Kerswell}}{{Kumar}
  et~al.}{1994}]{Kumar:94}
{Kumar} P.,  {Goldreich} P.,   {Kerswell} R.,  1994, \mn@doi [\apj]
  {10.1086/174159}, \href
  {https://ui.adsabs.harvard.edu/abs/1994ApJ...427..483K} {427, 483}

\bibitem[\protect\citeauthoryear{{LIGO Scientific Collaboration}, {Virgo
  Collaboration}  \& et al.}{{LIGO Scientific Collaboration}
  et~al.}{2017}]{GW170817}
{LIGO Scientific Collaboration} {Virgo Collaboration}  et al. 2017, \mn@doi
  [\prl] {10.1103/PhysRevLett.119.161101}, \href
  {https://ui.adsabs.harvard.edu/abs/2017PhRvL.119p1101A} {119, 161101}

\bibitem[\protect\citeauthoryear{{LIGO Scientific Collaboration}, {Virgo
  Collaboration}  \& et al.}{{LIGO Scientific Collaboration}
  et~al.}{2018}]{GW170817eos}
{LIGO Scientific Collaboration} {Virgo Collaboration}  et al. 2018, \mn@doi
  [\prl] {10.1103/PhysRevLett.121.161101}, \href
  {https://ui.adsabs.harvard.edu/abs/2018PhRvL.121p1101A} {121, 161101}

\bibitem[\protect\citeauthoryear{{LIGO Scientific Collaboration}, {Virgo
  Collaboration}  \& et al.}{{LIGO Scientific Collaboration}
  et~al.}{2019a}]{GW170817prop}
{LIGO Scientific Collaboration} {Virgo Collaboration}  et al. 2019a, \mn@doi
  [Physical Review X] {10.1103/PhysRevX.9.011001}, \href
  {https://ui.adsabs.harvard.edu/abs/2019PhRvX...9a1001A} {9, 011001}

\bibitem[\protect\citeauthoryear{{LIGO Scientific Collaboration}, {Virgo
  Collaboration}, {Weinberg}  \& et al.}{{LIGO Scientific Collaboration}
  et~al.}{2019b}]{GW170817pg}
{LIGO Scientific Collaboration} {Virgo Collaboration} {Weinberg} N.~N.,   et
  al. 2019b, \mn@doi [\prl] {10.1103/PhysRevLett.122.061104}, \href
  {https://ui.adsabs.harvard.edu/abs/2019PhRvL.122f1104A} {122, 061104}

\bibitem[\protect\citeauthoryear{{LIGO Scientific Collaboration}, {Virgo
  Collaboration}  \& et al.}{{LIGO Scientific Collaboration}
  et~al.}{2020}]{GW190425}
{LIGO Scientific Collaboration} {Virgo Collaboration}  et al. 2020, \mn@doi
  [\apjl] {10.3847/2041-8213/ab75f5}, \href
  {https://ui.adsabs.harvard.edu/abs/2020ApJ...892L...3A} {892, L3}

\bibitem[\protect\citeauthoryear{{Lackey} \& {Wade}}{{Lackey} \&
  {Wade}}{2015}]{Lackey:15}
{Lackey} B.~D.,  {Wade} L.,  2015, \mn@doi [\prd] {10.1103/PhysRevD.91.043002},
  \href {https://ui.adsabs.harvard.edu/abs/2015PhRvD..91d3002L} {91, 043002}

\bibitem[\protect\citeauthoryear{{Lai}}{{Lai}}{1994}]{Lai:94c}
{Lai} D.,  1994, \mn@doi [\mnras] {10.1093/mnras/270.3.611}, \href
  {https://ui.adsabs.harvard.edu/abs/1994MNRAS.270..611L} {270, 611}

\bibitem[\protect\citeauthoryear{{Lai}}{{Lai}}{1996}]{Lai:96}
{Lai} D.,  1996, \mn@doi [\prl] {10.1103/PhysRevLett.76.4878}, \href
  {https://ui.adsabs.harvard.edu/abs/1996PhRvL..76.4878L} {76, 4878}

\bibitem[\protect\citeauthoryear{{Lai}, {Rasio}  \& {Shapiro}}{{Lai}
  et~al.}{1993}]{Lai:93}
{Lai} D.,  {Rasio} F.~A.,   {Shapiro} S.~L.,  1993, \mn@doi [\apjs]
  {10.1086/191822}, \href
  {https://ui.adsabs.harvard.edu/abs/1993ApJS...88..205L} {88, 205}

\bibitem[\protect\citeauthoryear{{Lai}, {Rasio}  \& {Shapiro}}{{Lai}
  et~al.}{1994a}]{Lai:94a}
{Lai} D.,  {Rasio} F.~A.,   {Shapiro} S.~L.,  1994a, \mn@doi [\apj]
  {10.1086/173606}, \href
  {https://ui.adsabs.harvard.edu/abs/1994ApJ...420..811L} {420, 811}

\bibitem[\protect\citeauthoryear{{Lai}, {Rasio}  \& {Shapiro}}{{Lai}
  et~al.}{1994b}]{Lai:94b}
{Lai} D.,  {Rasio} F.~A.,   {Shapiro} S.~L.,  1994b, \mn@doi [\apj]
  {10.1086/173812}, \href
  {https://ui.adsabs.harvard.edu/abs/1994ApJ...423..344L} {423, 344}

\bibitem[\protect\citeauthoryear{Landau \& Lifshitz}{Landau \&
  Lifshitz}{1982}]{Landau:82}
Landau L.,  Lifshitz E.,  1982, Mechanics: Volume 1.
No.~v. 1, Elsevier Science, \url
  {https://books.google.com/books?id=bE-9tUH2J2wC}

\bibitem[\protect\citeauthoryear{{Landry} \& {Essick}}{{Landry} \&
  {Essick}}{2019}]{Landry:19}
{Landry} P.,  {Essick} R.,  2019, \mn@doi [\prd] {10.1103/PhysRevD.99.084049},
  \href {https://ui.adsabs.harvard.edu/abs/2019PhRvD..99h4049L} {99, 084049}

\bibitem[\protect\citeauthoryear{{Ma}, {Yu}  \& {Chen}}{{Ma}
  et~al.}{2020}]{Ma:20}
{Ma} S.,  {Yu} H.,   {Chen} Y.,  2020, \mn@doi [\prd]
  {10.1103/PhysRevD.101.123020}, \href
  {https://ui.adsabs.harvard.edu/abs/2020PhRvD.101l3020M} {101, 123020}

\bibitem[\protect\citeauthoryear{{Ma}, {Yu}  \& {Chen}}{{Ma}
  et~al.}{2021}]{Ma:21}
{Ma} S.,  {Yu} H.,   {Chen} Y.,  2021, \mn@doi [\prd]
  {10.1103/PhysRevD.103.063020}, \href
  {https://ui.adsabs.harvard.edu/abs/2021PhRvD.103f3020M} {103, 063020}

\bibitem[\protect\citeauthoryear{{Matas} et~al.,}{{Matas}
  et~al.}{2020}]{Matas:20}
{Matas} A.,  et~al., 2020, \mn@doi [\prd] {10.1103/PhysRevD.102.043023}, \href
  {https://ui.adsabs.harvard.edu/abs/2020PhRvD.102d3023M} {102, 043023}

\bibitem[\protect\citeauthoryear{{Messenger} \& {Read}}{{Messenger} \&
  {Read}}{2012}]{Messenger:12}
{Messenger} C.,  {Read} J.,  2012, \mn@doi [\prl]
  {10.1103/PhysRevLett.108.091101}, \href
  {https://ui.adsabs.harvard.edu/abs/2012PhRvL.108i1101M} {108, 091101}

\bibitem[\protect\citeauthoryear{{Nagar} et~al.,}{{Nagar}
  et~al.}{2018}]{Nagar:18}
{Nagar} A.,  et~al., 2018, \mn@doi [\prd] {10.1103/PhysRevD.98.104052}, \href
  {https://ui.adsabs.harvard.edu/abs/2018PhRvD..98j4052N} {98, 104052}

\bibitem[\protect\citeauthoryear{{Pan}, {Lyu}, {Bonga}, {Ortiz}  \&
  {Yang}}{{Pan} et~al.}{2020}]{Pan:20}
{Pan} Z.,  {Lyu} Z.,  {Bonga} B.,  {Ortiz} N.,   {Yang} H.,  2020, \mn@doi
  [\prl] {10.1103/PhysRevLett.125.201102}, \href
  {https://ui.adsabs.harvard.edu/abs/2020PhRvL.125t1102P} {125, 201102}

\bibitem[\protect\citeauthoryear{{Parisi} \& {Sturani}}{{Parisi} \&
  {Sturani}}{2018}]{Parisi:18}
{Parisi} A.,  {Sturani} R.,  2018, \mn@doi [\prd] {10.1103/PhysRevD.97.043015},
  \href {https://ui.adsabs.harvard.edu/abs/2018PhRvD..97d3015P} {97, 043015}

\bibitem[\protect\citeauthoryear{{Passamonti}, {Andersson}  \&
  {Pnigouras}}{{Passamonti} et~al.}{2021}]{Passamonti:21}
{Passamonti} A.,  {Andersson} N.,   {Pnigouras} P.,  2021, \mn@doi [\mnras]
  {10.1093/mnras/stab870}, \href
  {https://ui.adsabs.harvard.edu/abs/2021MNRAS.504.1273P} {504, 1273}

\bibitem[\protect\citeauthoryear{{Passamonti}, {Andersson}  \&
  {Pnigouras}}{{Passamonti} et~al.}{2022}]{Passamonti:22}
{Passamonti} A.,  {Andersson} N.,   {Pnigouras} P.,  2022, \mn@doi [\mnras]
  {10.1093/mnras/stac1380}, \href
  {https://ui.adsabs.harvard.edu/abs/2022MNRAS.514.1494P} {514, 1494}

\bibitem[\protect\citeauthoryear{{Poisson}}{{Poisson}}{2020}]{Poisson:20}
{Poisson} E.,  2020, \mn@doi [\prd] {10.1103/PhysRevD.101.104028}, \href
  {https://ui.adsabs.harvard.edu/abs/2020PhRvD.101j4028P} {101, 104028}

\bibitem[\protect\citeauthoryear{{Poisson} \& {Will}}{{Poisson} \&
  {Will}}{2014}]{Poisson:14}
{Poisson} E.,  {Will} C.~M.,  2014, {Gravity}.
Cambridge University Press

\bibitem[\protect\citeauthoryear{{Pratten}, {Schmidt}  \& {Williams}}{{Pratten}
  et~al.}{2022}]{Pratten:22}
{Pratten} G.,  {Schmidt} P.,   {Williams} N.,  2022, \mn@doi [\prl]
  {10.1103/PhysRevLett.129.081102}, \href
  {https://ui.adsabs.harvard.edu/abs/2022PhRvL.129h1102P} {129, 081102}

\bibitem[\protect\citeauthoryear{{Read}, {Markakis}, {Shibata}, {Ury{\={u}}},
  {Creighton}  \& {Friedman}}{{Read} et~al.}{2009}]{Read:09}
{Read} J.~S.,  {Markakis} C.,  {Shibata} M.,  {Ury{\={u}}} K.,  {Creighton} J.
  D.~E.,   {Friedman} J.~L.,  2009, \mn@doi [\prd]
  {10.1103/PhysRevD.79.124033}, \href
  {https://ui.adsabs.harvard.edu/abs/2009PhRvD..79l4033R} {79, 124033}

\bibitem[\protect\citeauthoryear{{Reisenegger} \& {Goldreich}}{{Reisenegger} \&
  {Goldreich}}{1994}]{Reisenegger:94}
{Reisenegger} A.,  {Goldreich} P.,  1994, \mn@doi [\apj] {10.1086/174105},
  \href {https://ui.adsabs.harvard.edu/abs/1994ApJ...426..688R} {426, 688}

\bibitem[\protect\citeauthoryear{{Saffer} \& {Yagi}}{{Saffer} \&
  {Yagi}}{2021}]{Saffer:21}
{Saffer} A.,  {Yagi} K.,  2021, \mn@doi [\prd] {10.1103/PhysRevD.104.124052},
  \href {https://ui.adsabs.harvard.edu/abs/2021PhRvD.104l4052S} {104, 124052}

\bibitem[\protect\citeauthoryear{{Sathyaprakash} et~al.,}{{Sathyaprakash}
  et~al.}{2012}]{Sathyaprakash:12}
{Sathyaprakash} B.,  et~al., 2012, \mn@doi [Classical and Quantum Gravity]
  {10.1088/0264-9381/29/12/124013}, \href
  {http://adsabs.harvard.edu/abs/2012CQGra..29l4013S} {29, 124013}

\bibitem[\protect\citeauthoryear{{Schenk}, {Arras}, {Flanagan}, {Teukolsky}  \&
  {Wasserman}}{{Schenk} et~al.}{2002}]{Schenk:02}
{Schenk} A.~K.,  {Arras} P.,  {Flanagan} {\'E}.~{\'E}.,  {Teukolsky} S.~A.,
  {Wasserman} I.,  2002, \mn@doi [\prd] {10.1103/PhysRevD.65.024001}, \href
  {https://ui.adsabs.harvard.edu/abs/2002PhRvD..65b4001S} {65, 024001}

\bibitem[\protect\citeauthoryear{{Steinhoff}, {Hinderer}, {Buonanno}  \&
  {Taracchini}}{{Steinhoff} et~al.}{2016}]{Steinhoff:16}
{Steinhoff} J.,  {Hinderer} T.,  {Buonanno} A.,   {Taracchini} A.,  2016,
  \mn@doi [\prd] {10.1103/PhysRevD.94.104028}, \href
  {https://ui.adsabs.harvard.edu/abs/2016PhRvD..94j4028S} {94, 104028}

\bibitem[\protect\citeauthoryear{{Steinhoff}, {Hinderer}, {Dietrich}  \&
  {Foucart}}{{Steinhoff} et~al.}{2021}]{Steinhoff:21}
{Steinhoff} J.,  {Hinderer} T.,  {Dietrich} T.,   {Foucart} F.,  2021, \mn@doi
  [Physical Review Research] {10.1103/PhysRevResearch.3.033129}, \href
  {https://ui.adsabs.harvard.edu/abs/2021PhRvR...3c3129S} {3, 033129}

\bibitem[\protect\citeauthoryear{{Thorne}}{{Thorne}}{1980}]{Thorne:80}
{Thorne} K.~S.,  1980, \mn@doi [Reviews of Modern Physics]
  {10.1103/RevModPhys.52.299}, \href
  {https://ui.adsabs.harvard.edu/abs/1980RvMP...52..299T} {52, 299}

\bibitem[\protect\citeauthoryear{{Townsend} \& {Teitler}}{{Townsend} \&
  {Teitler}}{2013}]{Townsend:13}
{Townsend} R.~H.~D.,  {Teitler} S.~A.,  2013, \mn@doi [\mnras]
  {10.1093/mnras/stt1533}, \href
  {https://ui.adsabs.harvard.edu/abs/2013MNRAS.435.3406T} {435, 3406}

\bibitem[\protect\citeauthoryear{{Townsend}, {Goldstein}  \&
  {Zweibel}}{{Townsend} et~al.}{2018}]{Townsend:18}
{Townsend} R.~H.~D.,  {Goldstein} J.,   {Zweibel} E.~G.,  2018, \mn@doi
  [\mnras] {10.1093/mnras/stx3142}, \href
  {https://ui.adsabs.harvard.edu/abs/2018MNRAS.475..879T} {475, 879}

\bibitem[\protect\citeauthoryear{{Tsang}, {Read}, {Hinderer}, {Piro}  \&
  {Bondarescu}}{{Tsang} et~al.}{2012}]{Tsang:12}
{Tsang} D.,  {Read} J.~S.,  {Hinderer} T.,  {Piro} A.~L.,   {Bondarescu} R.,
  2012, \mn@doi [\prl] {10.1103/PhysRevLett.108.011102}, \href
  {https://ui.adsabs.harvard.edu/abs/2012PhRvL.108a1102T} {108, 011102}

\bibitem[\protect\citeauthoryear{{Van Hoolst}}{{Van
  Hoolst}}{1994}]{VanHoolst:94}
{Van Hoolst} T.,  1994, \aap, \href
  {https://ui.adsabs.harvard.edu/abs/1994A&A...286..879V} {286, 879}

\bibitem[\protect\citeauthoryear{{Venumadhav}, {Zimmerman}  \&
  {Hirata}}{{Venumadhav} et~al.}{2014}]{Venumadhav:14}
{Venumadhav} T.,  {Zimmerman} A.,   {Hirata} C.~M.,  2014, \mn@doi [ApJ]
  {10.1088/0004-637X/781/1/23}, \href
  {http://adsabs.harvard.edu/abs/2014ApJ...781...23V} {781, 23}

\bibitem[\protect\citeauthoryear{{Vick} \& {Lai}}{{Vick} \&
  {Lai}}{2019}]{Vick:19}
{Vick} M.,  {Lai} D.,  2019, \mn@doi [\prd] {10.1103/PhysRevD.100.063001},
  \href {https://ui.adsabs.harvard.edu/abs/2019PhRvD.100f3001V} {100, 063001}

\bibitem[\protect\citeauthoryear{{Wang} \& {Lai}}{{Wang} \&
  {Lai}}{2020}]{Wang:20}
{Wang} J.-S.,  {Lai} D.,  2020, \mn@doi [\prd] {10.1103/PhysRevD.102.083005},
  \href {https://ui.adsabs.harvard.edu/abs/2020PhRvD.102h3005W} {102, 083005}

\bibitem[\protect\citeauthoryear{{Weinberg}}{{Weinberg}}{2016}]{Weinberg:16}
{Weinberg} N.~N.,  2016, \mn@doi [ApJ] {10.3847/0004-637X/819/2/109}, \href
  {http://adsabs.harvard.edu/abs/2016ApJ...819..109W} {819, 109}

\bibitem[\protect\citeauthoryear{{Weinberg}, {Arras}, {Quataert}  \&
  {Burkart}}{{Weinberg} et~al.}{2012}]{Weinberg:12}
{Weinberg} N.~N.,  {Arras} P.,  {Quataert} E.,   {Burkart} J.,  2012, \mn@doi
  [\apj] {10.1088/0004-637X/751/2/136}, \href
  {https://ui.adsabs.harvard.edu/abs/2012ApJ...751..136W} {751, 136}

\bibitem[\protect\citeauthoryear{{Weinberg}, {Arras}  \& {Burkart}}{{Weinberg}
  et~al.}{2013}]{Weinberg:13}
{Weinberg} N.~N.,  {Arras} P.,   {Burkart} J.,  2013, \mn@doi [\apj]
  {10.1088/0004-637X/769/2/121}, \href
  {https://ui.adsabs.harvard.edu/abs/2013ApJ...769..121W} {769, 121}

\bibitem[\protect\citeauthoryear{Wu}{Wu}{1998}]{Wu:98}
Wu Y.,  1998, PhD thesis, California Institute of Technology,
  \mn@doi{10.7907/nc60-8v65}, \url {https://thesis.library.caltech.edu/3736/}

\bibitem[\protect\citeauthoryear{{Xu} \& {Lai}}{{Xu} \& {Lai}}{2017}]{Xu:17}
{Xu} W.,  {Lai} D.,  2017, \mn@doi [\prd] {10.1103/PhysRevD.96.083005}, \href
  {https://ui.adsabs.harvard.edu/abs/2017PhRvD..96h3005X} {96, 083005}

\bibitem[\protect\citeauthoryear{{Yagi} \& {Yunes}}{{Yagi} \&
  {Yunes}}{2013}]{Yagi:13}
{Yagi} K.,  {Yunes} N.,  2013, \mn@doi [Science] {10.1126/science.1236462},
  \href {https://ui.adsabs.harvard.edu/abs/2013Sci...341..365Y} {341, 365}

\bibitem[\protect\citeauthoryear{{Yakovlev}, {Levenfish}  \&
  {Shibanov}}{{Yakovlev} et~al.}{1999}]{Yakovlev:99}
{Yakovlev} D.~G.,  {Levenfish} K.~P.,   {Shibanov} Y.~A.,  1999, \mn@doi
  [Physics Uspekhi] {10.1070/PU1999v042n08ABEH000556}, \href
  {https://ui.adsabs.harvard.edu/abs/1999PhyU...42..737Y} {42, 737}

\bibitem[\protect\citeauthoryear{{Yang}}{{Yang}}{2019}]{Yang:19}
{Yang} H.,  2019, \mn@doi [\prd] {10.1103/PhysRevD.100.064023}, \href
  {https://ui.adsabs.harvard.edu/abs/2019PhRvD.100f4023Y} {100, 064023}

\bibitem[\protect\citeauthoryear{{Yang}, {East}, {Paschalidis}, {Pretorius}  \&
  {Mendes}}{{Yang} et~al.}{2018}]{Yang:18}
{Yang} H.,  {East} W.~E.,  {Paschalidis} V.,  {Pretorius} F.,   {Mendes} R.
  F.~P.,  2018, \mn@doi [\prd] {10.1103/PhysRevD.98.044007}, \href
  {https://ui.adsabs.harvard.edu/abs/2018PhRvD..98d4007Y} {98, 044007}

\bibitem[\protect\citeauthoryear{{Yu} \& {Weinberg}}{{Yu} \&
  {Weinberg}}{2017a}]{Yu:17a}
{Yu} H.,  {Weinberg} N.~N.,  2017a, \mn@doi [\mnras] {10.1093/mnras/stw2552},
  \href {https://ui.adsabs.harvard.edu/abs/2017MNRAS.464.2622Y} {464, 2622}

\bibitem[\protect\citeauthoryear{{Yu} \& {Weinberg}}{{Yu} \&
  {Weinberg}}{2017b}]{Yu:17b}
{Yu} H.,  {Weinberg} N.~N.,  2017b, \mn@doi [\mnras] {10.1093/mnras/stx1188},
  \href {https://ui.adsabs.harvard.edu/abs/2017MNRAS.470..350Y} {470, 350}

\bibitem[\protect\citeauthoryear{{Yu}, {Weinberg}  \& {Fuller}}{{Yu}
  et~al.}{2020}]{Yu:20a}
{Yu} H.,  {Weinberg} N.~N.,   {Fuller} J.,  2020, \mn@doi [\mnras]
  {10.1093/mnras/staa1858}, \href
  {https://ui.adsabs.harvard.edu/abs/2020MNRAS.496.5482Y} {496, 5482}

\bibitem[\protect\citeauthoryear{{Yu}, {Weinberg}  \& {Arras}}{{Yu}
  et~al.}{2021}]{Yu:21}
{Yu} H.,  {Weinberg} N.~N.,   {Arras} P.,  2021, \mn@doi [\apj]
  {10.3847/1538-4357/ac0a79}, \href
  {https://ui.adsabs.harvard.edu/abs/2021ApJ...917...31Y} {917, 31}

\bibitem[\protect\citeauthoryear{{Yu}, {Weinberg}  \& {Arras}}{{Yu}
  et~al.}{2022}]{Yu:22a}
{Yu} H.,  {Weinberg} N.~N.,   {Arras} P.,  2022, \mn@doi [\apj]
  {10.3847/1538-4357/ac5627}, \href
  {https://ui.adsabs.harvard.edu/abs/2022ApJ...928..140Y} {928, 140}

\makeatother
\end{thebibliography}




\appendix
\onecolumn

\section{A toy model demonstrating the \lowercase{f}-mode frequency shifts}
\label{appx:toy}

\begin{figure}
   \centering
   \includegraphics[width=0.55\textwidth]{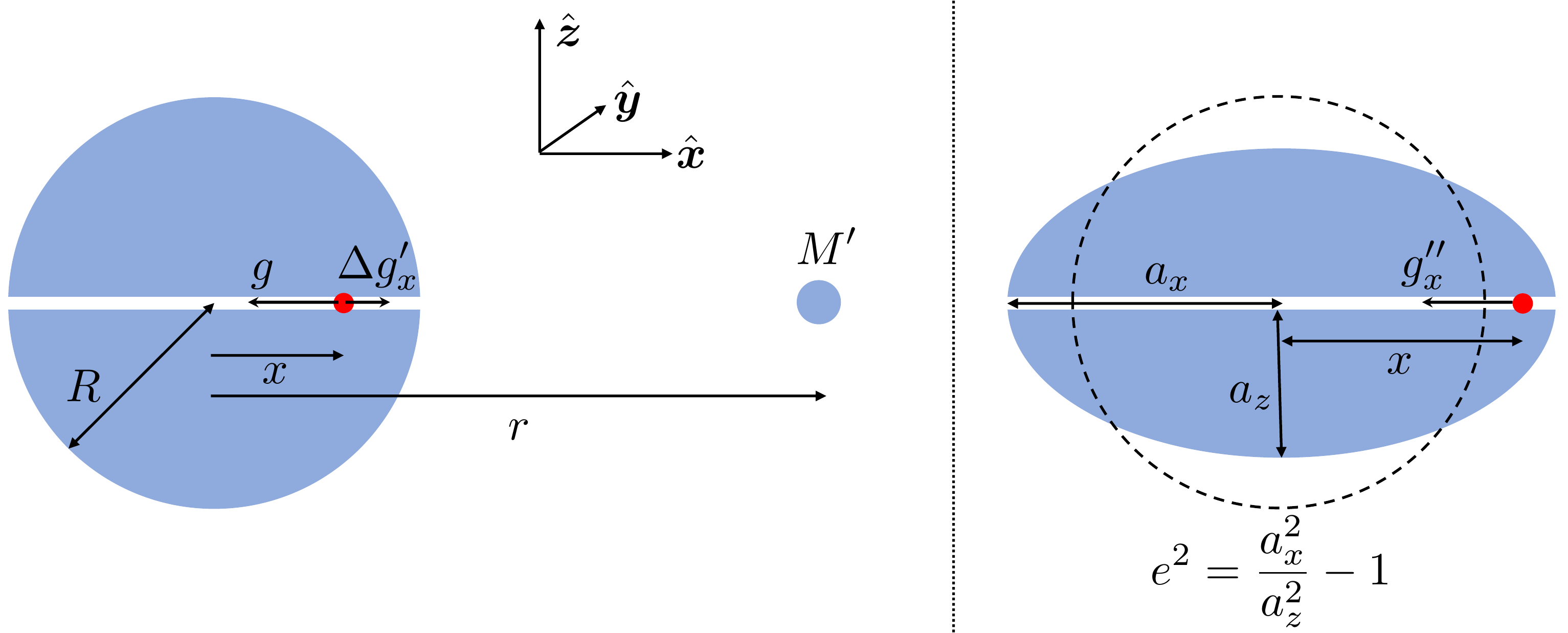} 
   \caption{A toy model demonstrating the origin of the f-mode frequency shift. A test particle released in a hole through a star's center will oscillate in the hole as a harmonic oscillator with the same frequency as the f-mode (up to an order unity constant). On the left, the oscillation frequency of the test particle is reduced by the tidal acceleration $\Delta g'_x$ from a companion of mass $M'$ [similar to Eq.~(\ref{eq:domega_a_leading})]. On the right, the oscillation frequency is reduced by the tidal deformation of the background star caused by the f-modes, corresponding to the oscillator's anharmonicity (Appendix~\ref{appx:anharm}). 
   }
   \label{fig:cartoon}
\end{figure}

The frequency shift of the f-modes derived in Eq.~(\ref{eq:domega_a_leading}) can be understood from a toy model illustrated in the left part of Fig.~\ref{fig:cartoon}. 
Imagine we drill a hole through the center of a uniform-density star in the equatorial plane where the $l=2$, $|m|=2$ f-modes mainly reside. 
If we drop a test particle  into the hole (red dot in Fig.~\ref{fig:cartoon}), the particle will oscillate inside the hole as a harmonic oscillator. When the particle is at $x$, it feels an inward acceleration $g=4\pi \rho x/3$, and its equation of motion is given by
$\ddot{x} = -4\pi \rho x/3=-\omega^2 x$. The oscillation frequency is therefore $\omega=\sqrt{4\pi \rho/3}$, the same as the f-mode frequency up to an order unity constant. Indeed, both oscillations are similar in nature as they are both characterized by the dynamical frequency of the star. 

If now a companion of mass $M'$ is present in the equatorial plane and the separation vector is along the direction of the hole (in the $x$ direction), then $M'$ will produce a tidal acceleration $\Delta g'_x=2\left(M'/r^2\right) (x/r)$ in the opposite direction of $g$. In this case, the test particle still behaves as a harmonic oscillator, though its equation of motion is now modified by $\Delta g'_x$ as
\begin{equation}
    \ddot{x} = -(g - \Delta g'_x) = -\frac{4\pi}{3} \rho \left[1 - 2\left(\frac{M'}{M}\right)\left(\frac{R}{r}\right)^3\right] x
    =-\omega^2 \left(1 + 2\frac{\Delta \omega_x}{\omega}\right)x. 
\end{equation}
If instead, the hole is perpendicular to the orbital vector but still in the equatorial plane (i.e., along the $y$ direction), then a test particle oscillating in it  experiences a tidal acceleration $|\Delta g'_y| = \left(M'/r^2\right) (y/r)$ pointing towards the center of the star. In this case, the oscillation is increased by $\Delta \omega_y/\omega_y= (M'/M)(R/r)^3/2=-\Delta \omega_x/\omega/2$ due to $M'$. Since the $l=2$, $|m|=2$ f-modes mainly oscillate in the equatorial plane, it experiences a frequency shift that is approximately the mean of these two results,
\begin{equation}
    \frac{\Delta \omega}{\omega} \sim \frac{1}{2}\frac{\Delta \omega_x + \Delta \omega_y}{\omega} = -\frac{1}{4}\frac{M'}{M}\left(\frac{R}{r}\right)^3 = -\frac{1}{4}\frac{M'}{M}\frac{\Omega^2}{M_{\rm t}} R^3<0.
    \label{eq:toy_domega_eq}
\end{equation}
Up to a constant of order unity, this agrees with the  leading-order frequency shift of the $f$-mode found in Section~\ref{sec:mode_nl} and given by  Eq.~(\ref{eq:domega_a_leading}). In other words, the presence of the companion's tidal field reduces the eigenfrequency of the f-mode, thereby enhancing the finite frequency response to the tidal drive. 

If we further let the star to be deformed by the tide into an ellipsoid (with $a_x>R$ and $a_y=a_z<R$; see the right part of Fig.~\ref{fig:cartoon}), then the gravitational acceleration $g$ should be replaced by $g''_x=(1-2e^2/5)g$ along the $x$ direction,  where $e=\sqrt{a_x^2/a_z^2-1}\propto \xi^r(R)/R\propto |C_a|$ is the eccentricity of the ellipsoid. The reduction of the inward gravitational acceleration will also cause a frequency shift of the test particle's oscillation (see, e.g., ~\citealt{Poisson:14}), 
\begin{equation}
    \frac{\Delta \omega_x}{\omega} = \frac{1}{2}\frac{{g''_x}-g}{g} = -\frac{1}{5}e^2 \propto |C_a|^2 \propto \left(\frac{R}{r}\right)^6. 
    \label{eq:toy_anharm}
\end{equation}
The oscillation frequency along the $y$ direction will increase since the gravitational acceleration along the $y$ direction is $g''_y = (1+e^2/5)g$. Nonetheless, $\Delta \omega_y=-\Delta \omega_x/2$, and therefore, on average the $l=|m|=2$ f-modes will experience a negative frequency shift $\Delta \omega/\omega\sim -1/e^2<0$.
Such a frequency shift can also be understood from the fact that the tidal deformation tends to reduce the density of the star~\cite{Lai:96}, hence reducing the f-mode frequency $\propto \sqrt{\rho}$. 
We note the frequency shift due to this effect is formally a higher-order correction than the shift induced directly by the companion's tidal acceleration in Eq.~(\ref{eq:toy_domega_eq}). In fact, Eq.~(\ref{eq:toy_anharm}) describes the anharmonicity of an oscillator. A detailed derivation of the anharmonicity from the modal expansion analysis is presented in Appendix~\ref{appx:anharm} below. 

\section{Anharmonic frequency shift}
\label{appx:anharm}

We extend the analysis in Section~\ref{sec:mode_nl} to demonstrate the appearance of the anharmonicity in the modal picture. We will also demonstrate the significance of including four-mode coupling terms [terms $\propto \eta_{abcd}$ in Eq.~(\ref{eq:ode_mode_amp_general})] in solving the numerical equations. 

We use the same convention adopted in Section~\ref{sec:mode_nl} and use $(a, b, c)$ to specifically denote $l=2$ modes with $(m_a, m_b, m_c)=(2, -2, 0)$. Since only the $a$ mode will experience the most significant dynamical tide effect near the merger when $2\Omega \lesssim \omega_a$, we thus solve $C_b$ and $C_c$ in terms of $C_a$ as well as their linear solutions $B_b$ and $B_c$.
We have
\begin{align}
    &C_b\simeq B_b +  \frac{\omega}{\omega + 2\Omega}\left(V_{20}+ 4 \kappa_2 {\rm Re}[B_c] \right)C_a^\ast + \Delta C_b, \\
    &C_c\simeq B_c + 2\kappa_2 |C_a|^2 + 2V_{22} {\rm Re}\left[C_a\right] + 4\kappa_{2} {\rm Re}\left[C_a B_b\right] + \Delta C_c,
\end{align}
where 
\begin{align}
    & \Delta C_b \simeq \frac{\omega}{\omega+2\Omega}
      \left(V_{20}B_b + 4\kappa_2 {\rm Re}\left[B_c\right]B_b + 2V_{22}{\rm Re}\left[B_c\right]\right), \\
    & \Delta C_c \simeq 2 \left(V_{22}{\rm Re}\left[B_b\right] + V_{00}{\rm Re}\left[B_c\right]  +\kappa_2 |B_b|^2 + \kappa_0 {\rm Re}\left[B_c B_c\right]+\kappa_0 |B_c|^2\right).
\end{align}

Plugging $C_b$ and $C_c$ back to the equation of $C_a$, we have
\begin{equation}
    \dot{C}_a + i(\omega_a + \Delta \omega_a^{(\rm 3m)} -m_a\dot{\phi}) C_a = i\omega_a [V_a+ \Delta V_a^{(\rm 3m)}],
\end{equation}
where 
\begin{align}
    \Delta V_a^{(\rm 3m)} 
    =& V_{20}B_b^\ast + 2V_{22} {\rm Re}[B_c] + 4 \kappa_2 B_b^\ast {\rm Re}[B_c] 
    + 4 \kappa_2(V_{22} + 2 \kappa_2 B_b^\ast)|C_a|^2 \nonumber \\
    & + 4 |V_{22} + 2 \kappa_{22} B_b|^2{\rm Re}[C_a] 
    + V_{20} \Delta C_b^\ast + 2 V_{22}{\rm Re}[\Delta C_c] 
    + 4\kappa_2 B_b^\ast {\rm Re}[\Delta C_c] + 4 \kappa_2 {\rm Re}[B_c]\Delta B_b^\ast,
    \label{eq:dV_a_3md}
\end{align}
and
\begin{align}
    \left(\frac{\Delta \omega_a}{\omega_a}\right)^{(\rm 3m)} 
    =& -\left(V_{20} + 4\kappa_2{\rm Re}[B_c]\right) -8\kappa_2^2|C_a|^2 \nonumber\\
    &-8\kappa_2 V_{22}{\rm Re}[C_a] - 16 \kappa_2^2 \left[B_b C_a\right] -\frac{\omega_a}{\omega_a+2\dot{\phi}}(V_{20}+4\kappa_2{\rm Re}[B_c])^2 - 4 \kappa_2  {\rm Re}[\Delta C_c].
    \label{eq:domega_a_3md}
\end{align}
As shown in Fig.~\ref{fig:domega_vs_dVa}, our focus will be on the frequency shift. In Eq.~(\ref{eq:domega_a_3md}), the first term is the result we quote in Eq.~(\ref{eq:domega_a_leading}) as it formally scales as $(R/r)^3$ while the rest of the terms $\propto (R/r)^6$. Note that this term exists because the system is continuously forced by the tide and $|V_{20}|\sim |C_c|\sim |C_a|$. 

On the other hand, we may have $|C_a|\gg  |V_{20}|\sim |C_c|$ if mode $a$ is resonantly excited due to, e.g., a rotating NS~\citep{Ma:20, Steinhoff:21} and/or orbital eccentricity~\citep{Yang:19}. In this limit, the $-8\kappa_2^2 |C_a|^2$ term may dominant the frequency shift. Since $|C_a|^2$ corresponds to the energy of mode $a$, we notice that it corresponds to the anharmonicity of a \emph{free} oscillator~\citep{Landau:82}. As discussed in \citet{Yu:21}, mode $a$ can couple to not only $l=2$ modes but also $l=0$ and $l=4$ ones. The total anharmonic frequency shift of mode $a$ due to three-mode interaction can be written as 
\begin{equation}
    \left(\frac{\Delta \omega_a}{\omega_a}\right)^{(\rm 3m)} \simeq -\left(\sum_{d, m_d=-4}^{l_d=4}\frac{2\omega_d}{4\Omega + \omega_d}\kappa_{aad}^2 + \sum_{e, m_e=0}^{l_e=0,2,4}4\kappa_{aa^\ast e}^2\right) \big{|}C_a\big{|}^2.
    \label{eq:domega_a_3md_Ea}
\end{equation}
We have dropped terms that do not scale as $|C_a|^2$ since we have assumed mode $a$ is approximated by a free oscillator with its amplitude much greater than other modes and the equilibrium tide.

Formally at the same order, four-mode coupling could also contribute to the anharmonic frequency shift.\footnote{In the original analysis of \citet{Yu:21} (and in \citealt{Kumar:94}), the four-mode contribution was ignored. Additionally, there was a numerical error that overestimates the contribution of p-modes to the frequency shift. An erratum to \citet{Yu:21} is under preparation at the moment of preparing this work.} 
We only explicitly write out the equation for $C_a$. 
\begin{align}
    \dot{C}_a + i(\omega_a-m_a\dot{\phi}) C_a &= \text{(linear and three-mode terms)}+i\omega_a (
      3 \eta_{aaa^\ast a^\ast} |C_a|^2 C_a \nonumber \\
    & + 6\eta_{aaa^\ast b} |C_a|^2 C_b^\ast + 3\eta_{aa^\ast a^\ast b^\ast}C_a^2 C_b  
    + 6\eta_{aa^\ast b b^\ast}C_a |C_b|^2 + 3\eta_{aab b} C_a^\ast (C_b^\ast)^2 \nonumber \\
    & + 6\eta_{aa^\ast cc^\ast} C_a |C_c|^2 
    + 3\eta_{aa^\ast cc} C_a (C_c^\ast)^2 + 3\eta_{aa^\ast c^\ast c^\ast} C_a C_c^2 \nonumber \\
    & + 3\eta_{abbb^\ast} |C_b|^2 C_b + 6\eta_{ab cc^\ast} C_b |C_c|^2 + 3\eta_{ab cc} C_b (C_c^\ast)^2 + 3\eta_{ab c^\ast c^\ast} C_b C_c^2
    ).
\end{align}
By collecting terms  $\propto C_a$ on the right-hand side, we can read out directly the four-mode contributions to the anharmonic frequency shift. 
\begin{align}
    \left(\frac{\Delta \omega_a}{\omega_a}\right)^{(\rm 4m)} \simeq 
    - 3 \eta_{22} |C_a|^2 - 3 \eta_{22} C_a C_b
    - 6 \eta_{22} |C_b|^2 -12 \eta_{20} |C_c|^2 \simeq - 3 \eta_{22} |C_a|^2,
    \label{eq:domega_a_4md}
\end{align}
where we have made the approximation that $C_c\simeq C_c^\ast$ and defined
\begin{align*}
    &\eta_{22} = \eta_{aaa^\ast a^\ast} \text{ and similar terms}, \\
    &\eta_{20} = \eta_{aa^\ast cc^\ast} \text{ and similar terms}.  
\end{align*}
Their values are in Table~\ref{tab:coup_coeff}. In the second equality in Eq.~(\ref{eq:domega_a_4md}), we dropped terms that do not scale as $|C_a|^2$ since the anharmonic effect could be significant only if $|C_a|\gg|C_{b,c}|$. This is also why we ignore the four-mode counterpart of the $U_{ablm}$ term as it does not contribute to $\left({\Delta \omega_a}/{\omega_a}\right)$ with terms $\propto |C_a|^2$. 

Note that when $\Omega \ll |\omega_d|$, the three-mode interaction always lowers the frequency $\omega_a$ [Eq.~(\ref{eq:domega_a_3md_Ea})]. On the other hand,  the four-mode interaction increases $\omega_a$ (as $\eta_{22}<0$), making $\Delta \omega_a$ 70\% smaller than the value predicted using three-mode coupling only. Numerically, we find $\Delta \omega_a/\omega_a\simeq -2.7 |C_a|^2$ using the coupling coefficients presented in Table~\ref{tab:coup_coeff}. To arrive at these numbers, we note that the Cowling approximation should not be assumed for f-modes as terms due to perturbed gravity could significantly modify the coupling coefficients. We thus extend \citet{Weinberg:16} and show explicitly in Appendix~\ref{appx:kap4_del_Phi} the expressions for the perturbed gravity terms in $\eta_{abcd}$. 

\begin{figure*}
\subfloat{
	\begin{minipage}[c][0.915\width]{0.48\textwidth}
	   \centering
	   \includegraphics[width=\textwidth]{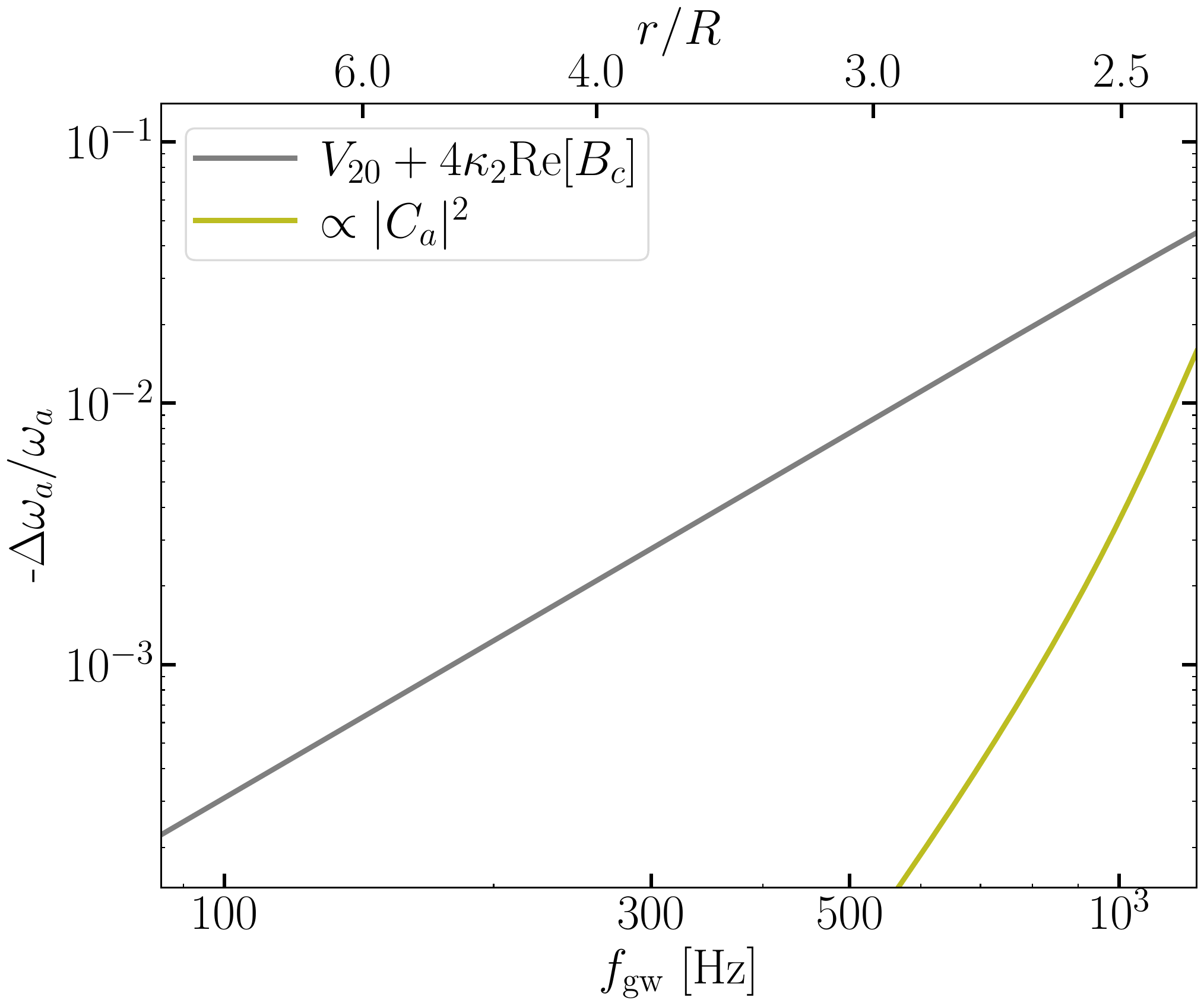}
	\end{minipage}}
 \hfill 	
  \subfloat{
	\begin{minipage}[c][0.915\width]{0.48\textwidth}
	   \centering
	   \includegraphics[width=\textwidth]{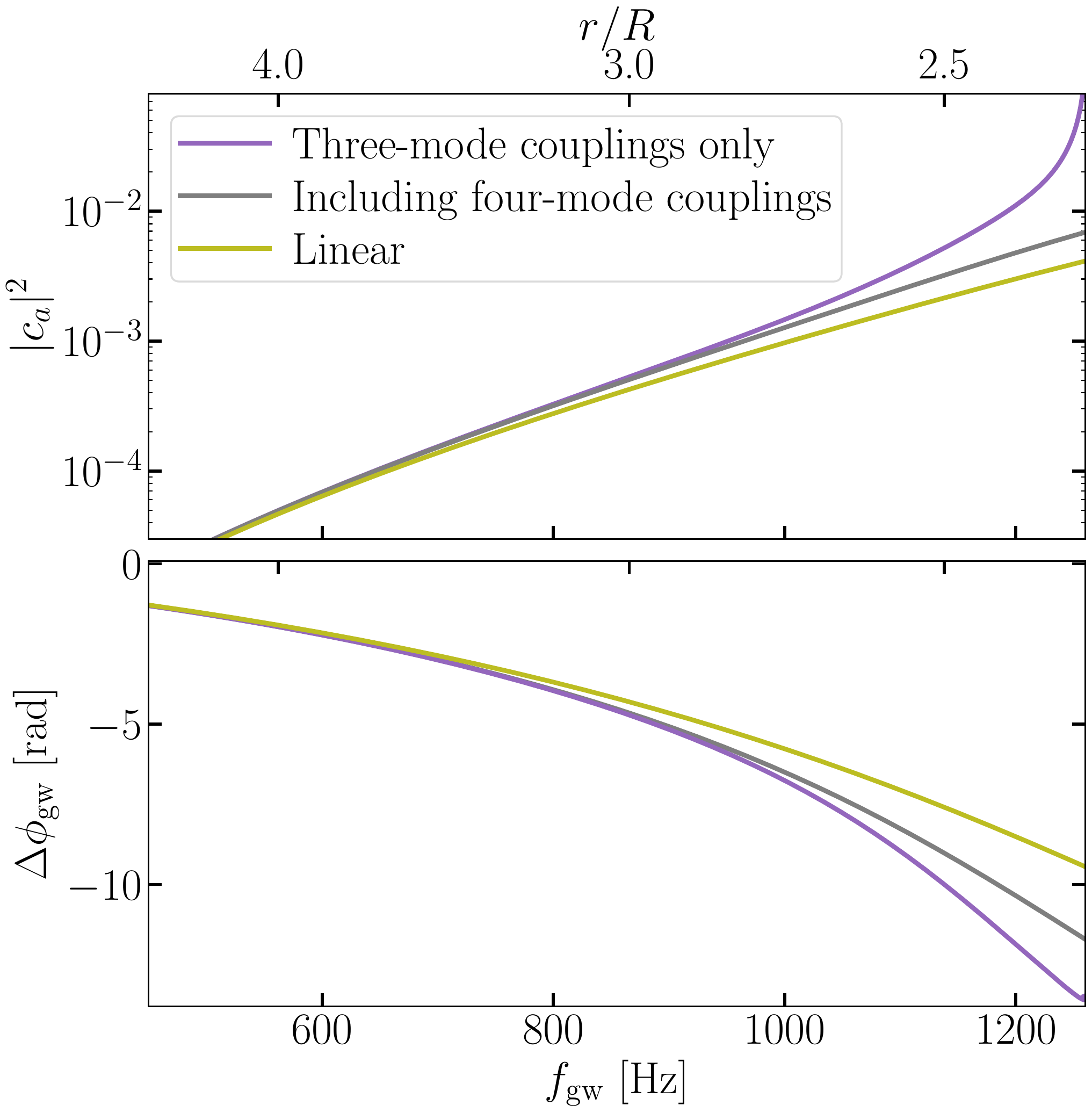}
	\end{minipage}}
\caption{Left: comparison of the non-linear frequency shift. The gray trace is the result from Eq.~(\ref{eq:domega_a_leading}) which we use in the main text, while the olive curve is due to the anharmonicity [the sum of terms $\propto |C_a|^2$ in Eqs.~(\ref{eq:domega_a_3md_Ea}) and (\ref{eq:domega_a_4md})]. Because the finite-frequency effect is moderate in non-rotating, Newtonian NSs, the anharmonicity is small and can be ignored. Right: comparison of mode energy (of the $l_a=m_a=2$ mode) and GW phase shift with (gray) and without (purple) four-mode coupling terms. Near the final merger, the three-mode-only result in fact experiences an artificial run away. It thus indicates the significance of incorporating four-mode couplings in the numerical integration. 
}
\label{fig:anharm}
\end{figure*}

In the left panel of Fig.~\ref{fig:anharm}, we compare the anharmonic frequency shift (i.e., frequency shift proportional to the energy of the mode) with Eq.~(\ref{eq:domega_a_leading}). Because $\omega_a/(\omega_a-2\Omega)\simeq 2$ when $f_{\rm gw}=1000\,{\rm Hz}$, we have $|C_a|\sim |C_c|$ and thus Eq.~(\ref{eq:domega_a_leading}) dominates throughout the evolution. On the other hand, it is possible for us to enter the regime where the anharmonicity becomes more significant if $|C_a|$ is enhanced relative to other modes due to NS rotation and/or orbital eccentricity (together with relativistic redshifts; see the discussion in Section~\ref{sec:discussion}). We defer to future studies to explore this possibility.

Meanwhile, we also emphasize that the partial cancellation between the three-mode and four-mode coupling terms as illustrated in our discussion on $\Delta \omega_a/\omega_a$ [Eqs.~(\ref{eq:domega_a_3md_Ea}) and (\ref{eq:domega_a_4md})] is in fact a key reason why our analytical result could be accurate with just one iteration of perturbative calculation (Section~\ref{sec:mode_nl}). In the right panel of Fig.~\ref{fig:anharm}, we compare the evolution of mode energy (top panel) and GW phase shift (bottom panel) with and without four-mode coupling terms. In particular, we obtain the purple curve by setting $\eta_{abcd}$ to 0 when numerically integrating the differential equations. Note that near the merger, the mode energy runs away unphysically. The reason is illustrated in appendix D of \citet{Wu:98} which we briefly recap below. Consider a toy model where the dynamics of a three-mode system could be describe by a potential $\Psi$ with
\begin{equation}
    \Psi = E (A^2 - A^3),
\end{equation}
where $A$ corresponds to mode amplitude and $E>0$ is a constant [cf. Eq.~(\ref{eq:E_mode})]. For small oscillations with $|A|\ll 1$, the motion is bound with small corrections from the three-mode interaction. However, if $A\gg 2/3$, the system could climb over the potential well and escape to $A\to\infty$, as seen in the purple curves in Fig.~\ref{fig:anharm}. The four-mode interaction stabilizes the system by adding an $A^4$ piece in the potential.

\section{Mass quadrupole and Love number}
\label{appx:quadrupole}

In this appendix, we present the relation between the mass quadrupole tensor and the quadrupole of a mode with angular quantum number $(l_a, m_a)$. This will make the connection between the tidal overlap $I_a$ and the love number $k_2$ more transparent. It is also useful for deriving the Burke-Thorne dissipation terms in the differential equations. 

We first consider the mass quadrupole of the NS induced by the tide. Focusing on the quadrupole with spherical harmonic degree $(l, m)$ and using a Lagrangian picture with perturbed quantities denoted by the prime symbol, $\vect{x'}=\vect{x} + \vect{\xi}$,  we have
\begin{align}
    Q_{lm}^{\rm ns}
    &= \int \diff^3 x' \rho (x') x'^l Y^\ast_{lm}(\theta', \phi') - \int \diff^3 x \rho (x) x^l Y^\ast_{lm}(\theta, \phi),\nonumber \\
    &= \int \diff^3 x \rho(x) \left\{\vect{\xi} \cdot \nabla\left[x^l Y^\ast_{lm}(\theta, \phi)\right] + \frac{1}{2} \vect{\xi} \cdot \left(\vect{\xi}\cdot \nabla\right)\nabla\left[x^l Y^\ast_{lm}(\theta, \phi)\right]  \right\}, \nonumber \\
    &= \left(\sum_a^{m_a=m} I_a c_a  + \frac{1}{2}\sum_{ab}^{m_a+m_b+m=0} J_{ablm} c_a^\ast c_b^\ast \right) MR^l, 
    \label{eq:Qlm}
\end{align}
where in the second line we have used $\rho(x')\diff^3 x' = \rho(x)\diff^3 x$ for mass conservation and then expanded $x^l Y_{lm}^\ast$ around $\vect{x}$ to second order in $\vect{\xi}$. Then in the third line, we have first expanded $\vect{\xi}$ into eigenmodes and then used directly the definition of $I_a$ and $J_{ablm}$. Note that the first, linear term selects out modes with $(l_a, m_a)=(l, m)$ and the second, non-linear term has contributions from modes with $m_a+m_b+m=0$, $l_a+l_b+l={\rm even}$, and $|l_a-l_b|\leq l \leq l_a+l_b$. 

To get the quadrupole in the Cartesian coordinate, we follow \citet{Poisson:14} and use the tensor spherical harmonic $\mathcal{Y}$ defined through $Y_{lm}(\theta, \phi) = {\mathcal{Y}_{lm}^{\langle i_1 ... i_l \rangle}}^\ast n_{\langle i_1 ... i_l \rangle}(\theta, \phi)$, where $n_{\langle i_1 ... i_l\rangle}$ is a tensor formed by unit vectors $n^i$, with $\vect{n}=\left[\sin\theta\cos\phi, \sin\theta\sin\phi, \cos\theta\right]^T$. The angular bracket denotes taking the STF part. Since the background geometry is Euclidean, we have $n^i=n_i$. This allows us to relate the quadrupole evaluated for a particular spherical harmonic (which is directly obtained from our modal decomposition) to that in the Cartesian coordinate (which is convenient for computing, e.g., the Burke-Thorne terms) as
\begin{equation}
Q^{\langle i_1 ... i_l \rangle}_{\rm ns} = N_l \sum_m {\mathcal{Y}_{lm}^{\langle i_1 ... i_l \rangle}}^\ast Q_{lm}^{\rm ns}
\end{equation}
where $N_l = 4\pi l!/(2l+1)!!$. Note further $\mathcal{Y}_{l,-m}^{\langle i_i ... i_l \rangle} = (-1)^m {\mathcal{Y}_{lm}^{\langle i_1 ... i_l\rangle}}^\ast$. 
Restricting to $l=2$, we have 
\begin{equation}
    \mathcal{Y}_{22} = \sqrt{\frac{15}{32\pi}}
    \begin{bmatrix}
    1 & -i & 0\\
    -i & -1 & 0 \\
    0 & 0 & 0
    \end{bmatrix}
    \text{ and }
    \mathcal{Y}_{20} = \sqrt{\frac{5}{16\pi}}
    \begin{bmatrix}
    -1 & 0 & 0\\
    0 & -1 & 0 \\
    0 & 0 & 2
    \end{bmatrix},
\end{equation}
together with $N_2 = 8\pi /15$. 

We are most interested in the $l=2$ quadrupole, which can be written as
\begin{align}
    \frac{Q^{\langle i j \rangle}_{\rm ns}}{MR^2} 
    &= N_2 \sum_m
    \left(
    \sum_{a,\omega_a>0}^{m_a=m} 2{\rm Re}\left[{\mathcal{Y}_{2m_a}^{\langle ij \rangle}}^\ast I_a c_a \right]
    + \sum_{ab, \omega_a>0}^{m_a+m_b=-m}{\rm Re}\left[{\mathcal{Y}_{2m}^{\langle ij \rangle}}^\ast J_{ab2m} c_a^\ast c_b^\ast \right]
    \right).
    \label{eq:Qlm2Qij}
\end{align}
The first term describes the linear contribution from $l_a=2$ modes and the second piece corresponds to the non-linear correction (note that $a$ runs over only positive-frequency modes while $b$ runs over both signs of frequencies). 

Eq.~(\ref{eq:Qlm2Qij}) can be especially helpful for us to see the connection between the modal expansion used in our analysis and the Love number, $k_2$, which is commonly used by the GW community. Formally, $k_2$ is defined by 
\begin{equation}
    Q_{ij}^{\rm ns} = -\frac{2}{3}k_2R^5 \mathcal{E}_{ij},
    \label{eq:love_def}
\end{equation}
where 
\begin{equation}
    \mathcal{E}_{ij} = -M' \partial_{ij} \frac{1}{r}
\end{equation}
is the tidal potential. 
We note  \citep{Poisson:14}
\begin{equation}
    \partial_{i_1...i_l}\frac{1}{r}=\partial_{\langle i_1...i_l \rangle}\frac{1}{r}=(-1)^l(2l-1)!! \frac{n^{\rm orb}_{\langle i_1...i_l \rangle}}{r^{l+1}}, 
\end{equation}
and
\begin{align}
    n^{\langle i_1...i_l \rangle}_{\rm orb}&=N_l\sum_m {\mathcal{Y}_{lm}^\ast}^{<i_1 ... i_l>}Y_{lm}^\ast(\pi/2, \phi) =\sum_m \frac{l!}{(2l-1)!!}W_{lm} e^{-im\phi} {\mathcal{Y}_{lm}^\ast},
    \label{eq:n_iL}
\end{align}
where we have used $W_{lm}\equiv 4\pi(2l+1)^{-1} Y_{lm}(\pi/2, 0)$ and the fact that the binary motion is in the $\theta=\pi/2$ plane. 
Using Eq.~(\ref{eq:Qlm2Qij}) and grouping terms with the same $\mathcal{Y}_{lm}^{\langle ij \rangle}$, we can find the love number for each $m$
\begin{equation}
    k_{2m}=\frac{2\pi}{5}\frac{M'}{M}\left(\frac{r}{R}\right)^3 \frac{1}{W_{lm}}
    \left(\sum_{a}^{m_a=m} I_a C_a + 
    \frac{1}{2}\sum_{ab}^{m_a+m_b=-m} J_{ab2m} C_a^\ast C_b^\ast\right). 
\end{equation}
One can further plug in the leading-order solution of $C_a$ as described in Sections~\ref{sec:lin_sol} and \ref{sec:mode_nl} to obtain an effective Love number (for each harmonic $m$; see also \citealt{Andersson:20, Passamonti:22}). Using the linear, adiabatic solution of $C_a$, we have 
\begin{equation}
    k_{2m}=k_2=\frac{4\pi}{5}\sum_{a, \omega_a>0}^{m_a=m} I_a^2.
\end{equation}
Note that in the linear, adiabatic limit, the values of $k_{2m}$ are the same for different $m$'s and can be collectively denoted by a single number $k_2$. Note further that the summation over modes is strongly dominated by the $l_a=2$ f-mode (Table~\ref{tab:coup_coeff}). Using $I_a=0.32$, we find $k_2=0.26$ in the linear, adiabatic limit, which agrees well with the expected value~\citep{Poisson:14}.

The relations summarized in this Appendix will also be useful for computing the Burke-Throne dissipation terms following \citet{Flanagan:08}. In particular, the interaction between the orbital and tidal quadrupole modifies the PP Burke-Throne terms in three ways.

First, in eq. (6a) of \citet{Flanagan:08}, there will be a term arising from the quadrupole of the NS, 
\begin{equation}
    g^i_{\rm (gw, ns)}=-\frac{2}{5}r_j\frac{d^5}{dt^5}Q^{\langle ij \rangle}_{\rm ns},
\end{equation}
where 
$
    \vect{r} = r\vect{n}_{\rm orb} = r\left[\cos\phi, \sin\phi, 0 \right]. 
$

We compute the total quadrupole of the NS in the Cartesian coordinate in terms of each mode’s contribution using Eq.~(\ref{eq:Qlm2Qij}). To evaluate the temporal derivatives of $Q_{\rm ns}^{\langle ij\rangle}$, we note
\begin{align}
    &\frac{\diff^5 c_a}{\diff t^5} = \frac{\diff^5}{\diff t^5} [C_a\exp(-im\phi)]\simeq -i (m\Omega)^5 C_a \exp(-im\phi), \\
    &\frac{\diff^5 (c_a^\ast c_b^\ast)}{\diff t^5} = \frac{\diff^5}{\diff t^5} [C_a^\ast C_b^\ast\exp(-im\phi)]\simeq -i (m\Omega)^5 C_a^\ast C_b^\ast \exp(-im\phi), 
\end{align}
where in the second line we have used $m_a+m_b+m=0$ as required by the angular selection rule. We have dropped terms that are smaller than the dominant one by $\mathcal{O}\left(t_{\rm gw} \dot{\phi} \right)$. Here $t_{\rm gw}\equiv r/\dot{r}$ is the characteristic timescale for GW induced orbital decay. We then convert the Cartesian $f^i_{\rm gw, ns}$ back to spherical coordinates, leading to 
\begin{align}
    g_{\phi}^{\rm (gw, ns)}&\simeq 
    -\frac{128}{5}\sqrt{\frac{2\pi}{15}}M R^2 r \Omega^5 \sum_{m=\pm 2}
    \left(
        \sum_{a, \omega_a>0}^{m_a=m}I_a{\rm Re}\left[C_a\right] 
        +\frac{1}{2}\sum_{ab, \omega_a>0}^{m_a+m_b=-m} J_{ab2m}{\rm Re}\left[C_a C_b\right]
    \right). 
\end{align}
If we use the convention of Section~\ref{sec:mode_nl} and use  $(a, b, c)$ to specifically denote the positive-frequency modes with $(m_a, m_b, m_c)=(2, -2, 0)$ and $l_a=l_b=l_c=2$, we can further simplify the second term in the parenthesis as
\begin{align}
    &\frac{1}{2}\sum_{m=\pm2}\sum_{ab, \omega_a>0}^{m_a+m_b=-m} J_{ab2m} {\rm Re}\left[C_a C_b\right] = 2 J_2 {\rm Re}\left[C_a C_c + C_b C_c\right] 
     \simeq \frac{4\omega_a^2}{\omega_a^2-4\Omega^2}J_{2} W_{22}W_{20}\left(\frac{M'}{M}\right)^2 I_a^2 R^6 \frac{\Omega^4}{M_{\rm t}^2}.
\end{align}

Second, the tidal back-reaction modifies the derivatives of the orbital quadrupole, defined as
\begin{equation}
    Q_{\rm orb}^{\langle ij \rangle} = \mu r^2 n^{\langle ij \rangle}_{\rm orb} =  \frac{2}{3} \sum_m W_{2m} Q_{2m} {\mathcal{Y}_{2m}^{\langle ij \rangle}}^\ast,
    \label{eq:Q_orb}
\end{equation}
where we have used Eq.~(\ref{eq:n_iL}) and defined $Q_{2m}=\mu r^2 \exp[-i m \phi]$. 
To compute its derivatives, we keep replacing the derivatives of $r$, $\phi$, and $C_a$ by the conservative parts of their equation of motion~\citep{Flanagan:08}, 
\begin{align}
    \ddot{r} &\to r\dot{\phi}^2-\frac{M+M'}{r^2} + g_r^{\rm (tide)}, \\
    r\ddot{\phi} &\to -2\dot{r}\dot{\phi} + g_\phi^{\rm (tide)}.\\
    \dot{C}_a&\to -i (\omega_a-m_a\dot{\phi}) C_a + i\omega_a \left[\frac{M'}{M}W_{lm}\left(\frac{R}{r}\right)^{l+1} \left(I_a + \sum_{b,lm}J_{ablm}C_b^\ast\right)
    + \sum_{bc}\kappa_{abc}C_b^\ast C_c^\ast\right] 
\end{align}
Of particular interest is the appearance of the $g_r^{\rm (tide)}$ term [Eq.~(\ref{eq:ar_tide})], which modifies the $r-\Omega$ relation of the orbit [see also Section~\ref{sec:eq_config} and Eq.~(\ref{eq:r_vs_omega})]. As a result, in addition to the PP terms given by Eqs.~(\ref{eq:a_BT_r_pp}) and (\ref{eq:a_BT_phi_pp}), we need to add additional corrections given by
\begin{align}
    g_{\phi}^{\rm (gw, br)}\simeq &-\frac{96}{5} M M'\left(\frac{R}{r}\right)^2 \Omega^3 
        \sum_a^{\omega_a>0}\left[W_{l_a m_a}I_a {\rm Re}[C_a] 
        + \left(\frac{1}{2}\sum_{b,lm} W_{lm}J_{ablm}{\rm Re}[C_a C_b]\right)
        \right].\\
    g_{r}^{\rm (gw, br)}\simeq &0
\end{align}
It is interesting to note that 
\begin{equation}
    g_\phi^{\rm (gw, pp)} + g_\phi^{\rm (gw, br)} = -\frac{32}{5}\mu r^3\Omega^5,
    \label{eq:f_gw_pp_br_summed}
\end{equation}
a form one would intuitively expect. Note that here $r=r(\Omega)$ is given by the modified $r-\Omega$ relation in Eq.~(\ref{eq:r_vs_omega}). 

Lastly, the Burke-Thorne force also acts on the modes. To derive its expression, we can first consider the acceleration $\vect{a}^{\rm (gw)}$ it induces on a perturbed fluid element at $\vect{x'}=\vect{x} + \vect{\xi}$. First we note that 
\begin{equation}
    \vect{a}_{\rm gw}(\vect{x}') = \vect{a}_{\rm gw}(\vect{x}) + \vect{\xi}\cdot \nabla \vect{a}_{\rm gw}(\vect{x}).  
\end{equation}
Furthermore, 
\begin{equation}
    a^i_{\rm gw}(\vect{x}) = -\frac{2}{5} x^j \frac{\diff ^5}{\diff t^5} Q_{\rm orb}^{\langle ij \rangle}.
\end{equation}
To proceed, we first decompose the orbital quadrupole into tensor spherical harmonics using Eq.~(\ref{eq:Q_orb}) and note
\begin{align}
    x^j {\mathcal{Y}_{2m}^{\langle ij \rangle}}^\ast = (x^i x^j)_{;i} {\mathcal{Y}_{2m}^{\langle ij \rangle}}^\ast 
    = \left(x^2 n^{\langle ij \rangle } {\mathcal{Y}_{2m}^{\langle ij \rangle}}^\ast \right)_{;i}
    = (x^2 Y_{2m})_{;i},
\end{align}
where the semicolon symbol stands for covariant derivative and we have used the identities $n^{ij} {\mathcal{Y}_{2m}^{\langle ij \rangle}}^\ast = n^{\langle ij \rangle} {\mathcal{Y}_{2m}^{\langle ij \rangle}}^\ast = Y_{lm}$~\citep{Poisson:14}.
We are now ready to write
\begin{align}
    \vect{a}_{\rm gw}(\vect{x}') = -\frac{2}{15}\sum_m W_{2m} \left[\nabla \left(x^2 Y_{2m} \right) + \left(\vect{\xi}\cdot \nabla \right) \nabla \left(x^2 Y_{2m} \right) \right] \frac{\diff^5 Q_{2m}}{\diff t^5}.
\end{align}
Its effect on each mode can be obtained by first contracting $\vect{a}_{\rm gw}$ with $\vect{\xi}^\ast$ and then integrating over $\rho \diff^3 x $~\citep{Schenk:02}. 
We thus have (for $l_a=2$)
\begin{equation}
    \dot{c}_a + i\omega_a c_a = i\omega_a \left[\left(\text{conservative terms}\right) + Z_a \right],
\end{equation}
where 
\begin{equation}
    Z_a = -\frac{2}{15} W_{2m_a} \frac{R^3}{M}\left(I_a \frac{\diff ^5 Q_{2m_a}}{\diff t^5} + \sum_{b}^{m_a+m_b+m=0} J_{ab2m} c_b^\ast \frac{\diff ^5 Q_{2m}^\ast}{\diff t^5} \right).
\end{equation}
One can verify that when the non-linear tide piece is ignored, our result reduces to eq. (6b) of \citet{Flanagan:08}.  This can be seen by directly contracting both sides of eq. (6b) of \citet{Flanagan:08} with $\mathcal{Y}_{lm}$ and using the identity $\mathcal{Y}_{lm'}^\ast \mathcal{Y}_{lm} = \delta_{mm'}/N_l$~\citep{Thorne:80}. The result follows by further plugging the linear part of $Q_{lm}^{\rm ns}$ [Eq.~(\ref{eq:Qlm})] into the left-hand side of eq. (6b) of \citet{Flanagan:08}. 

Eq.~(\ref{eq:f_gw_pp_br_summed}) suggests that we have
\begin{equation}
    \frac{\diff^5 Q_{2m}}{\diff t^5} \simeq  -i(m\Omega)^5\mu r^2 e^{-im\phi}.
\end{equation}
We thus have
\begin{equation}
    Z_a =\begin{cases}
    i \frac{2}{15}W_{22}\frac{M'}{M_{\rm t}} \left(\frac{R}{r}\right)^3 (m_a r \Omega)^5 \left( I_a + \sum_b^{m_b=0} J_{ab2-m_a} c_b^\ast \right) e^{-im\phi},  \text{ for } m_a=\pm 2, \\
    i \frac{2}{15}W_{20}\frac{M'}{M_{\rm t}} \left(\frac{R}{r}\right)^3 (m_b r \Omega)^5  \left(\sum_b^{m_b=-m=\pm 2} J_{ab2-m_b} C_b^\ast \right), \text{ for } m_a=0.
    \end{cases}
\end{equation}

The effect of $Z_a$ is to create an imaginary part in $C_a$, 
\begin{equation}
    {\rm Im}\left[C_a\right] \simeq \frac{\omega_a}{\omega_a - m\Omega} {\rm Im}\left[Z_a e^{im\phi}\right], 
\end{equation}
which then leads to a tangential tidal acceleration due to mode $a$
\begin{equation}
    g_\phi^{({\rm gw},a)} \simeq -\frac{4}{15} m_a^6 W_{lm_a}^2 I_a^2 M' R^3 \left(\frac{R}{r}\right)^2
    \frac{\omega_a}{\omega_a-m_a\Omega} \Omega^5 + \text{(non-linear terms)}.
    \label{eq:f_phi_Za}
\end{equation}
We thus see that while $m_a=2$ and $m_a=-2$ modes have opposite signs for $Z_a$, their contributions to the orbital decay add coherently. It is also easy to show that the $m_a=0$ mode does not contribute to the orbital decay via this channel even at the non-linear order we are considering. 

Besides terms due to the interaction between tidal and orbital quadrupoles, there is also damping on the $f$ mode due to its quadrupole beating with itself. This leads to an additional term in $Z_a$ given by
\begin{equation}
    Z_a^{\rm (mode)} = -\frac{8\pi}{75}I_a^2 R^5 \frac{d^5 c_a}{dt^5}
    \simeq i\frac{8\pi}{75} W_{l_a m_a} I_a^3 \frac{M'}{M}\left(\frac{R}{r}\right)^3 \frac{\omega_a}{\omega_a - m_a \Omega} (m_a R \Omega)^5.
    \label{eq:Z_a_Qmode}
\end{equation}
This term is smaller than Eq.~(\ref{eq:Z_a_Qorb}) by a factor of $\mathcal{O}\left(R/r\right)^5$, which is smaller than the leading-order non-linear effects we consider that corrects the linear solution at the $(R/r)^3$ order. We thus ignore its effect in our discussions. Nonetheless, this term can be amplified if the mode is close to resonance with the orbit due to NS rotation and/or orbital eccentricity. See the discussions in Section~\ref{sec:discussion}.

\section{Four-mode coupling without the Cowling approximation}
\label{appx:kap4_del_Phi}
We can break the four-mode coupling into seven pieces \citep{VanHoolst:94}
\begin{equation}
    \eta_{abcd} = -\frac{{\rm I} + {\rm II} + {\rm III} + {\rm IV} + {\rm V} + {\rm VI} + {\rm VII}}{6E_0},
\end{equation}
where terms I-V are provided in appendix C \citet{Weinberg:16}. Here we compute terms VI and VII which are due to perturbed gravity. For the coupling among f-modes, we find that the perturbed gravity terms (VI and VII) are crucial as they can modify the results obtained under the Cowling approximation by $\sim 70\%$. In this appendix specifically, we will use $r$ to denote the radial coordinate of a fluid element inside the NS. It should not be confused with the orbital separation as we will only consider an isolated NS here. 

The first perturbed gravity term we need to evaluate is, 
\begin{align}
    {\rm VI} =& -\int \diff^3 x \rho \left[
    \xi_a^i \xi_b^j\left(\int \diff^3 x'\rho(x') \xi_c^{k'} \xi_d^{s'}|\vect{x}-\vect{x'}|^{-1}_{;k's'}\right)_{;ij} + \right. \nonumber \\
    &+
    \left.
    \xi_a^i \xi_c^j\left(\int \diff^3 x'\rho(x') \xi_b^{k'} \xi_d^{s'}|\vect{x}-\vect{x'}|^{-1}_{;k's'}\right)_{;ij} + 
    \xi_a^i \xi_d^j\left(\int \diff^3 x'\rho(x') \xi_b^{k'} \xi_c^{s'}|\vect{x}-\vect{x'}|^{-1}_{;k's'}\right)_{;ij} 
    \right],
\end{align}
where ``$;$'' stands for covariant derivative and a quantity with a primed index means it is evaluated with respect to $\vect{x}'$. We can expand 
\begin{equation}
    \frac{1}{|\vect{x}-\vect{x}'|}=\sum_{lm}\tilde{r_l}(x, x')Y_{lm}^\ast (\theta', \phi') Y_{lm}(\theta, \phi), 
\end{equation}
where 
\begin{equation}
    \tilde{r_l}(r, r')=\frac{4\pi}{2l+1}\times 
    \begin{cases}
    &\frac{r'^l}{r^{l+1}} \quad \text{if $r'\leq r$},\\
    &\frac{r^l}{r'^{l+1}} \quad \text{if $r'>r$}.
    \end{cases}
\end{equation}

Following \citet{Weinberg:12}, we use a covariant basis with vectors $\epsilon_i=h_i e_i$, where $h_r=1$, $h_\theta=r$, $h_\phi=r\sin\theta$, and $e_{r, \theta, \phi}$ are unit vectors along the $r, \theta, \phi$ directions. The nonzero components of the metric are $(g_{rr}, g_{\theta\theta}, g_{\phi\phi})=(1, r^2, r^2\sin^2\theta)$. The Lagrangian displacement vector for an eigenmode can be written as\footnote{Note that we use $\xi_a^i$ to indicate the $i$ component of $\vect{\xi}_a$. In the coordinate we consider, the coordinate index corresponds to $i=(r, \theta, \phi)$. On the other hand, we use $a_r$ and $a_h$ to indicate the radial and tangential component of the Lagrangian displacement and the subscripts $r$ and $h$ do not corresponds to coordinate indices.} 
\begin{equation}
    \vect{\xi}_a = \left[\xi_a^r, \xi_a^\theta, \xi_a^\phi\right]=\left[a_rY_a, \frac{a_h}{r}\frac{\partial Y_a}{\partial \theta}, \frac{a_h}{r\sin^2\theta}\frac{\partial Y_a}{\partial \phi}\right]. 
\end{equation}

Consider a specific harmonic, and focus on the inner integral first (i.e., primed coordinate). 
We have terms like (in the right hand side, all terms are evaluated in the primed coordinate)
\begin{align}
    & \xi_c^{r'} \xi_d^{r'} (\tilde{r_l} Y_{lm}^\ast)_{;r'r'} =  
      c_r d_r\left(\frac{\partial^2}{\partial r'^2}\tilde{r_l}\right)  Y_c Y_d Y_{lm}^\ast, \\
    & \xi_c^{r'} \xi_d^{\theta'}  (\tilde{r_l} Y_{lm}^\ast)_{;r'\theta'} = 
      c_r d_h
      \left(
        \frac{1}{r'}\frac{\partial \tilde{r_l}}{\partial r'} - \frac{\tilde{r_l}}{{r'}^2}
      \right) 
      Y_c \frac{\partial Y_d}{\partial \theta'} \frac{Y_{lm}^\ast}{\partial \theta'}, \\
    & \xi_c^{r'} \xi_d^{\phi'} (\tilde{r_l} Y_{lm}^\ast)_{;r'\phi'} = 
      c_r d_h
      \left(
        \frac{1}{r'}\frac{\partial \tilde{r_l}}{\partial r'} - \frac{\tilde{r_l}}{{r'}^2}
      \right) 
      \frac{Y_c}{\sin^2 \theta'}  \frac{\partial Y_d}{\partial \phi'} \frac{\partial Y_{lm}^\ast}{\partial \phi'}, \\
    & \xi_c^{\theta'} \xi_d^{\theta'} (\tilde{r_l} Y_{lm}^\ast)_{;\theta'\theta'} = 
       c_h d_h 
       \left(
        \frac{\tilde{r_l}}{r'^2}
        \frac{\partial Y_c}{\partial \theta'} \frac{\partial Y_d}{\partial \theta'} 
        \frac{\partial Y_{lm}^\ast}{\partial \theta'^2} 
        + \frac{1}{r'}\frac{\partial \tilde{r_l}}{\partial r'} 
        \frac{\partial Y_c}{\partial \theta'} \frac{\partial Y_d}{\partial \theta'} 
        Y_{lm}^\ast
      \right),\\
    & \xi_c^{\theta'} \xi_d^{\phi'} (\tilde{r_l} Y_{lm}^\ast)_{;\theta'\phi'} = 
        c_h d_h \frac{\tilde{r_l}}{r'^2}  \frac{1}{\sin^2 \theta'} 
        \frac{\partial Y_c}{\partial \theta'}\frac{\partial Y_d}{\partial \phi'}
        \left(
          \frac{\partial^2 Y_{lm}^\ast}{\partial \theta'\partial \phi'}
         -\frac{\cos\theta'}{\sin\theta'}\frac{\partial Y_{lm}^\ast}{\partial \phi'}
        \right), \\
    & \xi_c^{\phi'} \xi_d^{\phi'} (\tilde{r_l} Y_{lm}^\ast)_{;\phi'\phi'} = 
      c_h d_h 
      \frac{\partial Y_c}{\partial \phi'}\frac{\partial Y_d}{\partial \phi'}
      \left(
        \frac{\tilde{r_l}}{r'^2} \frac{1}{\sin^4 \theta'}
        \left(\frac{\partial^2 Y_{lm}^\ast}{\partial \phi'^2}
             +\sin\theta'\cos\theta' \frac{\partial Y_{lm}^\ast}{\partial \theta}
        \right)
        +\frac{1}{r'}\frac{\partial \tilde{r_l}}{\partial r'} 
         \frac{1}{\sin^2 \theta'}Y_{lm}^\ast
      \right).
\end{align}
Thus 
\begin{align}
    &\int \diff^3 x' \rho(x') c^{k'}d^{s'}|\vect{x}-\vect{x'}|^{-1}_{;k's'} 
    =  \sum_{l}^{m=m_c+m_d} (-1)^m\frac{4\pi}{2l+1} Y_{lm}(\theta, \phi) \left[{\rm VI}^{<r}_{cd}(r) + {\rm VI}^{>r}_{cd}(r)\right],
\end{align}
where we have used $Y_{lm}^\ast = (-1)^m Y_{l-m}$, and the angular selection role requires $m=m_c+m_d=-(m_a+m_b)$. 
The radial part is defined as
\begin{align}
    &{\rm VI}^{<r}_{cd}(r) = r^{-(l+1)}\int_0^r \diff r' r'^l \rho(r')
    \left[
    l(l-1) c_{r'}d_{r'} T_{cdl-m} + (l-1) c_{r'}d_{h'} F_{c, dl-m} + (l-1) c_{h'}d_{r'} F_{d, cl-m}
    \right. \nonumber \\
    &\hspace{4.4cm}\left.
    + c_{h'}d_{h'} \left(G_{l-m,cd}+lF_{l-m, cd}\right)
    \right], \\
    &{\rm VI}^{>r}_{cd}(r) = r^l \int_r^R \diff r' r'^{-(l+1)} \rho(r') 
    \left[
    (l+1)(l+2) c_{r'}d_{r'} T_{cdl-m} - (l+2) c_{r'}d_{h'} F_{c, dl-m} - (l+2) c_{h'}d_{r'} F_{d, cl-m}
    \right. \nonumber \\
    &\hspace{4.4cm}\left.+ c_{h'}d_{h'} \left(G_{l-m,cd}-(l+1)F_{l-m, cd}\right)
    \right].
\end{align}
We further define ${\rm VI}_{cd}(r) ={\rm VI}^{<r}_{cd}(r) + {\rm VI}^{>r}_{cd}(r).$
The angular parts have been integrated following \citet{Weinberg:12}
\begin{align}
    &T_{abc} = \int \diff \Omega Y_a Y_b Y_c, \\
    &F_{a, bc} = \int \diff \Omega Y_a \nabla^i Y_b \nabla_j Y_c = \frac{T_{abc}}{\Lambda_b^2 + \Lambda_c^2 - \Lambda_a^2}, \\
    &G_{a, bc} = \int \diff \Omega g^{ik}g^{js}\nabla_i \nabla_j Y_a \nabla_k Y_b \nabla_s Y_c = \frac{T_{abc}}{4}[\Lambda_a^4-(\Lambda_b^2 - \Lambda_c^2)^2],
\end{align}
where $\Lambda_a^2 = l_a(l_a+1)$. Paired subscripts not separated by a comma are symmetric in those indices.

The outer integral can be evaluated similarly, 
\begin{align}
    &\int \diff^3 x \rho \xi_a^i \xi_b^j\left(\int \diff^3 x'\rho(x') \xi_c^{k'} \xi_d^{s'}|\vect{x}-\vect{x'}|^{-1}_{;k's'}\right)_{;ij} \nonumber \\
    =&\sum_l^{m=m_c+m_d} (-1)^m \frac{4\pi}{2l+1} \int \diff x^3 \rho \xi_a^i\xi_b^j\left[{\rm VI}_{cd}(r)Y_{lm}\right]_{;ij} \nonumber \\
    =&\sum_{l}^{m=m_c+m_d}(-1)^m\frac{4\pi}{2l+1}\int \diff r r^2 \rho(r)  
    \Bigg{[} 
     a_r b_r \left(\frac{\partial^2 {\rm VI}_{cd} }{\partial r^2}\right) T_{ablm}
    + a_r b_h \left(
        \frac{1}{r}\frac{\partial {\rm VI}_{cd}}{\partial r} - \frac{{\rm VI}_{cd}}{{r}^2}
      \right) F_{a,blm}
    + a_h b_r \left(
        \frac{1}{r}\frac{\partial {\rm VI}_{cd}}{\partial r} - \frac{{\rm VI}_{cd}}{r^2}
      \right) F_{b,alm}  \nonumber \\
   & \hspace{3.63cm} +   a_h b_h \left(\frac{{\rm VI}_{cd}}{r^2} G_{lm, ab} 
                    + \frac{1}{r}\frac{\partial{\rm VI}_{cd}}{\partial r} F_{lm,ab} \right)
    \Bigg{]}.
\end{align}

The other perturbed gravity term we need to evaluate is
\begin{equation}
    {\rm VII} = \int \diff^3 x \rho \left[
      \xi_a^i \xi_b^j \xi_c^k \delta \Phi_{d; ijk} 
    + \xi_a^i \xi_b^j \xi_d^k \delta \Phi_{c; ijk}
    + \xi_a^i \xi_c^j \xi_d^k \delta \Phi_{b; ijk}
    + \xi_b^i \xi_c^j \xi_d^k \delta \Phi_{a; ijk}
    \right],
\end{equation}
where $\delta \Phi_{d}$ is the Eulerian perturbation of the gravitational potential induced by mode $d$ and it is given by $\delta \Phi_d(r, \theta, \phi) = \delta \phi_d(r) Y_d(\theta, \phi)$. 

We will have terms
\begin{align}
    \left(\delta \phi_d Y_d\right)_{;rrr} &= \frac{\partial^3 \delta \phi_d}{\partial r^3} Y_d,
    \\
    \left(\delta \phi_d Y_d\right)_{;rr\theta}&=
    \left(\frac{\partial^2 \delta \phi_d}{\partial r^2} 
          - \frac{2}{r}\frac{\partial \delta \phi_d}{\partial r} 
          + \frac{2}{r^2}\delta \phi_d\right)
    \frac{\partial Y_d}{\partial \theta}
    \\
    \left(\delta \phi_d Y_d\right)_{;rr\phi}&=
    \left(\frac{\partial^2 \delta \phi_d}{\partial r^2} 
         - \frac{2}{r}\frac{\partial \delta \phi_d}{\partial r} 
         + \frac{2}{r^2}\delta \phi_d\right)
    \frac{\partial Y_d}{\partial \phi}
    \\
    \left(\delta \phi_d Y_d\right)_{;r\theta\theta}
    &=r \frac{\partial^2\delta \phi_d}{\partial r^2} Y_d 
    + \frac{\partial \delta \phi_d}{\partial r} 
      \left(
        \frac{\partial^2 Y_d}{\partial \theta^2} - Y_d
      \right)
    - \frac{2}{r}\delta \phi_d \frac{\partial^2 Y_d}{\partial \theta^2}
    \\
    \left(\delta \phi_d Y_d\right)_{;r\theta\phi}
    &=\frac{\partial \delta \phi_d}{\partial r}
      \left(
        \frac{\partial^2 Y_d}{\partial \theta\partial \phi}
      - \frac{\cos\theta}{\sin\theta}\frac{\partial Y_d}{\partial \phi}    
      \right)
    + \frac{2\delta \phi_d}{r}
      \left(
        -\frac{\partial^2 Y_d}{\partial \theta \partial \phi} 
        +\frac{\cos\theta}{\sin\theta}\frac{\partial Y_d}{\partial \phi}
      \right)
    \\
    \left(\delta \phi_d Y_d\right)_{;r\phi\phi}
    &=r\frac{\partial^2\delta \phi_d }{\partial r^2}\sin^2\theta Y_d
     + \frac{\partial \delta \phi_d}{\partial r}
      \left(
        \frac{\partial^2 Y_d}{\partial \phi^2} 
        + \sin \theta \cos\theta \frac{\partial Y_d}{\partial \theta}
        - \sin^2\theta Y_d
      \right)
     - \frac{2\delta \phi_d}{r}
       \left(
           \frac{\partial^2 Y_d}{\partial \phi^2}
         + \sin\theta\cos\theta \frac{\partial Y_d}{\partial \theta}
       \right)
    \\
    \left(\delta \phi_d Y_d\right)_{;\theta\theta\theta}&=
      3 r \frac{\delta \phi_d}{\partial r}\frac{\partial Y_d}{\partial \theta}
      + \delta\phi_d
      \left(\frac{\partial^3 Y_d}{\partial \theta^3} - 2\frac{\partial Y_d}{\partial \theta}\right) 
    \\
    \left(\delta \phi_d Y_d\right)_{;\theta\theta\phi}&=
      r \frac{\partial \delta \phi_d}{\partial r}\frac{\partial Y_d}{\partial \phi}
    + \delta \phi_d 
    \left(\frac{\partial^3 Y}{\partial \theta^2 \partial \phi} 
        - 2 \frac{\cos\theta}{\sin\theta}\frac{\partial^2 Y}{\partial \theta \partial \phi} 
        + 2\frac{\cos^2\theta}{\sin^2\theta}\frac{\partial Y}{\partial \phi}\right)
    \\
    \left(\delta \phi_d Y_d\right)_{;\theta\phi\phi}&=
      r\frac{\partial \delta \phi_d}{\partial r}\sin^2\theta \frac{\partial Y}{\partial \theta}
      + \delta \phi_d
        \left(
          \frac{\partial^3 Y_d}{\partial \theta \partial^2 \phi} 
        + \sin\theta\cos\theta \frac{\partial^2 Y}{\partial \theta^2} 
        - 2\frac{\cos\theta}{\sin\theta}\frac{\partial^2Y}{\partial \phi^2}
        -\frac{\partial Y_d}{\partial \theta}
        \right)
    \\
    \left(\delta \phi_d Y_d\right)_{;\phi\phi\phi}&=
      3 r\frac{\partial \delta \phi_d}{\partial r}\sin^2\theta \frac{\partial Y}{\partial \phi}
      + \delta \phi_d 
      \left(
        \frac{\partial^3 Y_d}{\partial \phi^3} 
      + 3\sin\theta \cos\theta \frac{\partial^2 Y_d}{\partial \theta \partial \phi}
      - 2\frac{\partial Y_d}{\partial \phi} 
      \right)
\end{align}

Because the background is Euclidean, covariant derivatives commute and the results are symmetric with respect to permutations of indices. 
Thus
\begin{align}
     &\int d\Omega \xi_a^i \xi_b^j \xi_c^k \delta \Phi_{d; ijk} 
     \nonumber \\
     &= a_r b_r c_r \frac{\partial^3 \delta \phi_d}{\partial r^3} T_{abcd} \\
     &+\Big{[}
       a_r b_r c_h F_{ab, cd}^{(2)}
     + a_r b_h c_r F_{ac, bd}^{(2)} 
     + a_h b_r c_r F_{bc,ad}^{(2)}  
     \nonumber \\ 
     &\quad\ 
      + a_r b_h c_h F_{ad, bc}^{(2)}
      + a_h b_r c_h F_{bd, ac}^{(2)}
      + a_h b_h c_r F_{cd, ab}^{(2)}
     \Big{]}\frac{1}{r}\frac{\partial^2 \delta \phi_d}{\partial r^2}\\
     &+\Big{[}
       - 2 a_r b_r c_h F_{ab,cd}^{(2)}
       - 2 a_r b_h c_r F_{ac,bd}^{(2)} 
       - 2 a_h b_r c_r F_{bc,ad}^{(2)}  
       \nonumber \\
     &\quad\ 
          + a_r b_h c_h \left(S_{a, bc, d} - (\Lambda_d^2+1)F_{ad, bc}^{(2)} \right)
          + a_h b_r c_h \left(S_{b, ac, d} - (\Lambda_d^2+1)F_{bd, ac}^{(2)} \right)   
          + a_h b_h c_r \left(S_{c, ab, d} - (\Lambda_d^2+1)F_{cd, ab}^{(2)} \right) 
      \nonumber \\
     &\quad + a_h b_h c_h \left(G_{ab,cd}^{(22)} + G_{ac,bd}^{(22)} + G_{ad,bc}^{(22)}\right)
     \Big{]}\frac{1}{r^2}\frac{\partial \delta \phi_d}{\partial r}
     \\
     &+\Big{[}
          2 a_r b_r c_h F_{ab, cd}^{(2)}
        + 2 a_r b_h c_r F_{ac, bd}^{(2)}
        + 2 a_h b_r c_r F_{bc, ad}^{(2)}
     \nonumber \\
     &\quad -2 a_r b_h c_h \left(S_{a, bc, d}-\Lambda_d^2 F_{ad,bc}^{(2)}\right)
            -2 a_h b_r c_h \left(S_{b, ac, d}-\Lambda_d^2 F_{bd,ac}^{(2)}\right)
            -2 a_h b_h c_r \left(S_{c, ab, d}-\Lambda_d^2 F_{cd,ab}^{(2)}\right)
    \nonumber \\
     &\quad\ + a_h b_h c_h R_{abc, d}
     \Big{]} \frac{\delta \phi_d}{r^3},
\end{align}
where we have defined the following angular integrals~\citep{Weinberg:16}, 
\begin{align}
    & f_{ab}^{(1)} = Y_aY_b,\ f_{ab}^{(2)}=\nabla Y_a\cdot \nabla Y_b, \ f_{ab}^{(3)} = \nabla_i\nabla^j Y_a \nabla_j \nabla^i Y_b, \\
    & F_{ab,cd}^{(i)}=\int \diff \Omega Y_a Y_b f_{cd}^{(i)}, \\
    & G_{ab,cd}^{(i)}=\int \diff \Omega f_{ab}^{(i)} f_{cd}^{(i)}, \\
    & S_{a,bc,d}=\int \diff \Omega Y_a \nabla_i Y_b \nabla^j Y_c \nabla_j \nabla^i Y_d + \Lambda_d^2 F_{ad,bc}^{(2)}, 
\end{align}
together with a new integral we introduce in this work,
\begin{align}
    R_{abc, d}&=\int \diff \Omega \nabla^i Y_a \nabla^j Y_b \nabla^k Y_c \nabla_i\nabla_j\nabla_k Y_d.  
\end{align}

\bsp	
\label{lastpage}
\end{document}